\documentclass[aps,prd,10pt,showpacs,amsmath,amssymb,nofootinbib]{revtex4-1}
\usepackage[latin1]{inputenc}
\usepackage{axodraw4j}
\usepackage{color}
\usepackage{graphicx}
\usepackage{dcolumn}
\usepackage{bm}
\usepackage{slashed}

\begin{document}

\def\k{{\mathbf k}}
\def\q{{\mathbf q}}
\def\P{{\mathbf P}}
\def\K{{\mathbf K}}
\def\L{{\mathbf L}}
\def\x{{\mathbf x}}
\def\r{{\mathbf r}}

\title{Dipole factorization for DIS at NLO I:\\
 Loop correction to the $\gamma^*_{T,L}\rightarrow q\bar{q}$ light-front wave-functions}
\author{Guillaume Beuf}
\email{beuf@ectstar.eu}
\affiliation{European Centre for Theoretical Studies in Nuclear Physics and Related Areas (ECT*)
and Fondazione Bruno Kessler, Strada delle Tabarelle 286, \\
 I-38123 Villazzano (TN), Italy\\}

\begin{abstract}
The one-loop QCD corrections to the light-front wave-function for the quark-antiquark Fock state inside a transverse or longitudinal off-shell photon are explicitly calculated, both in full momentum space and in mixed space (a.k.a. dipole space).
These results provide one of the main contributions to virtual NLO corrections to many DIS observables (inclusive or not) in the dipole factorization formalism at low Bjorken $x$.

In a follow-up article, these one-loop corrections are combined with earlier results on the wave-function for the quark-antiquark-gluon Fock state, in order to get the full set of NLO corrections to the DIS structure functions $F_2$ and $F_L$ in the dipole factorization formalism, valid at low Bjorken $x$.
\end{abstract}


\maketitle


\section{Introduction}

At low Bjorken $x$ ($x_{Bj}$), various deep inelastic scattering (DIS) observables can be studied using the dipole factorization formalism, in particular inclusive DIS structure functions \cite{Bjorken:1970ah,Nikolaev:1990ja} and diffractive DIS structure functions \cite{Nikolaev:1991et} as well as exclusive vector meson vector meson production \cite{Kopeliovich:1991pu} and deeply virtual Compton scattering. This formalism is motivated by light-front perturbation theory \cite{Kogut:1969xa,Bjorken:1970ah}, which provides the light-front wave-functions (LFWF) for the fluctuation of the exchanged photon into a quark-antiquark dipole, and relies on the eikonal approximation, which allows to describe the interaction of the quark-antiquark dipole on the target, in the high-energy limit relevant for DIS at low $x_{Bj}$.
The dipole factorization has a remarkable versatility. Not only does it allow one to study in a unified way inclusive, diffractive and exclusive observables, but it allows one to include and study various dynamical effects from QCD. For that reason, a large part of the theoretical and phenomenological work in the literature related to the results (at low $x_{Bj}$) from the HERA collider are based on the dipole factorization.

Pushing further the dipole picture, it has been possible
to rederive in a more intuitive way \cite{Mueller:1993rr,Mueller:1994jq,Mueller:1994gb} the BFKL equation \cite{Lipatov:1976zz,Kuraev:1977fs,Balitsky:1978ic}, which allows one to resum large logarithms arising in perturbation theory for high-energy (or equivalently low $x_{Bj}$) semi-hard scattering processes between two dilute objects in QCD, within the leading logarithmic approximation (LL).
Moreover, thanks to the use of the eikonal approximation, it is trivial to include coherent multiple scattering effects within the dipole factorization, via Wilson lines, in the case of a dense target. When such effects are relevant, one enters into the gluon saturation regime of QCD \cite{Gribov:1984tu,Mueller:1985wy,McLerran:1993ni,McLerran:1993ka,McLerran:1994vd}, also called Color Glass Condensate (CGC).
In the presence of gluon saturation effects, the BFKL equation has to be replaced by a nonlinear evolution in order to perform the high-energy LL resummation, which is the B-JIMWLK evolution \cite{Balitsky:1995ub, Jalilian-Marian:1997jx,Jalilian-Marian:1997gr,Jalilian-Marian:1997dw,
Kovner:2000pt,Weigert:2000gi,Iancu:2000hn,Iancu:2001ad,Ferreiro:2001qy}, or in a mean-field approximation the Balitsky-Kovchegov (BK) equation \cite{Balitsky:1995ub,Kovchegov:1999yj,Kovchegov:1999ua}.
Using the dipole factorization in conjunction with the BK equation (modified to include running QCD coupling effects \cite{Balitsky:2006wa,Kovchegov:2006vj}), it has been possible to obtain successful fits \cite{Albacete:2009fh,Albacete:2010sy,Lappi:2013zma} of the dipole-target amplitude on the DIS data from HERA, which can then be used for to make predictions for many observables in ep, eA, pp, pA and AA collisions at high energy.

All the works discussed so far have been performed in the leading order (LO) approximation, supplemented by LL resummation. However, in order to make precise quantitative studies in QCD, one should go one step further in perturbation theory, including next-to-leading order (NLO) contributions, preferably together with the resummation of next-to-leading logarithms (NLL) at high energy.
The NLO corrections to DIS have been studied in refs. \cite{Balitsky:2010ze,Beuf:2011xd,Balitsky:2012bs}. For the single-inclusive hadron production at forward rapidity in pA collisions within the so-called hybrid factorization formalism \cite{Dumitru:2001jn,Dumitru:2002qt,Dumitru:2005gt}, the NLO corrections have been calculated\footnote{However, the first complete NLO results \cite{Chirilli:2011km,Chirilli:2012jd} for that observable led to absurd numerical results \cite{Stasto:2013cha}. Then, it has been understood \cite{Kang:2014lha,Altinoluk:2014eka,Ducloue:2016shw} that the high-energy resummation was not done in a consistent way in the first calculation, leading to an oversubtraction of large logs. Moreover, it was suggested \cite{Altinoluk:2014eka,Watanabe:2015tja,Ducloue:2016shw} to switch from the usual high-energy LL resummation scheme formulated with $k^+$ ordering to another scheme with $k^-$ (also dubbed Ioffe time) ordering instead, presumably more suitable for that observable. With these two modifications, one obtains an improvement of the numerical results \cite{Watanabe:2015tja,Ducloue:2016shw}, but some problems remain and require further investigations.} in refs. \cite{Chirilli:2011km,Chirilli:2012jd}
(after some partial results in refs. \cite{Dumitru:2005gt,Altinoluk:2011qy}), including contributions from massless quarks and gluons. The massive quark contributions to the NLO corrections for that observable have also been calculated \cite{Altinoluk:2015vax}.

The NLL extensions of the BFKL equation \cite{Fadin:1998py,Ciafaloni:1998gs}, the BK equation \cite{Balitsky:2008zz,Balitsky:2009xg} and even the B-JIMWLK evolution \cite{Balitsky:2013fea,
Kovner:2013ona,Kovner:2014lca,Caron-Huot:2015bja} are now known\footnote{Note that a first insight into the NNLL resummation has been recently provided in ref. \cite{Caron-Huot:2016tzz}.}.
However, these evolution equations include large NLL corrections making them unstable. Those large corrections require a so-called collinear resummation, as done for example in ref. \cite{Salam:1998tj} in the case of the BFKL equation. Gluon saturation alone cannot cure the instabilities, as can be observed in numerical studies \cite{Avsar:2011ds,Lappi:2015fma}, so that collinear resummation is required as well for the B-JIMWLK and BK equations beyond strict LL accuracy. The dominant part of the collinear resummation (which amounts to improve the treatment of kinematics) has been performed for the BK equation within two different schemes in refs. \cite{Beuf:2014uia,Iancu:2015vea} (see also the earlier preliminary study \cite{Motyka:2009gi}). A prescription has also been proposed \cite{Iancu:2015joa} in the case of the BK equation for the sub-dominant part of the collinear resummation. Taking all this into account, it was found that the NLL BK becomes numerically stable \cite{Lappi:2016fmu}. It was also possible redo \cite{Iancu:2015joa,Albacete:2015xza} the fits to HERA data using the two partially collinear-resummed versions of the BK equation, but still without the full NLL contributions, for simplicity.

The NLO corrections to DIS within the dipole factorization are of two types. On the one hand, there is a new contribution coming from the fluctuation of the incoming virtual photon into a $q\bar{q}g$ Fock state, which then scatters on the target. On the other hand, there is the loop correction to the LFWF describing the splitting of the photon into a $q\bar{q}$ Fock state. Only the corrections of the first type, including the tree-level $\gamma^*_{T,L}\rightarrow q\bar{q}g$ LFWFs, have been calculated explicitly in refs. \cite{Balitsky:2010ze,Beuf:2011xd}, whereas the corrections of the second type have been guessed through different prescriptions.
In ref. \cite{Balitsky:2010ze} the calculations are done in covariant perturbation theory, and the results are presented in a very general way, but difficult to use in practice, due to the many integrations left to perform in order to obtain the DIS structure functions. Subsequently, these results have been reformulated in the case of DIS on a dilute target and presented in a $k_\perp$-factorized form suitable for BFKL phenomenology at NLO and NLL accuracy \cite{Balitsky:2012bs}.
By contrast, in ref. \cite{Beuf:2011xd}, the calculations are done in light-front perturbation theory, and the results are provided directly for DIS structure functions in the dipole factorization formalism, allowing in principle to extend the BK-based fits \cite{Albacete:2009fh,Albacete:2010sy,Lappi:2013zma,Iancu:2015joa,Albacete:2015xza}
to NLO accuracy. Due to the different formalisms used to make the calculations and to present the results, it has not been possible so far to compare the results from refs. \cite{Balitsky:2010ze,Balitsky:2012bs} and the ones from ref. \cite{Beuf:2011xd}. It has not yet been possible either to compare any of these with the earlier calculations of the photon impact factor at NLO \cite{Bartels:2000gt,Bartels:2001mv,Bartels:2002uz,Bartels:2004bi}, relevant for BFKL physics.
However, the results of \cite{Beuf:2011xd} for the $\gamma^*_{T,L}\rightarrow q\bar{q}g$ LFWFs have been confirmed by the calculation (within covariant perturbation theory) of the 3-jet production in DIS \cite{Boussarie:2014lxa}.

Unfortunately, I realized that the unitarity-based prescription that I used in \cite{Beuf:2011xd} to guess the $q\bar{q}$ contribution
to the NLO correction out of the $q\bar{q}g$ contribution is wrong, as explained in section \ref{sec:unitarity_rel} of the present paper. Instead the loop correction to the $\gamma^*_{T,L}\rightarrow q\bar{q}$ LFWFs have to be calculated explicitly (at least in the light-front perturbation theory formalism used in  \cite{Beuf:2011xd} and in the present study). The aim of the present article is to present and explain that calculation.

The loop corrections to the $\gamma^*_{T,L}\rightarrow q\bar{q}$ LFWFs have both UV and low $x$ (or more precisely low $k^+$) divergences.
The low $x$ divergence is there to induce part of the high-energy B-JIMWLK leading logs at the cross-section level.
By contrast, the UV divergence has to cancel against other contributions at the cross-section level,  because UV renormalization is irrelevant at the order of interest here\footnote{Indeed, the QED coupling is the only parameter appearing in the tree-level LFWF and thus potentially renormalized by NLO corrections. However, one-loop QCD corrections do not affect the running of the QED coupling.}.
The calculation is performed using dimensional regularization for the UV divergences and a $k^+$ cutoff regularization for the low $x$ divergence.
By comparison to earlier fixed-order NLO calculations in the gluon saturation/CGC context in light-front perturbation theory, the present calculation represents a significant step in complexity due to the appearance of various nontrivial transverse integrals and Dirac structures. It is found convenient to use tensor reduction techniques such as the Passarino-Veltman method \cite{Passarino:1978jh} in order to handle the transverse integrals.

The final results for the $\gamma^*_{T,L}\rightarrow q\bar{q}$ LFWF at one loop is given in momentum space by the eqs. \eqref{WF_NLO_T_1}, \eqref{VT_result_2} and \eqref{NT_result} in the transverse photon case, and by the eqs. \eqref{WF_NLO_L_1} and \eqref{VL_result_2} in the longitudinal photon case, and in mixed space (a.k.a. dipole space) by the eqs. \eqref{FT_WF_gamma_L_2} and \eqref{mixed_VL_result} in the longitudinal case and by the eqs. \eqref{FT_WF_gamma_T_2} and \eqref{mixed_VT_result}  in the transverse case.

In order to obtain the full NLO results for the DIS cross section, the $q\bar{q}$ contributions described by the LFWFs calculated here have to be combined with the $q\bar{q}g$ calculated in \cite{Beuf:2011xd}. However, both contributions have UV divergences which have to cancel each other. But the calculation has been done here in dimensional regularization, whereas the $q\bar{q}g$ contribution has been given only in $4$ dimensions in ref. \cite{Beuf:2011xd}, without an explicit UV regularization.
Hence, in the follow-up paper \cite{Beuf_paper_II}, the $\gamma^*_{T,L}\rightarrow q\bar{q}g$ LFWFs are recalculated in $D$ dimensions, and are combined with the results of the present paper. After demonstrating the cancellation of the UV divergences as well as of the accompanying regularization-scheme-dependent finite terms
between the $q\bar{q}$ and $q\bar{q}g$ contributions, the full NLO corrections to the $F_T$ and $F_L$ DIS structure functions are given in ref. \cite{Beuf_paper_II} in the dipole factorization form.

Note that the one-loop $\gamma^*_{T,L}\rightarrow q\bar{q}$ LFWFs calculated in the present article not only contribute to NLO corrections to DIS structure functions, but also provide an important subset of the NLO corrections to any DIS observable (inclusive, diffractive or exclusive)
\footnote{A few comments are in order concerning collinear divergences. First, LFWFs like the ones studied in the present paper describe the dynamics of the projectile only in the range $x^+<0$, before it reaches the target. Hence, such LFWFs can be in principle subject to initial-state collinear divergences, but not to final-state ones. For fully inclusive observables, like DIS structure functions, all the final-state dynamics (at $x^+>0$) decouples, thanks to the optical theorem. By contrast, less inclusive observables are typically sensitive to final-state dynamics, to be encoded in final-state LFWFs, different from the initial-state ones considered here, and which can be subject to final-state collinear divergences.
Second, for the photon (or lepton) initial-state LFWFs relevant for DIS, gluon loops or gluon emissions cannot lead to initial-state collinear divergences, since the only external leg is colorless. Indeed, the integral over the transverse momentum of the gluon will always be regulated in the IR by the $q\bar{q}$ relative transverse momentum (as well as by $Q^2$).
Third, initial-state collinear physics is nevertheless present. Indeed, the UV divergences encountered in the NLO calculations for DIS structure functions arise in the regime in which the first splitting ($\gamma\rightarrow q\bar{q}$) is much more collinear than the subsequent gluon emission. Hence, these UV divergences are related to the DGLAP evolution of the photon and not to UV renormalization. Note however that the photon DGLAP logs would become large and require resummation in the case of DIS structure functions only when $Q^2$ is much smaller than the saturation scale of the target. But in that photoproduction regime, our perturbative calculations should not be reliable anyway.}
which obeys a dipole factorization at low $x_{Bj}$. Moreover, all of the calculations related to DIS at NLO performed so far, in refs. \cite{Balitsky:2010ze,Beuf:2011xd,Balitsky:2012bs} and in the present paper and its follow-up \cite{Beuf_paper_II}, have considered massless quarks only. The extension of the present calculation to the case of massive quarks is possible within the same formalism, but it is left for further studies.

The plan of the present article is as follows.
The unitarity argument used in ref. \cite{Beuf:2011xd} (in section II.D there) is discussed in section \ref{sec:unitarity_rel}, and shown to be flawed.
Then, the tree-level expressions for the $\gamma^*_{T,L}\rightarrow q\bar{q}$ LFWFs are recalled in section \ref{sec:LO_WF}, in arbitrary space-time dimension $D$.
The one-loop self-energy graph for a quark or antiquark belonging to an arbitrary off-shell Fock state is calculated in section \ref{sec:self_energy}, since it appears several times as a subgraph for the one-loop $\gamma^*\rightarrow q\bar{q}$ LFWFs.
The full calculation of the $\gamma^*\rightarrow q\bar{q}$ LFWF at one loop in momentum space is presented in section \ref{sec:NLO_WF_T} for the transverse photon case and in section \ref{sec:NLO_WF_L} for the longitudinal photon case.
The section \ref{sec:FT_NLO_WF} is devoted to the Fourier transform to mixed space (a.k.a. dipole space) of the results of the two previous sections, leading to the main result of the present paper: the mixed-space expression for the one-loop LFWFs for the $q\bar{q}$ Fock sector inside a longitudinal or transverse virtual photon.

Supplemental material is provided in appendices. An elementary introduction to light-front perturbation theory is given in the appendix \ref{sec:formalism}. Some relations about Dirac spinors are derived in appendix \ref{sec:spinor_bilin}. The appendix \ref{sec:num_calc} provides details about the calculations of the numerators of the light-front diagrams encountered in the present paper. An introduction to the Passarino-Veltman tensor reduction method is given in appendix \ref{sec:PassVelt}.
The transverse-momentum scalar integrals, left after applying the Passarino-Veltman method, are studied and calculated in appendix \ref{sec:scalInt}. The appendix  \ref{sec:integrals} lists the integrals encountered when performing the $k^+$ integration in the loop. Finally, the integrals required to perform the Fourier transform to mixed-space of the LFWFs are calculated in appendix \ref{sec:FT_int}.

\emph{Note:} After the completion of the present calculation, I became aware of the independent work presented simultaneously in \cite{OtherToAppear}, in which are calculated NLO corrections to exclusive diffractive dijet production in DIS. The loop contributions calculated in the present paper provide in principle part of the NLO corrections calculated in \cite{OtherToAppear}. Note that the calculations in  \cite{OtherToAppear} are performed in covariant perturbation theory instead of light-front perturbation theory in the present article.


\section{On the problems with the unitarity prescription of ref. \cite{Beuf:2011xd} \label{sec:unitarity_rel}}

In order to understand what is the issue with the prescription used in the section II.D of ref. \cite{Beuf:2011xd}, it is sufficient to consider a simpler setup: the decomposition of the dressed state of an incoming real photon on a basis of Fock states. The Fock states are taken in momentum space, and the photon has a momentum $q$ and polarization $\lambda$. Following the formalism presented in appendix \ref{sec:formalism} (see also ref. \cite{Bjorken:1970ah}), one has
\begin{eqnarray}
|\gamma(\underline{q},\lambda)_H\rangle
&=& \sqrt{Z_{\gamma}} \bigg[
a_{\gamma}^{\dagger}(\underline{q},\lambda)\, |0\rangle
+\sum_{{\cal F} \neq \gamma} |{\cal F}\rangle\; \Psi_{\gamma\rightarrow {\cal F}}
\bigg]
\nonumber\\
&=& \sqrt{Z_{\gamma}} \bigg[
a_{\gamma}^{\dagger}(\underline{q},\lambda)\, |0\rangle
+\sum_{l \bar{l}\textrm{ states}}  \Psi_{\gamma\rightarrow l_0 \bar{l}_1} \;\; b_{l}^{\dagger}(\underline{k_0},h_0)\, d_{l}^{\dagger}(\underline{k_1},h_1)\, |0\rangle
\nonumber\\
&& \hspace{1.2cm}
+\sum_{q \bar{q}\textrm{ states}}  \Psi_{\gamma\rightarrow q_0 \bar{q}_1} \;\; b^{\dagger}(\underline{k_0},h_0,\alpha_0)\, d^{\dagger}(\underline{k_1},h_1,\alpha_1)\, |0\rangle
\nonumber\\
& & \hspace{1.2cm}
+\sum_{q \bar{q}g\textrm{ states}}  \Psi_{\gamma\rightarrow q_0 \bar{q}_1 g_2} \;\; b^{\dagger}(\underline{k_0},h_0,\alpha_0)\, d^{\dagger}(\underline{k_1},h_1,\alpha_1)\,\, a^{\dagger}(\underline{k_2},\lambda_2,a_2)\, |0\rangle
+\;\cdots
\bigg]
\, ,
\label{phys_photon_decomp_mom}
\end{eqnarray}
where, in the second step, the sum over all Fock states $|{\cal F}\rangle$ different from the one-photon Fock state has been split into contributions with a specific particle content: lepton-antilepton pair, quark-antiquark pair, quark-antiquark-gluon and so on. The terms left implicit in eq. \eqref{phys_photon_decomp_mom} appear only at higher order in perturbation theory.

The dressed photon state is normalized as
\begin{eqnarray}
\langle\gamma(\underline{q}',\lambda')_H
|\gamma(\underline{q},\lambda)_H\rangle
&=& (2q^+)(2\pi)^{D\!-\!1}\delta^{(D\!-\!1)}(\underline{q}'\!-\!\underline{q})\;
\delta_{\lambda',\lambda}
\label{phys_photon_norm}
\, .
\end{eqnarray}
From that relation as well as the orthonormality of the Fock state basis, one gets the unitarity relation
\begin{eqnarray}
(2q^+)(2\pi)^{D\!-\!1}\delta^{(D\!-\!1)}(\underline{q}'\!-\!\underline{q})\;
\delta_{\lambda',\lambda}\;
\frac{(1\!-\!Z_{\gamma})}{Z_{\gamma}}
&=&
{\sum_{l \bar{l}\textrm{ states}}}
\left({\Psi}_{\gamma'\rightarrow l_0 \bar{l}_1}\right)^{\dag} \;
{\Psi}_{\gamma\rightarrow l_0 \bar{l}_1}
 +{\sum_{q \bar{q}\textrm{ states}}}
 \left({\Psi}_{\gamma'\rightarrow q_0 \bar{q}_1}\right)^{\dag} \;
{\Psi}_{\gamma\rightarrow q_0 \bar{q}_1}
\nonumber\\
& & 
+{\sum_{q \bar{q}g\textrm{ states}}}  \left({\Psi}_{\gamma'\rightarrow q_0 \bar{q}_1 g_2}\right)^{\dag} \;
{\Psi}_{\gamma\rightarrow q_0 \bar{q}_1 g_2}
+O(\alpha_{em}\, \alpha_s^2)+O(\alpha_{em}^2)
\, ,
\label{phys_gluon_unitarity_rel}
\end{eqnarray}
where the $\gamma'$ labels signals LFWFs that corresponds to a photon state $(\underline{q}',\lambda')$ instead of $(\underline{q},\lambda)$. Due to transverse and light-front momentum conservation at each vertex, each term on the right-hand side of eq. \eqref{phys_gluon_unitarity_rel} contains an overall factor $(2q^+)(2\pi)^{D\!-\!1}\delta^{(D\!-\!1)}(\underline{q}'\!-\!\underline{q})$.

The unitarity relation \eqref{phys_gluon_unitarity_rel} can be written at any order in perturbation theory in both $\alpha_{em}$ and $\alpha_s$. In particular, the terms of order $\alpha_{em}\, \alpha_s$ exactly have to obey the relation
\begin{eqnarray}
(2q^+)(2\pi)^{D\!-\!1}\delta^{(D\!-\!1)}(\underline{q}'\!-\!\underline{q})\;
\delta_{\lambda',\lambda}\;
\Big(1\!-\!Z_{\gamma}\Big)_{\alpha_{em}\, \alpha_s}
&=&
 \left({\sum_{q \bar{q}\textrm{ states}}}
 \left({\Psi}_{\gamma'\rightarrow q_0 \bar{q}_1}\right)^{\dag} \;
{\Psi}_{\gamma\rightarrow q_0 \bar{q}_1}
\right)_{\alpha_{em}\, \alpha_s}
\nonumber\\
& & 
+\left({\sum_{q \bar{q}g\textrm{ states}}}
\left({\Psi}_{\gamma'\rightarrow q_0 \bar{q}_1 g_2}\right)^{\dag} \;
{\Psi}_{\gamma\rightarrow q_0 \bar{q}_1 g_2}
\right)_{\alpha_{em}\, \alpha_s}
\, .
\label{phys_gluon_unitarity_rel_pure_NLO}
\end{eqnarray}
Indeed, at this particular order, there cannot be contributions from the leptonic dipole Fock states.
The last term in eq. \eqref{phys_gluon_unitarity_rel_pure_NLO} contain only the tree-level contribution to the $q\bar{q}g$ Fock state LFWF, as calculated in ref. \cite{Beuf:2011xd}, but here inside an on-shell transverse photon.

The unitarity-based prescription used in ref. \cite{Beuf:2011xd} (see section II.D there) in order to obtain the loop correction to the $q\bar{q}$ Fock state LFWF amounts to using eq. \eqref{phys_gluon_unitarity_rel_pure_NLO} (or more precisely its mixed-space analog) while assuming that its left-hand side is identically zero.
A similar prescription had been used to obtain the $q\bar{q}$ virtual contributions to high-energy evolution equations, for example in refs. \cite{Mueller:1993rr,Balitsky:2008zz}. However, $(1\!-\!Z_{\gamma})$ receives a finite contribution at order $\alpha_{em}\, \alpha_s$, with neither UV nor low $x$ divergences. This does not invalidate the use of such a prescription in the case of high-energy evolution equations, where only low $x$ divergent terms are relevant. However, the nonzero contribution to $(1\!-\!Z_{\gamma})$ at order $\alpha_{em}\, \alpha_s$ implies that the unitarity prescription of ref. \cite{Beuf:2011xd} is flawed.
In practice, the unitarity relation \eqref{phys_gluon_unitarity_rel} can be used only to obtain $Z_{\gamma}$ at higher orders from the direct calculation of the right-hand side but not to give relations between the contributions of different Fock states in the right-hand side.

As the one-photon Fock state does not contribute to the scattering on a classical gluon shock-wave target, the knowledge of $Z_{\gamma}$ is not needed when calculating high-energy scattering on a dense target with gluon saturation, making the relation \eqref{phys_gluon_unitarity_rel} useless in that case.
By contrast, the $q\bar{q}$ Fock state obviously contributes to the cross section, and higher-order corrections to its LFWF have to be calculated directly, which motivates the present paper.

The discussion presented in this section generalizes straightforwardly from the case of real photon scattering to DIS observables (via an intermediate virtual photon), following the explanations presented in the appendix A.3 of ref. \cite{Beuf:2011xd}.


\section{Quark-antiquark components at LO\label{sec:LO_WF}}

Before starting the calculation of the loop correction to the $\gamma^*_{T,L}\rightarrow q\bar{q}$ LFWFs, let us rederive the tree-level contribution in $D$ dimensions, for completeness, using light-front perturbation theory as presented in appendix \ref{sec:formalism}.

\begin{figure}
\setbox1\hbox to 10cm{
 \includegraphics{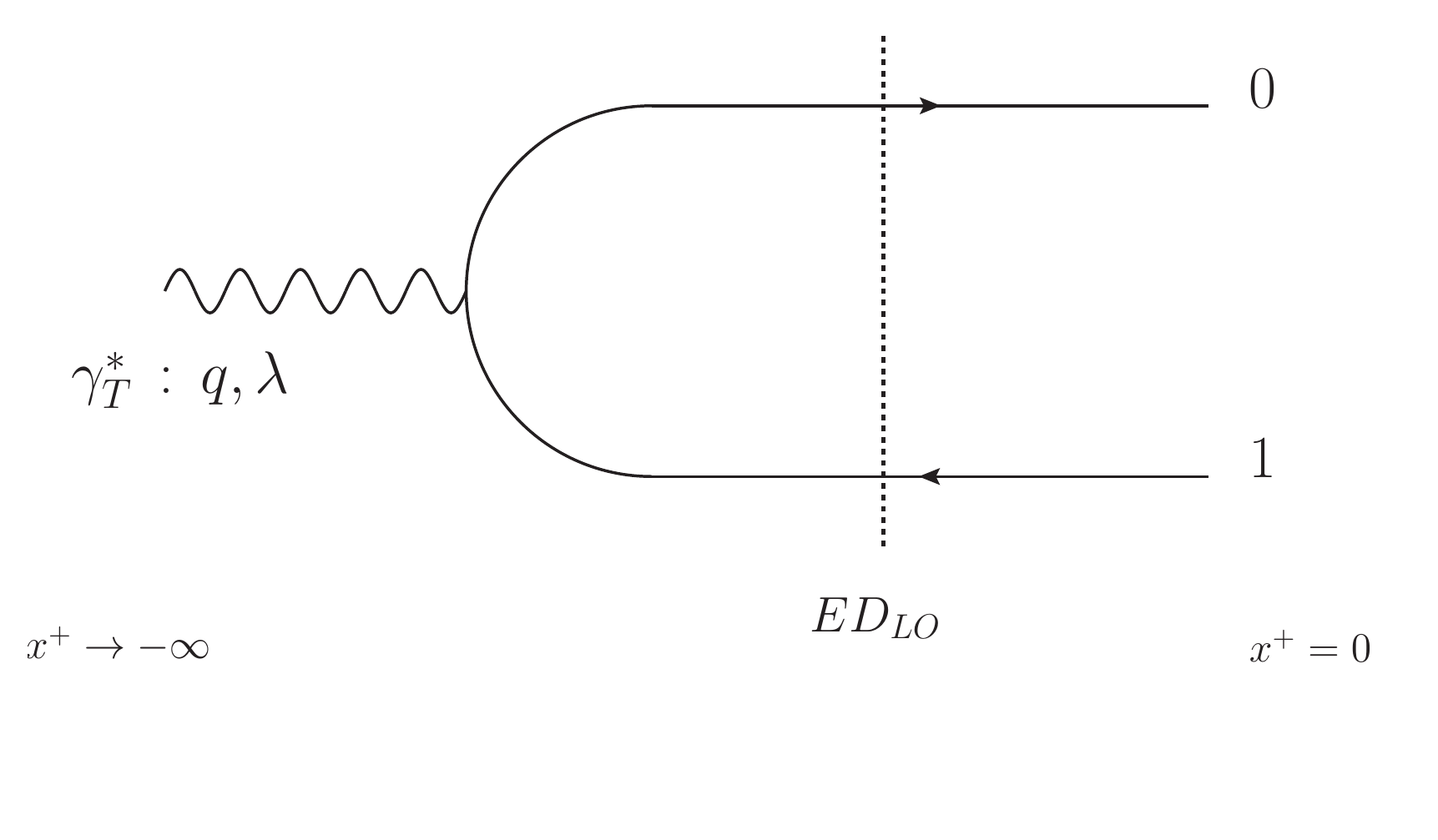}
}
\begin{center}
\hspace{-4cm}
\resizebox*{6cm}{!}{\box1}
\caption{\label{Fig:gammaT_LO}Tree-level contribution to the $q\bar{q}$ Fock component inside an incoming transverse photon.}
\end{center}
\end{figure}
For the transverse photon case, the relevant graph contributing to the LFWF is drawn in fig. \eqref{Fig:gammaT_LO}, and is written
\begin{eqnarray}
\Psi_{\gamma_T^{*}\rightarrow q_0+\bar{q}_1}^{LO}
&=&
\frac{
\langle 0| d_{1} b_0\, V_I(0)\, a_{\gamma}^{\dag} |0 \rangle
}{(ED_{LO})}
\nonumber\\
&=&
(2\pi)^{D-1}\delta^{(D-1)}(\underline{k_1}+\underline{k_{0}}\!-\!\underline{q})\;
\delta_{\alpha_{0} ,\,\alpha_{1}}\;
 \frac{\mu^{2-\frac{D}{2}}\, e\, e_f}{(ED_{LO})}\;\;\;
\overline{u}(0)\,\slashed{\epsilon}_{\lambda}\!(\underline{q})\;v(1)
\, .
\label{WF_T_LO}
\end{eqnarray}

Following the discussion in the appendix A.3 of ref. \cite{Beuf:2011xd}, it is possible to generalize the notion of LFWF to the case of an off-shell transverse photon, by assigning the off-shell value
\begin{eqnarray}
q^-&\equiv & \frac{(\q^2\!-\! Q^2)}{2q^+}
\, ,
\label{q_minus}
\end{eqnarray}
to the $q^-$ of this photon, appearing in each energy denominator for the perturbative expansion of the LFWF.

Hence, the energy denominator $(ED_{LO})$ appearing in eq. \eqref{WF_T_LO} is given by
\begin{eqnarray}
(ED_{LO})&\equiv & q^- -k_0^- -k_1^- +i \epsilon = \frac{(\q^2\!-\! Q^2)}{2q^+} - \frac{\k_0^2}{2k_0^+} - \frac{\k_1^2}{2k_1^+} +i \epsilon
\nonumber\\
&=& -\frac{ Q^2}{2q^+} - \frac{q^+}{2k_0^+ k_1^+}\; \left[\k_0\!-\!\frac{k_0^+}{q^+}\, \q \right]^2 +i \epsilon
\, .
\label{ED_LO_1}
\end{eqnarray}
Introducing the notation
\begin{eqnarray}
\P \equiv \k_0 -\frac{k_0^+}{q^+}\, \q = -\k_1 +\frac{k_1^+}{q^+}\, \q
\label{cv_k0_to_P}
\end{eqnarray}
for the transverse momentum of the quark relative to that of the photon, as well as
\begin{eqnarray}
\overline{Q}^2 \equiv \frac{k_0^+k_1^+}{(q^+)^2}\; {Q}^2
\, ,
\label{def_Qbar}
\end{eqnarray}
one obtains
\begin{eqnarray}
\frac{1}{(ED_{LO})}&= & - \left(\frac{2k_0^+ k_1^+}{q^+}\right)\;
   \frac{1}{\left[\P^2+\overline{Q}^2 -i \epsilon\right]}
\, .
\label{ED_LO_2}
\end{eqnarray}
In practice, only the case $Q^2>0$ will be considered, so that the $-i \epsilon$ can be dropped.

In light-front quantization, unphysical polarizations of the fields have been eliminated, in particular the longitudinal polarization of photons and gluons. They only appear implicitly inside nonlocal instantaneous vertices.
However, as argued in the appendix A.3 of ref. \cite{Beuf:2011xd}, in view of applications to DIS observables, it is possible and convenient to introduce an effective LFWF for longitudinal photons. At tree level, it is given by the graph analog to the one in fig. \eqref{Fig:gammaT_LO}, but with an incoming longitudinal photon, and an effective splitting vertex into $q\bar{q}$ given by
\begin{eqnarray}
V_{\gamma_L \rightarrow q_{0} \bar{q}_{1}}
&=& (2\pi)^{D-1}\delta^{(D-1)}(\underline{k_{0}}+\underline{k_{1}}\!-\!
\underline{q})\; \delta_{\alpha_{0},\, \alpha_{1}}\;
\mu^{2-\frac{D}{2}}\, e\, e_f\,  \frac{Q}{q^+}\;
\overline{u}(0)\,  \gamma^+\, v(1)
\, .
\label{gammaL_to_qqbar_vertex}
\end{eqnarray}
Then, one has
\begin{eqnarray}
\Psi_{\gamma_L^{*}\rightarrow q_0+\bar{q}_1}^{LO}
&=&
\frac{
V_{\gamma_L \rightarrow q_{0'} \bar{q}_{1'}}
}{(ED_{LO})}
\nonumber\\
&=&
(2\pi)^{D-1}\delta^{(D-1)}(\underline{k_1}+\underline{k_{0}}\!-\!\underline{q})\;
\delta_{\alpha_{0} ,\,\alpha_{1}}\;
 \frac{\mu^{2-\frac{D}{2}}\, e\, e_f}{(ED_{LO})}\;  \frac{Q}{q^+}\;\;\;
\overline{u}(0)\,\gamma^+\, v(1)
\, ,
\label{WF_L_LO}
\end{eqnarray}
with the energy denominator $(ED_{LO})$ given by eq. \eqref{ED_LO_2} again.


\section{Quark self-energy loop within an off-shell Fock state\label{sec:self_energy}}

As a further preliminary step, let us consider the one-loop self-energy type contribution for a quark belonging to a multiparticle off-shell Fock state.
That contribution shows up for example as a subgraph for the LFWF of $q\bar{q}$ inside either a transverse or longitudinal photon. More generally, this is a basic contribution which would appear in many calculations in QCD light-front perturbation theory at one loop or beyond. Hence, it is convenient to calculate this subgraph once for all, without specifying explicitly the rest of the diagram.

\begin{figure}
\setbox1\hbox to 10cm{
\includegraphics{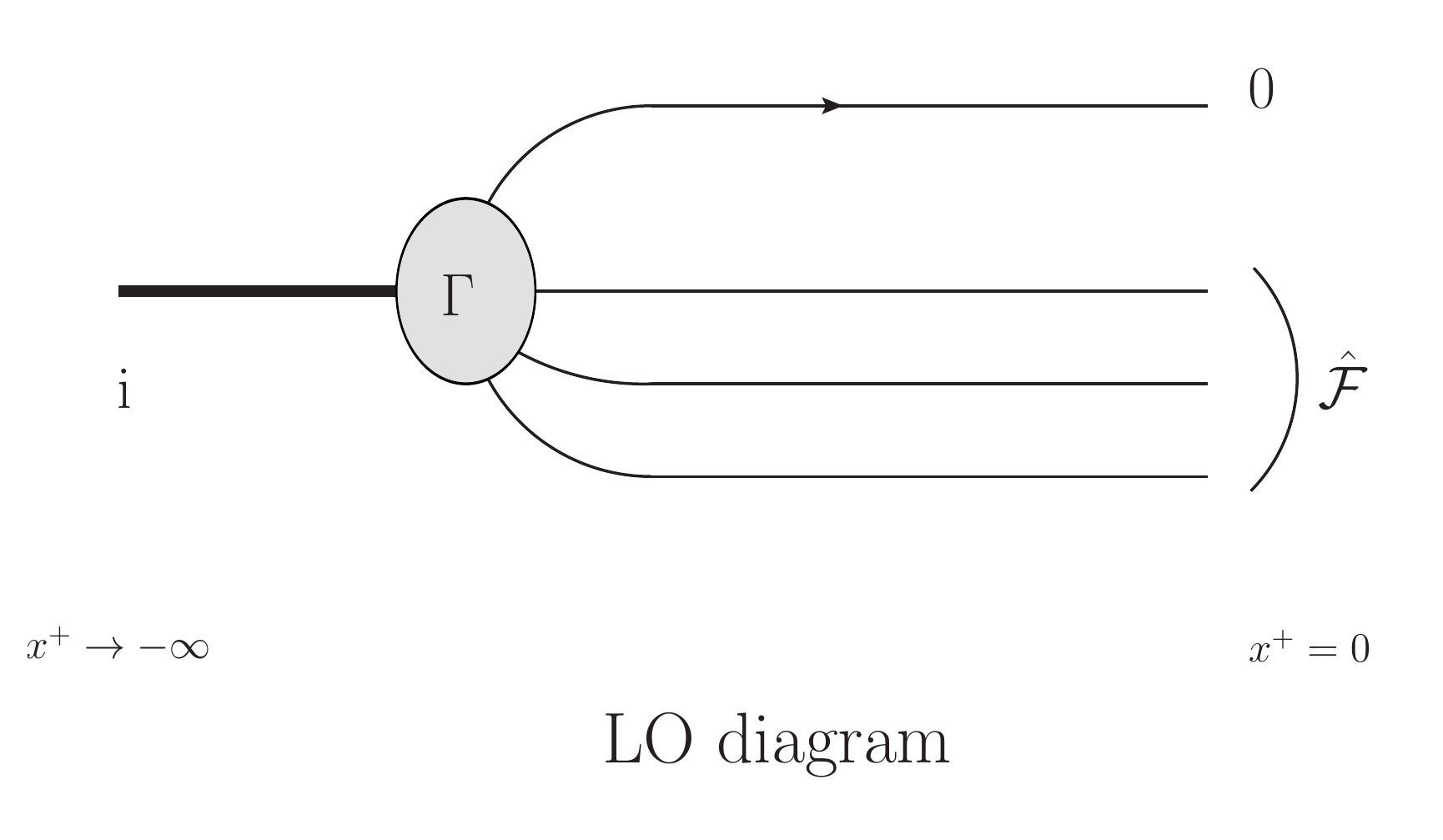}
}
\setbox2\hbox to 10cm{
\includegraphics{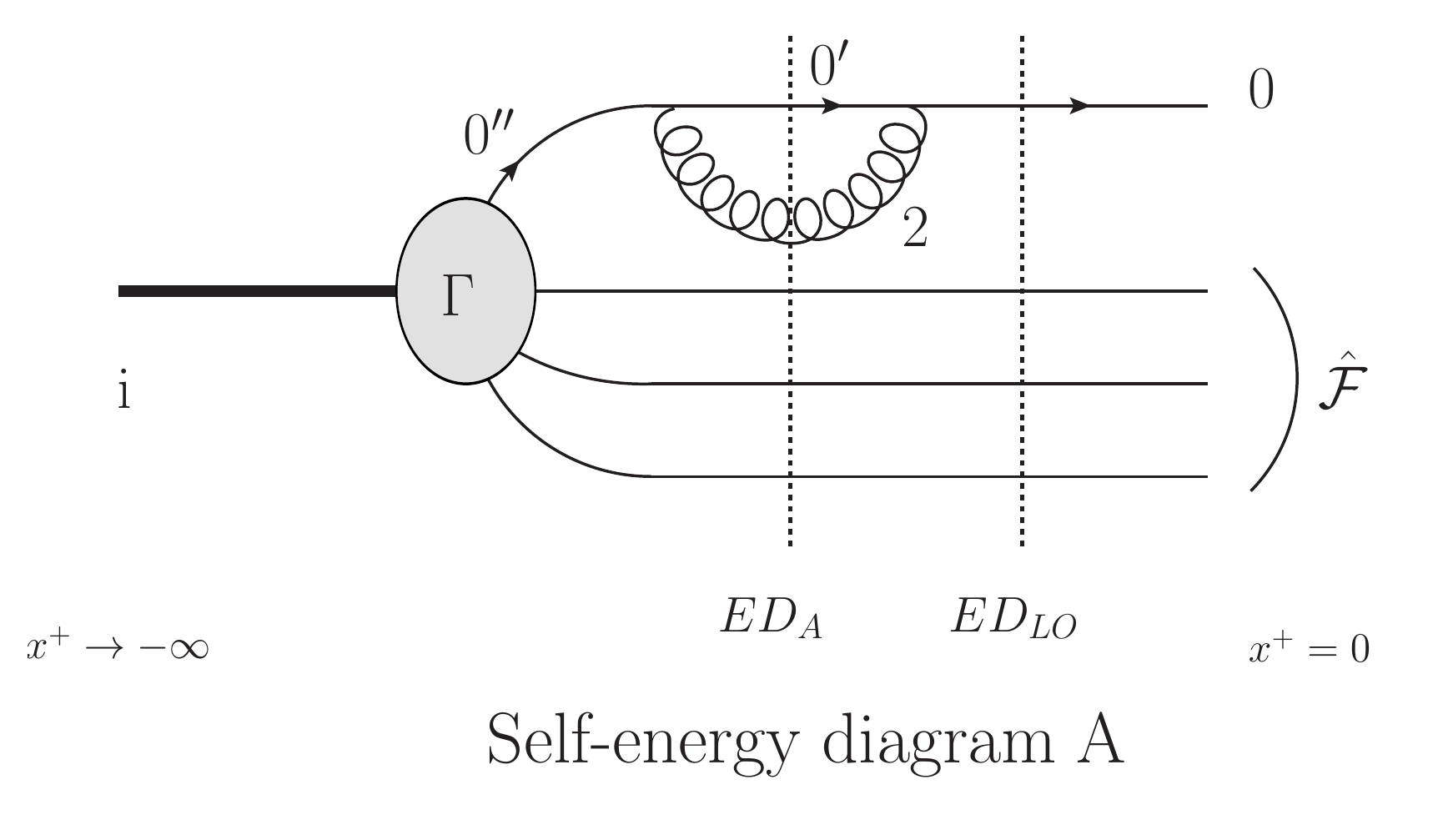}
}
\begin{center}
\resizebox*{10cm}{!}{\hspace{-7cm}\mbox{\box1 \hspace{9cm} \box2}}
\caption{\label{Fig:off_shell_SE} Left: LFWF (at a given order in perturbation theory) for a Fock state ${\cal F}$ inside a physical incoming state $i$. A quark is singled out from the Fock state ${\cal F}$, as ${\cal F}=q_0+\hat{\cal F}$.
Right: One-loop quark self-energy diagram correcting the contribution shown on the left.}
\end{center}
\end{figure}
Thus, let us consider the setup described in fig. \ref{Fig:off_shell_SE}. Let us call $\Gamma$ an arbitrary contribution (at some given finite order in perturbation theory) to the LFWF $\Psi_{i\rightarrow q_0+\hat{\cal F}}$, for a Fock state ${\cal F}=q_0+\hat{\cal F}$ including a quark, inside an incoming physical state $i$, as shown in the left-hand side of fig. \ref{Fig:off_shell_SE}. Then let us calculate the quark self-energy correction $A$ shown in the right-hand side of fig. \ref{Fig:off_shell_SE}.
In light-front perturbation theory, this diagram $A$
can be written as
\begin{eqnarray}
\Psi^A_{i\rightarrow q_0+\hat{\cal F}}= \sum_{q_{0''} \textrm{ states}}\;\;\;
 \sum_{q_{0'} g_2 \textrm{ states}}
\Psi^{\Gamma}_{i\rightarrow q_{0''}+\hat{\cal F}}\;
\frac{
\langle 0| b_0\, V_I(0)\, b_{0'}^{\dag} a_{2}^{\dag} |0 \rangle\;
\langle 0| a_{2} b_{0'}\, V_I(0)\, b_{0''}^{\dag} |0 \rangle
}{(ED_{LO})\; (ED_A)}
\, ,
\label{WF_A_1}
\end{eqnarray}
with the energy denominators $(ED_A)$ and $(ED_{LO})$ related to each other as
\begin{eqnarray}
(ED_A) &=& (ED_{LO}) -k_2^- -k_{0'}^-+k_0^-
= (ED_{LO}) -\frac{{\k_2}^2}{2\, k_2^+} -\frac{(\k_0\!-\!\k_2)^2}{2\, (k_0^+\!-\!k_2^+)}+\frac{{\k_0}^2}{2\, k_0^+}\nonumber\\
&=&(ED_{LO}) - \frac{k_0^+}{2\, k_2^+(k_0^+\!-\!k_2^+)}
\left[\k_2\!-\! \frac{k_2^+}{k_0^+}\, \k_0  \right]^2
\, .
\label{ED_A_vs_EDLO}
\end{eqnarray}
Note that one energy denominator $(ED_{LO})$ is implicitly included in $\Psi^{\Gamma}_{i\rightarrow q_{0}+\hat{\cal F}}$.
Using the result \eqref{ED_A_vs_EDLO} as well as the expressions for the $q\rightarrow qg$ and $qg\rightarrow q$ vertices from \eqref{q_qbar_g_vertices}, one obtains
\begin{eqnarray}
\Psi^A_{i\rightarrow q_0+\hat{\cal F}}
&=& \sum_{h_{0''}=\pm 1/2} \sum_{\alpha_{0''}}
 \int \frac{d^{D-1} \underline{k_{0''}}}{(2\pi)^{D-1} (2 k_{0''}^+)}\; \theta(k_{0''}^+)\;
\Psi^{\Gamma}_{i\rightarrow q_{0''}+\hat{\cal F}}\;\;\;
(2\pi)^{D-1}\delta^{(D-1)}(\underline{k_{0''}}\!-\!\underline{k_0})
\left(t^{a_2}t^{a_2}\right)_{\alpha_{0}\,\alpha_{0''}}\;
\frac{(\mu^2)^{2-\frac{D}{2}}\, g^2}{(ED_{LO})}\nonumber\\
&& \times
\int \frac{d^{D-1} \underline{k_2}}{(2\pi)^{D-1} (2 k_{2}^+)}\; \theta(k_{2}^+)\;
\int \frac{d^{D-1} \underline{k_{0'}}}{(2\pi)^{D-1} (2 k_{0'}^+)}\; \theta(k_{0'}^+)\;
\frac{(2\pi)^{D-1}\delta^{(D-1)}(\underline{k_2}+ \underline{k_{0'}}\!-\!\underline{k_{0}})\; \textrm{Num}_{A}}{
\left[(ED_{LO}) -  \frac{k_0^+}{2\, k_2^+(k_0^+\!-\!k_2^+)}
\left[\k_2\!-\! \frac{k_2^+}{k_0^+}\, \k_0  \right]^2\right]}
\, .
\label{WF_A_2}
\end{eqnarray}
The calculation of the numerator
\begin{eqnarray}
\textrm{Num}_{A}&=& \sum_{\textrm{phys. pol. }\lambda_2} \;\;\; \sum_{h_{0'}=\pm 1/2}  \overline{u}(0)\; \slashed{\epsilon}_{\lambda_2}\!(\underline{k_2})\; u(0')\;\;
 \overline{u}(0')\; \slashed{\epsilon}_{\lambda_2}^*\!(\underline{k_2})\; u(0'')\label{num_A_def}
\end{eqnarray}
is explained in detail in appendix \ref{sec:num_A}, and gives
\begin{eqnarray}
\textrm{Num}_{A}
&=& \delta_{h_{0''},\, h_0}\;\; \left\{4 \left(\frac{k_0^+}{k_2^+}\right)^2 +\frac{(D\!-\!2) k_0^+}{(k_0^+\!-\!k_2^+)} \right\}\; \left[\k_2\!-\! \frac{k_2^+}{k_0^+}\, \k_0  \right]^2
\, . \label{num_A_result}
\end{eqnarray}
The quarks $q_{0''}$ and $q_{0}$ are forced to have the same momentum $\underline{k_{0}}$ by momentum conservation at the vertices, the same helicity $h_0$ due to the expression \eqref{num_A_result} for the numerator $\textrm{Num}_{A}$, and the same color $\alpha_0$ due to the relation $\left(t^{a_2}t^{a_2}\right)_{\alpha_{0}\,\alpha_{0''}}= C_F\, \delta_{\alpha_{0},\,\alpha_{0''}}$. Hence, the contribution of the quark self-energy loop factorizes within the wave-function $\Psi^A_{i\rightarrow q_0+\hat{\cal F}}$, leading to an expression of the form
\begin{eqnarray}
\Psi^A_{i\rightarrow q_0+\hat{\cal F}}=
\Psi^{\Gamma}_{i\rightarrow q_{0}+\hat{\cal F}}\; \; \left[\frac{\alpha_s\, C_F}{2\pi}\right]\;\;
{\cal V}_{A}\, ,
\end{eqnarray}
pulling for later convenience the coupling and color factor out of the factor ${\cal V}_{A}$ associated with the loop. Performing the obvious change of variable
\begin{eqnarray}
\k_2 \mapsto \K \equiv \k_2\!-\! \frac{k_2^+}{k_0^+}\, \k_0\, ,
\end{eqnarray}
the form factor ${\cal V}_{A}$ can be written as
\begin{eqnarray}
{\cal V}_{A} &=&
\frac{8\pi^2\, (\mu^2)^{2-\frac{D}{2}}}{2k_0^+(ED_{LO})}\;
\int_{0}^{k_0^+}\!\!\!\!\!\frac{d {k_2^+}}{(2\pi)(2 k_{2}^+)}\;
\frac{1}{2(k_{0}^+\!-\! k_{2}^+)}\;
\left\{4 \left(\frac{k_0^+}{k_2^+}\right)^2 +\frac{(D\!-\!2) k_0^+}{(k_0^+\!-\!k_2^+)} \right\}\;
\nonumber\\
&&\quad \times \;\;
\int \frac{d^{D-2} \K}{(2\pi)^{D-2}}\;
\frac{\K^2}{\left[(ED_{LO}) -  \frac{k_0^+}{2\, k_2^+(k_0^+\!-\!k_2^+)}
\K^2\right]}
\nonumber\\
&=&- \frac{\pi\, (\mu^2)^{2-\frac{D}{2}}}{k_0^+(ED_{LO})}\;
\int_{0}^{k_0^+}\!\!\frac{d {k_2^+}}{k_{0}^+}\;
\left\{4 \left(\frac{k_0^+}{k_2^+}\right)^2 +\frac{(D\!-\!2) k_0^+}{(k_0^+\!-\!k_2^+)} \right\}\;
\nonumber\\
&&\quad \times \;\;
\left\{\int \frac{d^{D-2} \K}{(2\pi)^{D-2}}\;
+ \frac{2\, k_2^+(k_0^+\!-\!k_2^+)}{k_0^+} (ED_{LO})
\int \frac{d^{D-2} \K}{(2\pi)^{D-2}}\;
\frac{1}{\left[\K^2 -  \frac{2\, k_2^+(k_0^+\!-\!k_2^+)}{k_0^+}
(ED_{LO})\right]}
\right\}
\nonumber\\
&=& -\int_{0}^{k_0^+}\!\!\frac{d {k_2^+}}{k_{0}^+}\;
\left\{2 \left(\frac{k_0^+\!-\!k_2^+}{k_2^+}\right) +\frac{(D\!-\!2) k_2^+}{2k_0^+} \right\}\;
\mathcal{A}_0(\Delta_1)
\label{VA_1}
\end{eqnarray}
In the second step, the first integral in $\K$ has naively a quadratic UV divergence, but it vanishes identically in dimensional regularization. By contrast, the second integral in $\K$ has the form of the scalar integral $\mathcal{A}_0(\Delta)$ (see eq. \eqref{def_A0}) in the Passarino-Veltman method \cite{Passarino:1978jh}, explained in appendix \ref{sec:PassVelt}. In this particular case, the parameter appearing in the scalar integral $\mathcal{A}_0$ is
\begin{eqnarray}
\Delta_1 = -\frac{2\, k_2^+(k_0^+\!-\!k_2^+)}{k_0^+} (ED_{LO})\, .
\end{eqnarray}
The exact expression for $\mathcal{A}_0(\Delta)$ in $D$ dimensions is given in eq. \eqref{A0_exact_result}. As is well-known, dimensional regularization can be used to regulate both the UV and the soft and collinear divergences. Here, it turns out that dimensional regularization also regulates the low $x$ divergence, as can be found by using the result \eqref{A0_exact_result} and performing the integral over $k_2^+$ in ${\cal V}_{A}$ in eq.\eqref{VA_1}.
However, in the context of high-energy QCD, it is more convenient to regulate the low $x$ divergence with an explicit cutoff in $k^+$ (see discussion in refs. \cite{Balitsky:2008zz,Beuf:2014uia}), which facilitate the resummation of high-energy leading logarithms with the BK \cite{Balitsky:1995ub,Kovchegov:1999yj,Kovchegov:1999ua} or JIMWLK equations \cite{Jalilian-Marian:1997jx,Jalilian-Marian:1997gr,Jalilian-Marian:1997dw, Kovner:2000pt,Weigert:2000gi,Iancu:2000hn,Iancu:2001ad,Ferreiro:2001qy}, as well as next-to-leading logarithms \cite{Balitsky:2008zz,Balitsky:2009xg,Balitsky:2013fea, Kovner:2013ona,Kovner:2014lca}. Hence, using eq. \eqref{A0_expand}, the $D\rightarrow 4$ expansion is performed before taking the $k_2^+$ integral. One obtains
\begin{eqnarray}
{\cal V}_{A} &=& -\int_{0}^{k_0^+}\!\!\frac{d {k_2^+}}{k_{0}^+}\;
\left\{2 \left(\frac{k_0^+\!-\!k_2^+}{k_2^+}\right) +\frac{k_2^+}{k_0^+} \right\}\;
\Bigg\{
\Gamma\!\left(2\!-\! \frac{D}{2}\right)\;
\left[\frac{-2\, k_0^+ (ED_{LO})}{4\pi\, \mu^2}\right]^{\frac{D}{2}-2}
- \log\left(\frac{k_2^+}{k_0^+}\right)
- \log\left(\frac{k_0^+\!-\!k_2^+}{k_0^+}\right)\Bigg\}\;\;
\nonumber\\
&& \hspace{6cm}
+\int_{0}^{k_0^+}\!\!\frac{d {k_2^+}}{k_{0}^+}\; \left(\frac{k_2^+}{k_0^+}\right)\;\;
+ O\left(D\!-\!4\right)\, ,
\label{VA_2}
\end{eqnarray}
where a cutoff $k_2^+>k^+_{\min}$ is implicitly applied, when needed. In the spirit of the $\overline{MS}$ scheme, the universal constants are left unexpanded together with the UV pole. Note that, due to the $D$ dependence in the curly bracket in the expression \eqref{VA_1}, one gets an additional term, analog to the so-called rational terms in standard pQCD calculations of scattering amplitudes at higher orders (see for example the review \cite{Ellis:2011cr}). Then, performing the integration over $k_2^+$ using the integrals in appendix \ref{sec:integrals}, one gets
\begin{eqnarray}
{\cal V}_{A} &=& 2 \left[\log\left(\frac{k^+_{\min}}{k_0^+}\right) +\frac{3}{4} \right]\;
\Gamma\!\left(2\!-\! \frac{D}{2}\right)\;
\left[\frac{-2\, k_0^+ (ED_{LO})}{4\pi\, \mu^2}\right]^{\frac{D}{2}-2}
\nonumber\\
&& -\left[\log\left(\frac{k^+_{\min}}{k_0^+}\right)\right]^2
 -\frac{\pi^2}{3} + 3+\frac{1}{2}
 + O\left(D\!-\!4\right)\, .
 \label{VA_3}
\end{eqnarray}
The energy denominator $(ED_{LO})$ encodes the off-shellness of the Fock state
${\cal F}=q_0+\hat{\cal F}$, corresponding to a transient violation of $k^-$ conservation. In covariant perturbation theory, $k^-$ is always conserved and instead each intermediate particle is allowed to be off mass shell. The mapping between these two parametrizations of the off-shellness is such that the $k^-$ off-shellness of the Fock state ${\cal F}$ corresponds to a mass off-shellness of the quark $q_0$ with a virtuality scale $\sqrt{-2\, k_0^+ (ED_{LO})}$. Hence, it is not surprising to see this scale appearing in the logarithm accompanying the UV divergence in eq. \eqref{VA_3}.

In eq. \eqref{VA_3}, the last term, $1/2$, is the rational term induced by the $D$ dependence of the coefficient multiplying the UV-divergent scalar integral $\mathcal{A}_0(\Delta_1)$ in eq. \eqref{VA_1}. Besides the conventional dimensional regularization \cite{'tHooft:1972fi} used in this calculation, there exist other variants of dimensional regularization, such as dimensional reduction \cite{Siegel:1979wq} or the four-dimensional helicity scheme \cite{Bern:1991aq,Bern:2002zk}, which amounts to a different treatment of the numerator. These alternative schemes should lead to a different value for that rational term. Hence, the last term $1/2$ as well as all of the first line in eq. \eqref{VA_3} are definitely UV regularization scheme dependent, whereas the other three terms could be UV regularization scheme independent, especially the $-\pi^2/3$ term.

The double logarithmic low $x$ divergence in the second line of eq. \eqref{VA_3}
is unphysical and thus should cancel in the sum over the diagrams. By contrast, one expects single logarithmic low $x$ divergence to survive up to the cross-section level, where it can be dealt with by high-energy resummation at LL accuracy, using either the BFKL, BK or JIMWLK equation.

The same calculation could be done for the self-energy of an antiquark inside an off-shell Fock state. The only difference is that both the gluon emission and the gluon absorption vertices get a extra $(-1)$ factor, so that the total result is unchanged. Hence, the self-energy graph for an antiquark is also given by the same ${\cal V}_{A}$ factor given in eq. \eqref{VA_3}.


\section{Transverse photon: quark-antiquark Fock component at one loop\label{sec:NLO_WF_T}}

\begin{figure}
\setbox1\hbox to 10cm{
\includegraphics{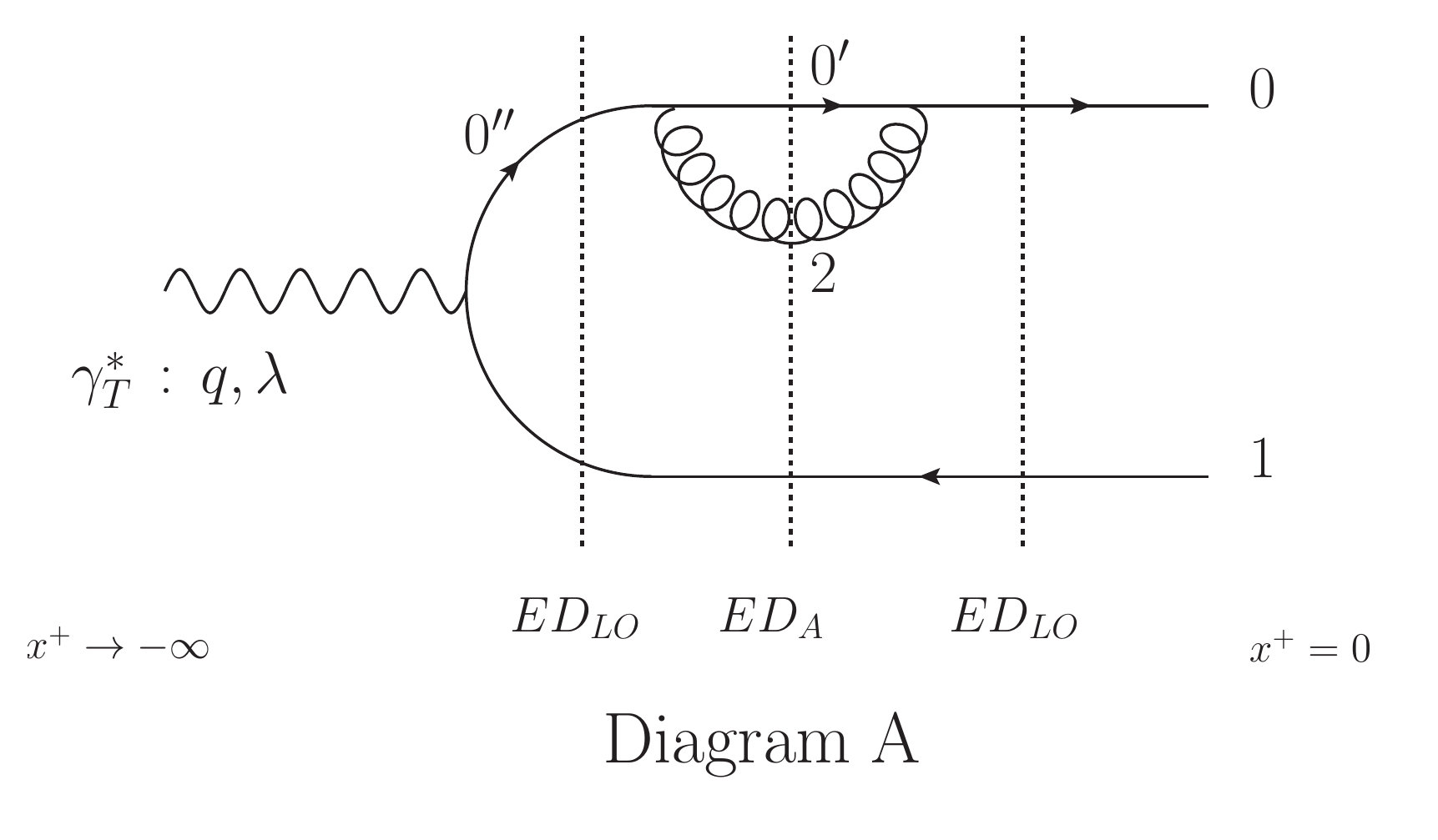}
}
\setbox2\hbox to 10cm{
\includegraphics{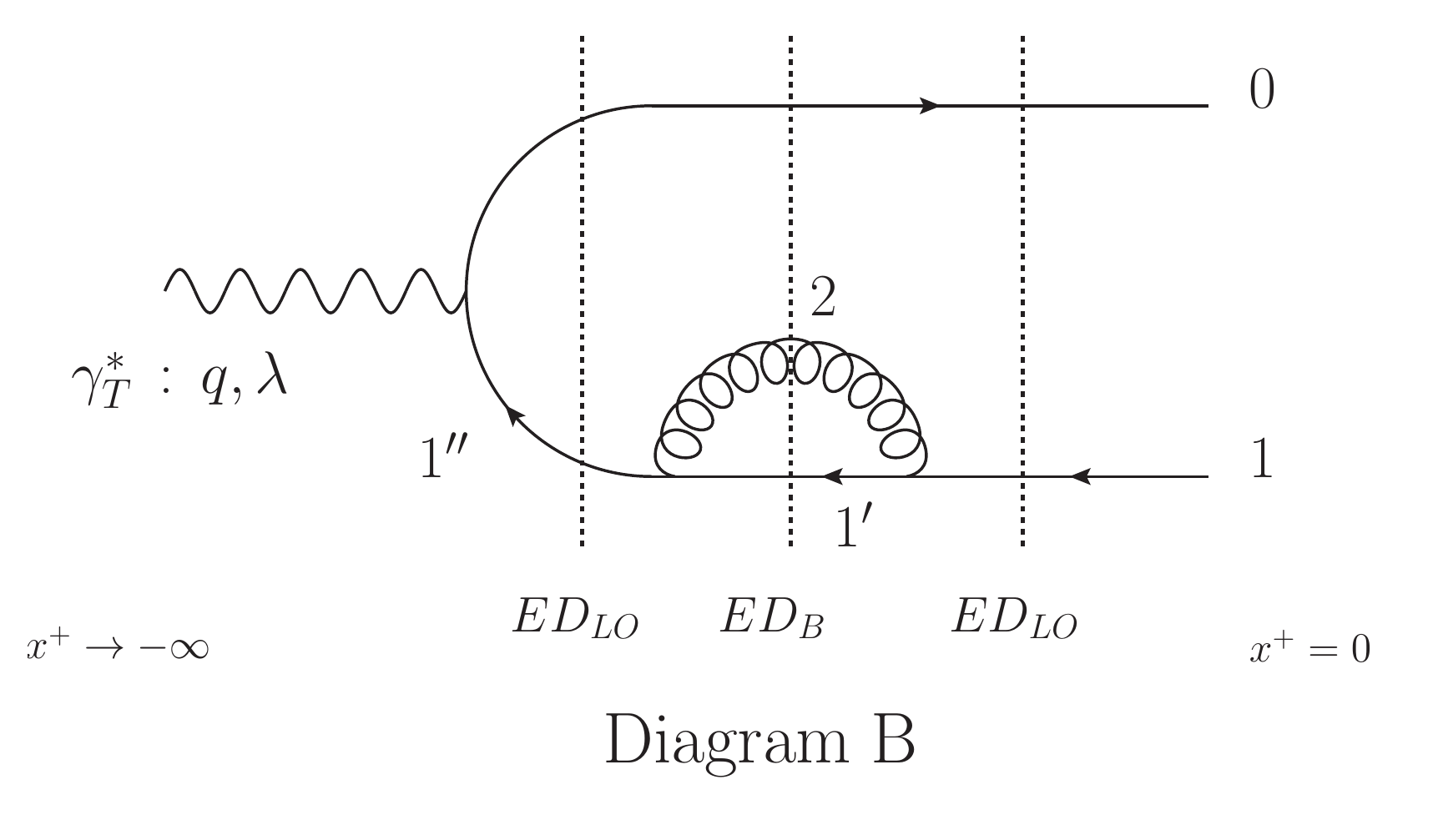}
}
\setbox3\hbox to 10cm{
\includegraphics{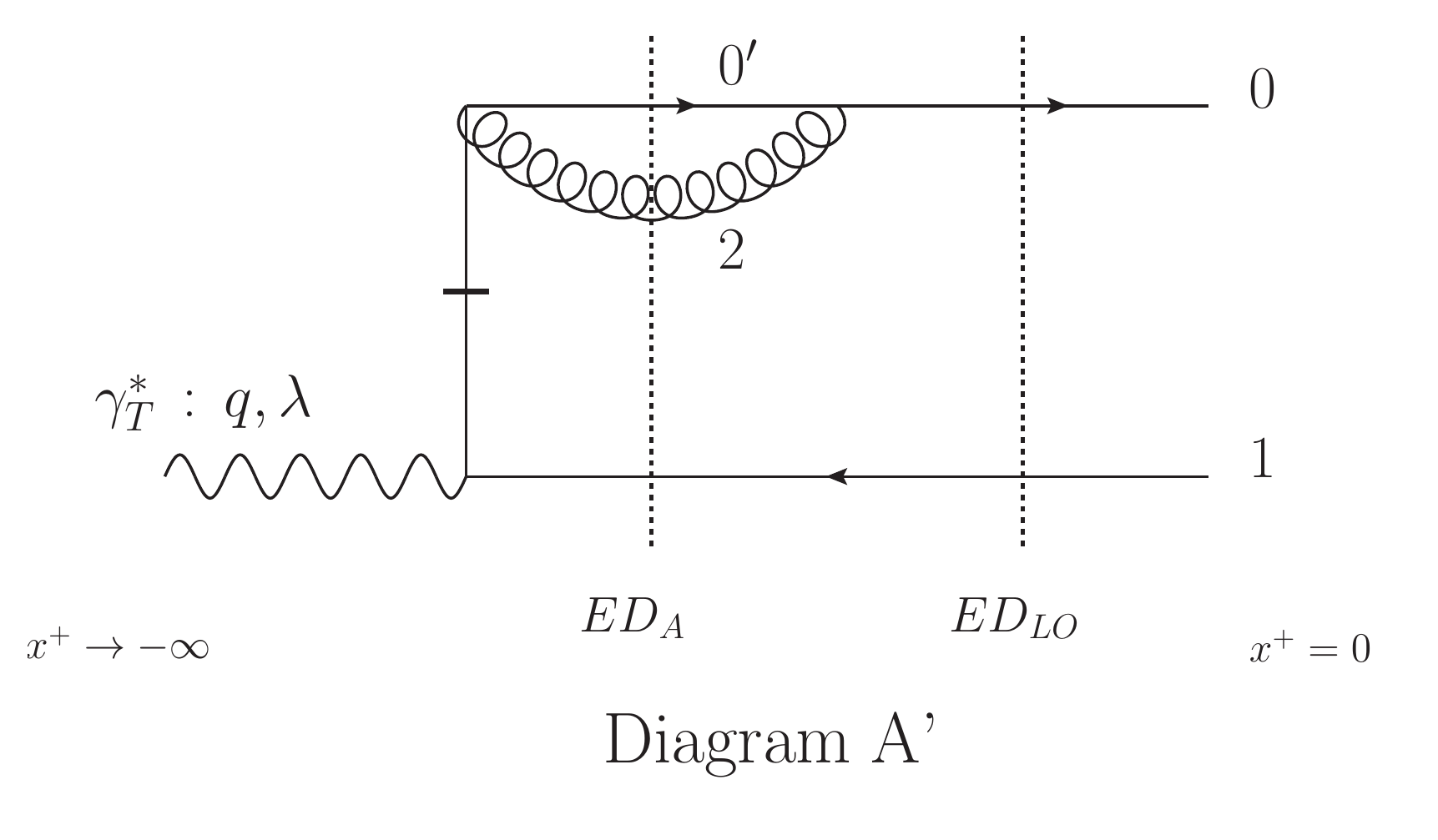}
}
\setbox4\hbox to 10cm{
\includegraphics{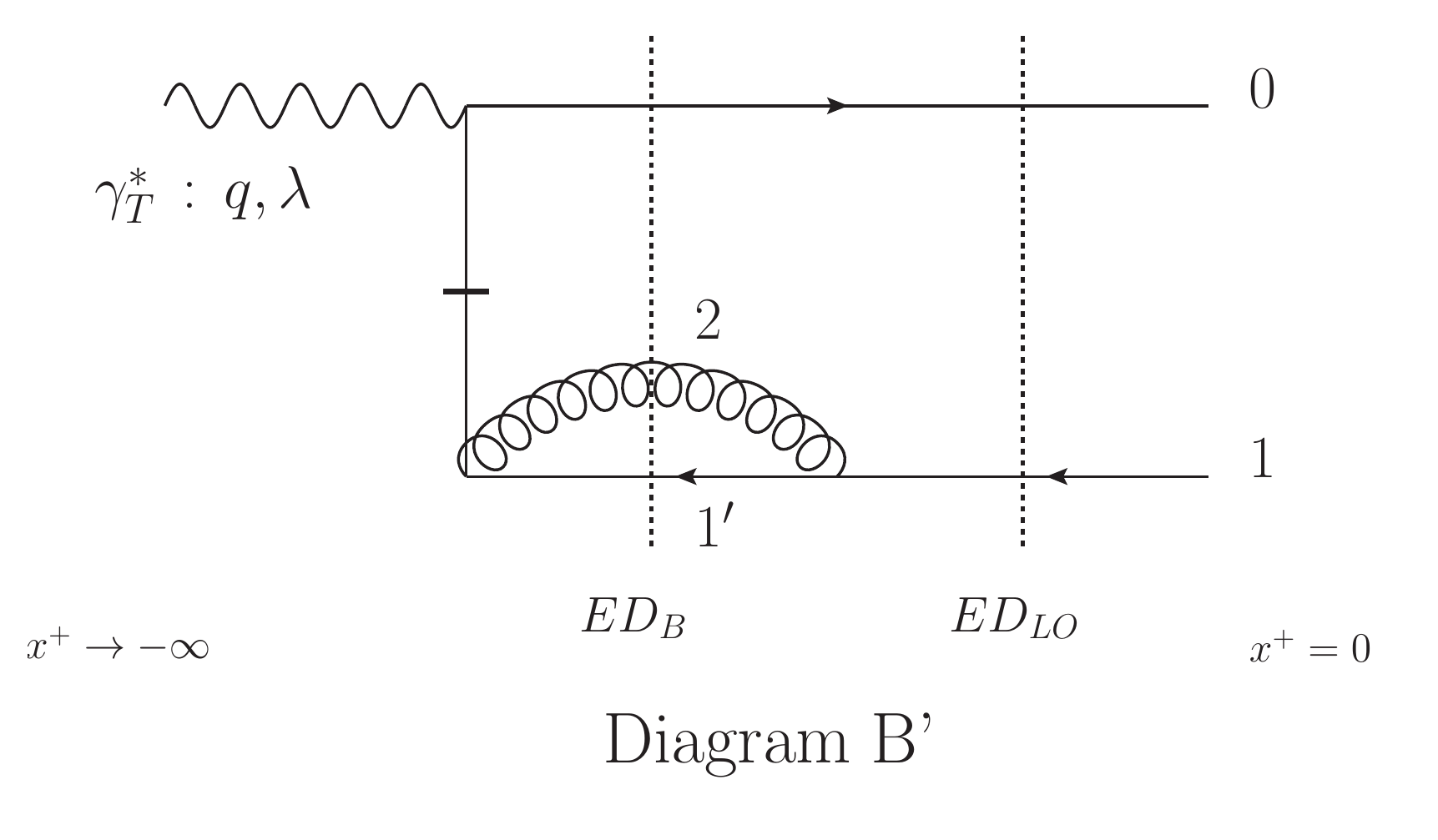}
}
\setbox5\hbox to 10cm{
\includegraphics{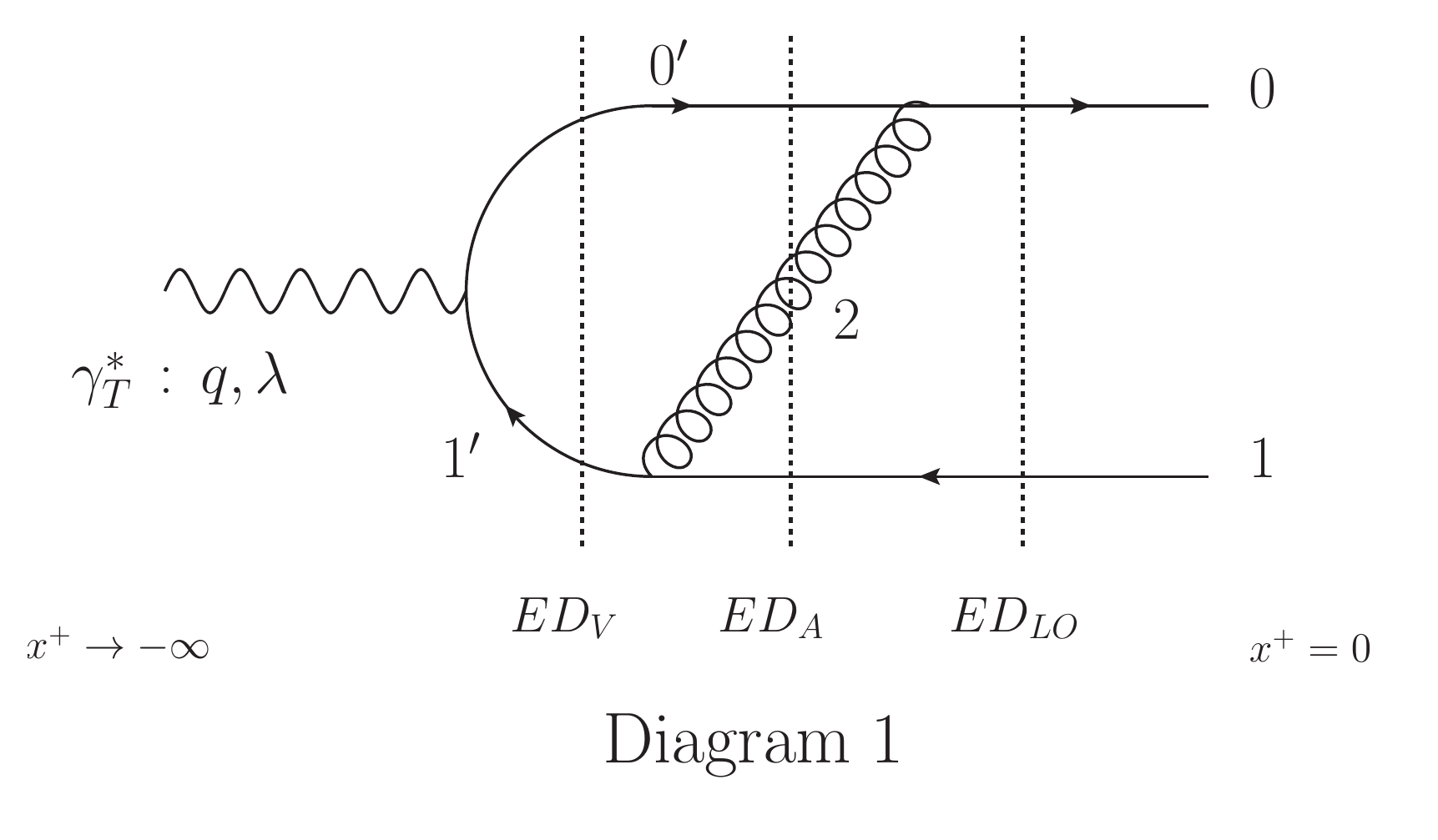}
}
\setbox6\hbox to 10cm{
\includegraphics{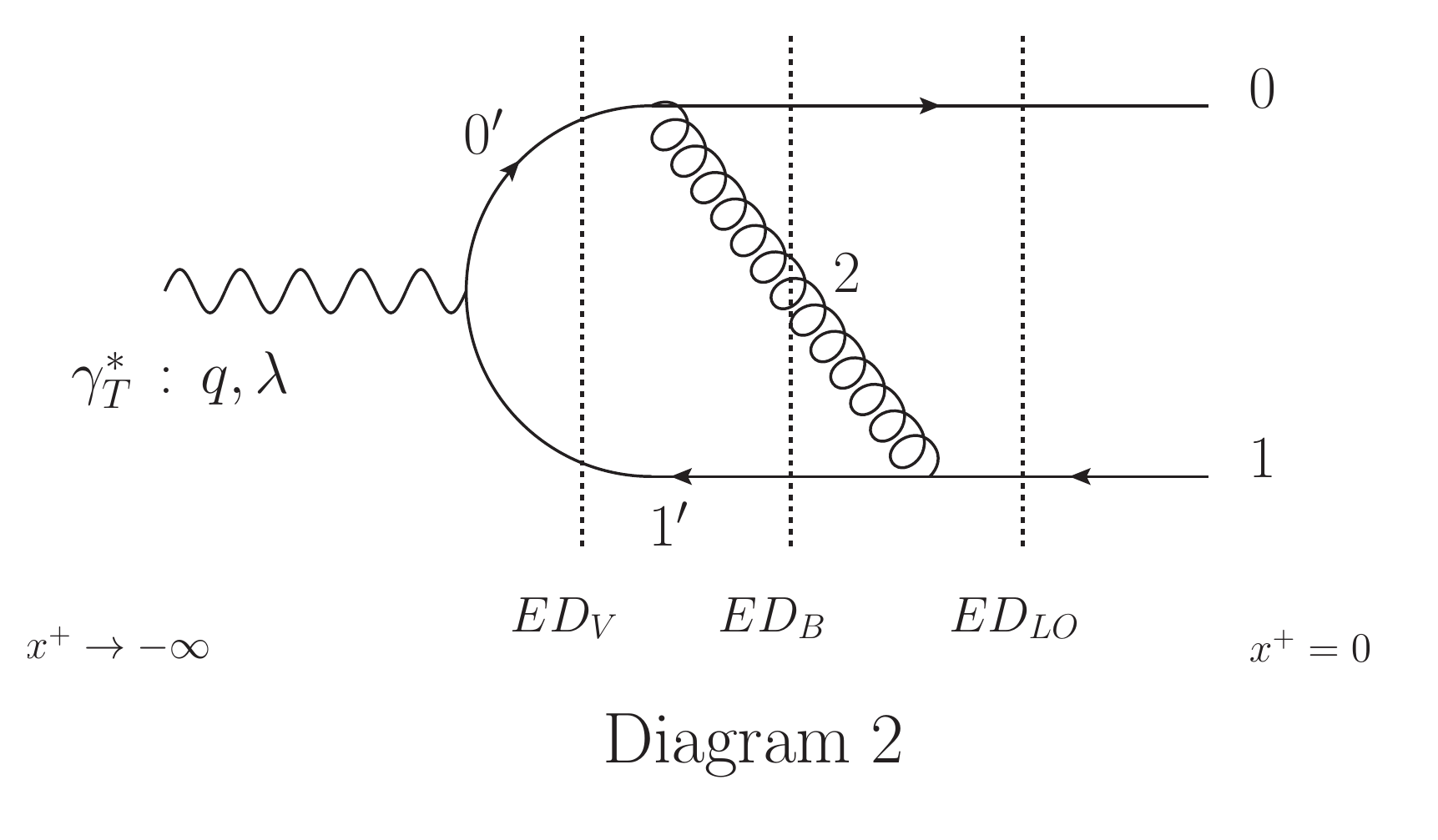}
}
\setbox7\hbox to 10cm{
\includegraphics{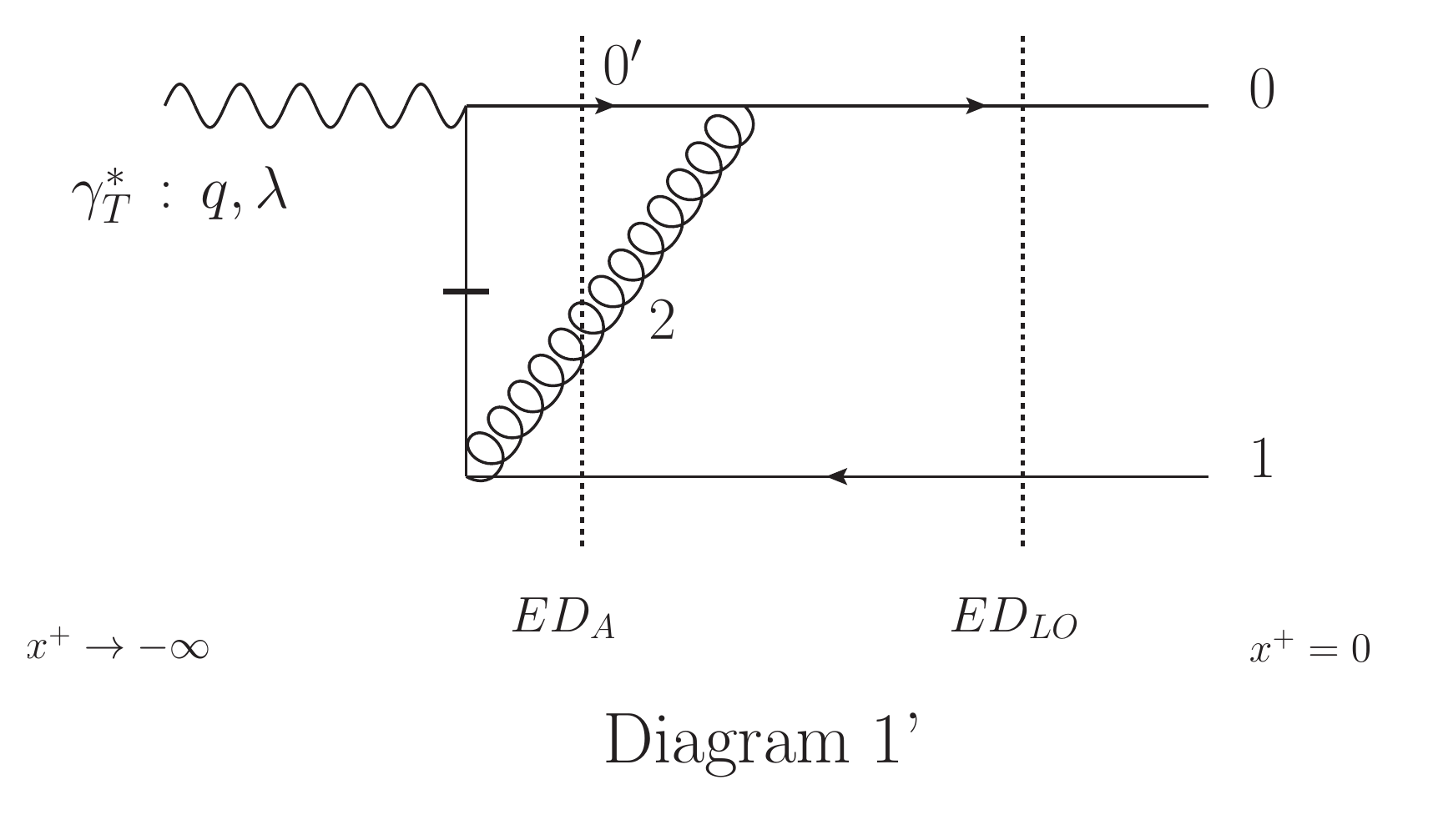}
}
\setbox8\hbox to 10cm{
\includegraphics{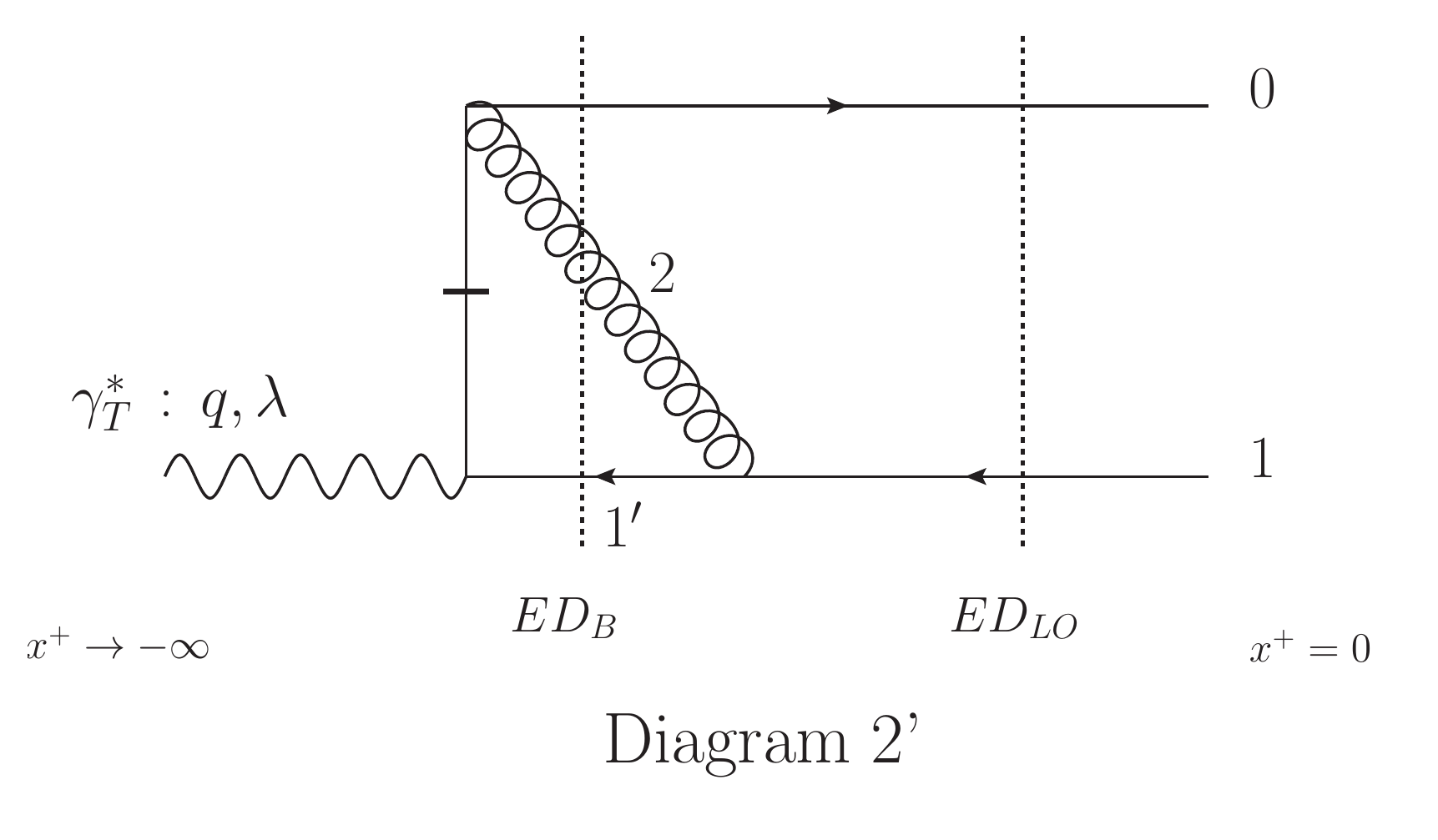}
}
\setbox9\hbox to 10cm{
\includegraphics{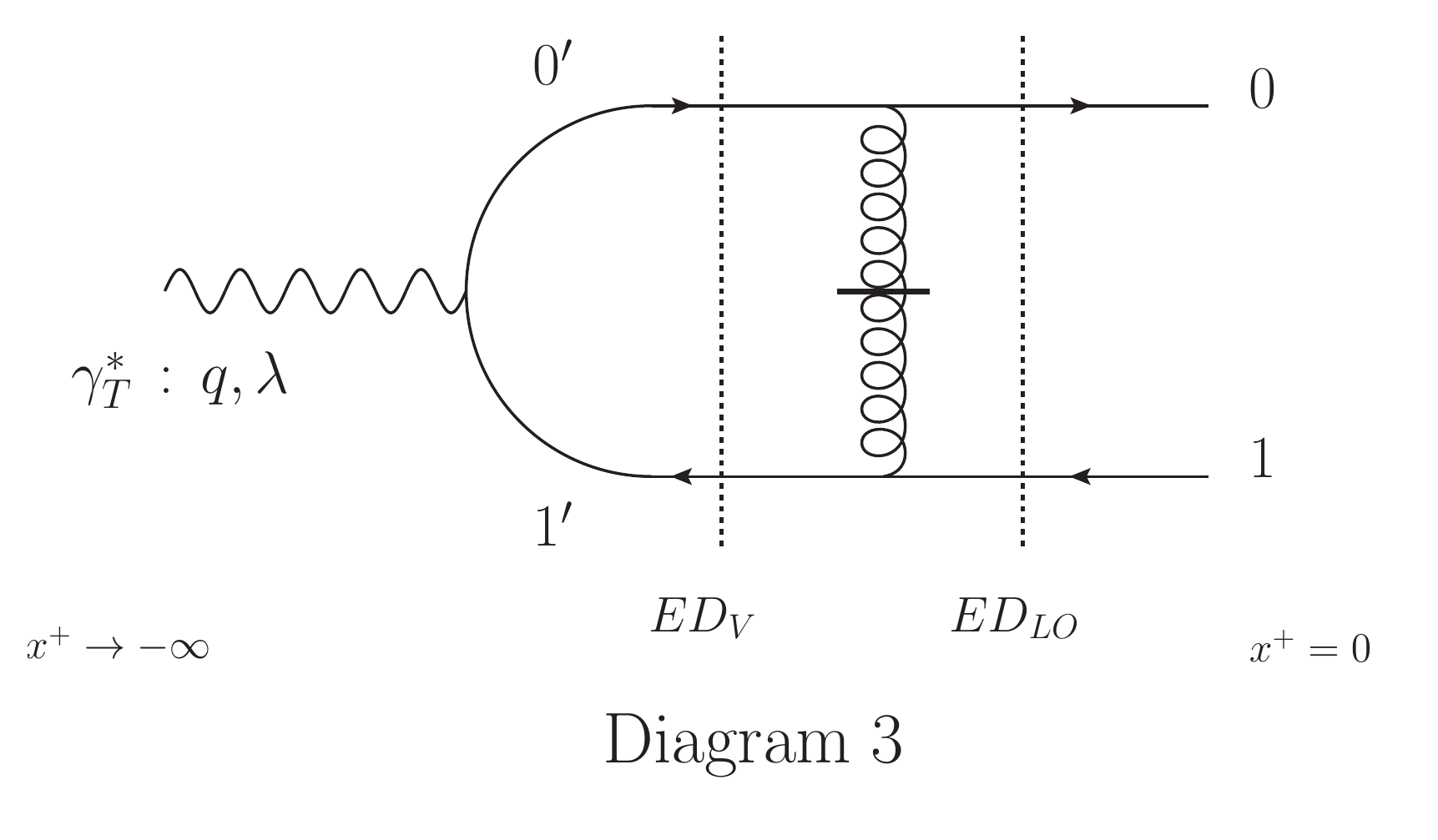}
}
\begin{center}
\resizebox*{10cm}{!}{\hspace{-7cm}\mbox{\box1 \hspace{9cm} \box2}}
\resizebox*{10cm}{!}{\hspace{-7cm}\mbox{\box3 \hspace{9cm} \box4}}
\resizebox*{10cm}{!}{\hspace{-7cm}\mbox{\box5 \hspace{9cm} \box6}}
\resizebox*{10cm}{!}{\hspace{-7cm}\mbox{\box7 \hspace{9cm} \box8}}
\resizebox*{10cm}{!}{\hspace{-7cm}\mbox{\hspace{9.5cm} \box9 \hspace{9.5cm}}}
\caption{\label{Fig:gammaT_NLO}One-gluon-loop diagrams contributing to the light-front wave-function of the $q\bar{q}$ Fock component inside an incoming transverse photon.}
\end{center}
\end{figure}
The light-front diagrams relevant\footnote{In general, one should also consider self-inertia graphs, associated with the mismatch between the expressions of the interaction part of the light-front Hamiltonian before and after applying the normal ordering prescription to it. However, in dimensional regularization, all of the self-inertia terms vanish (see for example \cite{Mustaki:1990im,Zhang:1993dd}).} for the calculation of the $\gamma^*_{T}\rightarrow q\bar{q}$ LFWF at one loop are presented in fig. \ref{Fig:gammaT_NLO}. It is instructive to note that all of these diagrams involve only four distinct energy denominators. One of them, $(ED_{LO})$ is the same as in the LO graph from fig. \ref{Fig:gammaT_LO}, and is given by eq. \eqref{ED_LO_1} or \eqref{ED_LO_2}.
Then, the energy denominator $(ED_A)$ is the same as in eq. \eqref{ED_A_vs_EDLO}, once $(ED_{LO})$ is fixed to the expression given in eq. \eqref{ED_LO_1}. Hence, one has
\begin{eqnarray}
(ED_{A}) &=& - \frac{k_0^+}{2\, k_2^+(k_0^+\!-\!k_2^+)}
\left[\k_2\!-\! \frac{k_2^+}{k_0^+}\, \k_0  \right]^2
- \frac{q^+}{2k_0^+ k_1^+}\; \left[\k_0\!-\!\frac{k_0^+}{q^+}\, \q \right]^2
-\frac{ Q^2}{2q^+}+i \epsilon
\, .
\label{ED_A_1}
\end{eqnarray}
The energy denominator $(ED_B)$ is symmetric to $(ED_A)$ by exchange of the roles of $\underline{k_{0}}$ and $\underline{k_1}$, so that
\begin{eqnarray}
(ED_{B}) &=& - \frac{k_1^+}{2\, k_2^+(k_1^+\!-\!k_2^+)}
\left[\k_2\!-\! \frac{k_2^+}{k_1^+}\, \k_1  \right]^2
- \frac{q^+}{2k_0^+ k_1^+}\; \left[\k_0\!-\!\frac{k_0^+}{q^+}\, \q \right]^2
-\frac{ Q^2}{2q^+}+i \epsilon
\, .
\label{ED_B_1}
\end{eqnarray}
Finally, the last new energy denominator is $(ED_V)$, which is analog
to $(ED_{LO})$, but with the Fock state $q_0 \bar{q}_1$ replaced by $q_{0'} \bar{q}_{1'}$. Hence
\begin{eqnarray}
(ED_{V}) &=& - \frac{q^+}{2k_{0'}^+ k_{1'}^+}\; \left[\k_{0'}\!-\!\frac{k_{0'}^+}{q^+}\, \q \right]^2
-\frac{ Q^2}{2q^+}+i \epsilon
\, .
\label{ED_V_1}
\end{eqnarray}

There is no diagram involving both $(ED_A)$ and $(ED_B)$. Hence, it is natural to split the diagrams between the ones involving $(ED_A)$, shown in the left side of fig. \ref{Fig:gammaT_NLO}, and those involving $(ED_B)$, shown in the right side of  fig. \ref{Fig:gammaT_NLO}. In the former ones, the momentum of the gluon is flowing into the quark, whereas in the latter ones, the momentum of the gluon is flowing into the antiquark. These two classes of diagrams thus correspond to two distinct kinematics.
Only the calculation of the first class of diagrams is necessary, thanks to the symmetry of the kinematics by exchange of the quark and the antiquark between the two classes of graphs.

The only exception is the diagram $3$, shown on the bottom of fig. \ref{Fig:gammaT_NLO}. It features neither $(ED_A)$ nor $(ED_B)$ but only $(ED_V)$. Moreover, in that diagram, the gluon momentum can flow either upwards into the quark or downwards into the antiquark. When needed, it is thus possible to split the diagram $3$ into a contribution of the first class and another one of the second class, according to the direction of the $k^+$ momentum flow along the instantaneous gluon line.

In the case of the first class of diagrams, it is convenient to parameterize the transverse momentum integration in the loop by the relative momentum $\K$ of the gluon with respect to the quark after the loop, defined by
\begin{eqnarray}
\K &=& \k_{2}\!-\!\frac{k_{2}^+}{k_{0}^+}\, \k_{0}
\, .
\label{def_K_in_type_A_graphs}
\end{eqnarray}
Then, using the notations \eqref{cv_k0_to_P} and \eqref{def_Qbar} as well, one can rewrite the energy denominators relevant for the first class of diagrams as
\begin{eqnarray}
\frac{1}{(ED_{A})} &=& - \frac{2\, k_2^+(k_0^+\!-\!k_2^+)}{k_0^+}\;
\frac{1}{\Big[\K^2+\Delta_1\Big]}
\label{ED_A_2}
\end{eqnarray}
and
\begin{eqnarray}
\frac{1}{(ED_{V})}  &=& - \frac{2\, (k_0^+\!-\!k_2^+) (k_1^+\!+\!k_2^+)}{q^+}\;
\frac{1}{\Big[(\K+\L)^2+\Delta_2\Big]}
\, ,
\label{ED_V_2}
\end{eqnarray}
where
\begin{eqnarray}
\Delta_1 &\equiv & \frac{q^+\, k_2^+\, (k_0^+\!-\! k_2^+)}{(k_0^+)^2 k_1^+}\; \Big[\P^2+ \overline{Q}^2\Big]
\label{def_Delta1}\\
\Delta_2 &\equiv & \frac{(k_0^+\!-\! k_2^+) (k_1^+\!+\! k_2^+)}{k_0^+\, k_1^+}\; \overline{Q}^2
= \frac{(k_0^+\!-\! k_2^+) (k_1^+\!+\! k_2^+)}{(q^+)^2}\; {Q}^2
\label{def_Delta2}\\
\L &\equiv & - \frac{(k_0^+\!-\! k_2^+)}{k_0^+}\; \P
= - (k_0^+\!-\! k_2^+) \left[\frac{\k_0}{k_0^+}\!-\!\frac{\q}{q^+}\right]
\, .
\label{def_L}
\end{eqnarray}

\subsection{Diagram A for transverse photon\label{sec:diag_A_T}}

According to the results of the section \ref{sec:self_energy}, the contribution of the diagram $A$ to the transverse photon wave function factorizes as
\begin{eqnarray}
\Psi_{\gamma_T^{*}\rightarrow q_0+\bar{q}_1}^{A}
&=&
\Psi_{\gamma_T^{*}\rightarrow q_0+\bar{q}_1}^{LO}\;
\times\, \left[\frac{\alpha_s\, C_F}{2\pi}\right]\,
{\cal V}_{A}^{T}
\, ,\label{WF_A_T}
\end{eqnarray}
where
\begin{eqnarray}
{\cal V}_{A}^{T}&=& \int_{0}^{k_0^+}\!\! \frac{d{k_2^+}}{k_0^+}\;
 \bigg[-\frac{2k_0^+}{k_2^+} +2 -(D\!-\!2)\frac{k_2^+}{2k_0^+}\bigg]\;
   \mathcal{A}_0(\Delta_1)
\label{VAT_decomp}
\, .
\end{eqnarray}
In eq. \eqref{VAT_decomp}, the value $\Delta_1$ for the parameter of the scalar integral $\mathcal{A}_0$ is the one given by eq. \eqref{def_Delta1}.
Then, following section \ref{sec:self_energy},
the form factor ${\cal V}_{A}^{T}$ can be written more explicitly as
\begin{eqnarray}
{\cal V}_{A}^{T}
&=&
\left[2\log\left(\frac{k^+_{\min}}{k_0^+}\right) +\frac{3}{2} \right]\,
\bigg[
\Gamma\!\left(2\!-\! \frac{D}{2}\right)\;
\left(\frac{\overline{Q}^2}{4\pi\, \mu^2}\right)^{\frac{D}{2}-2}
-\log\left(\frac{\P^2+\overline{Q}^2}{\overline{Q}^2}\right)
-\log\left(\frac{q^+}{k_1^+}\right)
\bigg]
\nonumber\\
&&
 -\left[\log\left(\frac{k^+_{\min}}{k_0^+}\right)\right]^2
 -\frac{\pi^2}{3} + 3+\frac{1}{2}
 + O\left(D\!-\!4\right)
\label{VAT_eval_1}
\, .
\end{eqnarray}


\subsection{Diagrams with instantaneous lines for transverse photon}

\subsubsection{Diagrams with instantaneous quark line}

The diagrams A' and 1' from fig. \ref{Fig:gammaT_NLO} together correspond to the contribution
\begin{eqnarray}
\Psi_{\gamma_T^{*}\rightarrow q_0+\bar{q}_1}^{A'+1'}=  \sum_{q_{0'} g_2 \textrm{ states}}
\frac{
\langle 0| b_0\, V_I(0)\, b_{0'}^{\dag} a_{2}^{\dag} |0 \rangle\;
\langle 0| a_{2} d_{1} b_{0'}\, V_I(0)\, a_{\gamma}^{\dag} |0 \rangle
}{(ED_{LO})\; (ED_A)}\, .
\label{WF_Aprime_1prime_1}
\end{eqnarray}
Using the expression \eqref{gamma_2_qqbarg_vertex}
for the $\gamma_{T}\rightarrow q\bar{q}g$ instantaneous vertex
, one finds
\begin{eqnarray}
\Psi_{\gamma_T^{*}\rightarrow q_0+\bar{q}_1}^{A'+1'}&=&
(2\pi)^{D-1}\delta^{(D-1)}(\underline{k_1}+\underline{k_{0}}\!-\!\underline{q})\;
\delta_{\alpha_{0} ,\,\alpha_{1}}\;
 \frac{\mu^{2-\frac{D}{2}}\, e\, e_f}{(ED_{LO})}\;
 \frac{(\mu^2)^{2-\frac{D}{2}}\, g^2\, C_F}{2}\nonumber\\
 &&\;\; \times\;\;
  \int \frac{d^{D-1} \underline{k_2}}{(2\pi)^{D-1} (2 k_{2}^+)}\; \theta(k_{2}^+)\;
 \frac{\theta(k_{0}^+\!-\!k_{2}^+)}{2(k_{0}^+\!-\!k_{2}^+)}\;
 \frac{\left\{ -\frac{\textrm{Num}_{A'}^T}{k_0^+}
+\frac{\textrm{Num}_{1'}^T}{(k_1^++k_2^+)}
\right\}}{\left[(ED_{LO}) - \frac{k_0^+}{2\, k_2^+(k_0^+\!-\!k_2^+)}
\left[\k_2\!-\! \frac{k_2^+}{k_0^+}\, \k_0  \right]^2\right]}
\, .
\label{WF_Aprime_1prime_2}
\end{eqnarray}
The calculation of the numerators
\begin{eqnarray}
\textrm{Num}_{A'}^T&=& \sum_{\textrm{phys. pol. }\lambda_2} \;\;\; \sum_{h_{0'}=\pm 1/2}  \overline{u}(0)\; \slashed{\epsilon}_{\lambda_2}\!(\underline{k_2})\; u(0')\;\;
 \overline{u}(0')\; \gamma^+ \slashed{\epsilon}_{\lambda_2}^*\!(\underline{k_2})\;
 \slashed{\epsilon}_{\lambda}\!(\underline{q})\; v(1)\label{num_Aprime_def}
\end{eqnarray}
and
\begin{eqnarray}
\textrm{Num}_{1'}^T&=& \sum_{\textrm{phys. pol. }\lambda_2} \;\;\; \sum_{h_{0'}=\pm 1/2}  \overline{u}(0)\; \slashed{\epsilon}_{\lambda_2}\!(\underline{k_2})\; u(0')\;\;
 \overline{u}(0')\; \gamma^+  \slashed{\epsilon}_{\lambda}\!(\underline{q})\;
 \slashed{\epsilon}_{\lambda_2}^*\!(\underline{k_2})\;
 v(1)\label{num_1prime_def}
\end{eqnarray}
is explained in detail in appendix \ref{sec:num_calc}, and gives
\begin{eqnarray}
\textrm{Num}_{A'}^T
&=& \left[\k_2^j\!-\! \frac{k_2^+}{k_0^+}\, \k_0^j  \right]\;
\left\{-2\,\frac{k_0^+}{k_2^+} +D\!-\!2\right\}\;
\overline{u}(0)\, \gamma^+ \gamma^j  \slashed{\epsilon}_{\lambda}\!(\underline{q})\; v(1)
\label{num_Aprime_result}
\end{eqnarray}
and
\begin{eqnarray}
\textrm{Num}_{1'}^T
&=& \left[\k_2^j\!-\! \frac{k_2^+}{k_0^+}\, \k_0^j  \right]\;
\left\{-2\,\frac{k_0^+}{k_2^+}\; \overline{u}(0)\, \gamma^+   \slashed{\epsilon}_{\lambda}\!(\underline{q})\; \gamma^j v(1)
-(D\!-\!4)\;\overline{u}(0)\, \gamma^+ \gamma^j  \slashed{\epsilon}_{\lambda}\!(\underline{q})\; v(1)\right\}
\, .
\label{num_1prime_result}
\end{eqnarray}

Hence, after the change of variable \eqref{def_K_in_type_A_graphs},
one finds that the diagrams $A'$ and $1'$ are both proportional to the trivial integral
\begin{eqnarray}
\int \frac{d^{D-2} \K}{(2\pi)^{D-2}}\;  \frac{\K^j}{\K^2 +\Delta_1} =0
\label{trivial_Aj_int}
\, ,
\end{eqnarray}
and thus they vanish:
\begin{eqnarray}
\Psi_{\gamma_T^{*}\rightarrow q_0+\bar{q}_1}^{A'}=
\Psi_{\gamma_T^{*}\rightarrow q_0+\bar{q}_1}^{1'}=0
\, .
\end{eqnarray}
By symmetry between the quark and the antiquark, one finds that the
diagrams $B'$ and $2'$ cannot contribute either:
\begin{eqnarray}
\Psi_{\gamma_T^{*}\rightarrow q_0+\bar{q}_1}^{B'}=
\Psi_{\gamma_T^{*}\rightarrow q_0+\bar{q}_1}^{2'}=0
\, .
\end{eqnarray}

\subsubsection{Diagram with instantaneous gluon line}

The diagram 3 from fig. \ref{Fig:gammaT_NLO} contributes to the $q\bar{q}$ LFWF inside a transverse photon as
\begin{eqnarray}
\Psi_{\gamma_T^{*}\rightarrow q_0+\bar{q}_1}^{3}=  \sum_{q_{0'} \bar{q}_{1'} \textrm{ states}}
\frac{
\langle 0| d_{1} b_0\, V_I(0)\, b_{0'}^{\dag} d_{1'}^{\dag} |0 \rangle\;
\langle 0| d_{1'}  b_{0'}\, V_I(0)\, a_{\gamma}^{\dag} |0 \rangle
}{(ED_{LO})\; (ED_V)}
\, .
\label{WF_3_T_1}
\end{eqnarray}
Using the expression \eqref{qqbar_to_qqbar_vertex}
for the instantaneous Coulomb interaction vertex between a quark and an antiquark, one gets
\begin{eqnarray}
\Psi_{\gamma_T^{*}\rightarrow q_0+\bar{q}_1}^{3}&=&
(2\pi)^{D-1}\delta^{(D-1)}(\underline{k_1}+\underline{k_{0}}\!-\!\underline{q})\;
\delta_{\alpha_{0} ,\,\alpha_{1}}\;
 \frac{\mu^{2-\frac{D}{2}}\, e\, e_f}{(ED_{LO})}\;
(\mu^2)^{2-\frac{D}{2}}\, g^2\, C_F\nonumber\\
 &&\;\; \times\;\;
  \int \frac{d^{D-1} \underline{k_{0'}}}{(2\pi)^{D-1} (2 k_{0'}^+)}\; \theta(k_{0'}^+)\;
 \frac{\theta(q^+\!-\!k_{0'}^+)}{2(q^+\!-\!k_{0'}^+)}\;\;
  \frac{(-1)}{(k_{0}^+\!-\!k_{0'}^+)^2}\;\;
 \frac{\textrm{Num}_{3}^T}{(ED_V)}
 \, .
\label{WF_3_T_2}
\end{eqnarray}
The calculation of the numerator
\begin{eqnarray}
\textrm{Num}_{3}^T&=&  \sum_{h_{0'},h_{1'}=\pm 1/2}
\overline{u}(0)\, \gamma^+ u(0')\;\;
\overline{u}(0')\; \slashed{\epsilon}_{\lambda}\!(\underline{q})\; v(1')\;\;
\overline{v}(1')\, \gamma^+ v(1)
\label{num_3_T_def}
\end{eqnarray}
in performed in appendix \ref{sec:num_3_T}, and gives
\begin{eqnarray}
\textrm{Num}_{3}^T
&=& \left[\k_{0'}^j\!-\! \frac{k_{0'}^+}{q^+}\, \q^j  \right]\;
\Big\{2\,k_{0'}^+\; \overline{u}(0)\, \gamma^+   \slashed{\epsilon}_{\lambda}\!(\underline{q})\; \gamma^j v(1)
- 2\,k_{1'}^+\; \overline{u}(0)\, \gamma^+ \gamma^j  \slashed{\epsilon}_{\lambda}\!(\underline{q})\; v(1)\Big\}
\, .
\label{num_3_T_result}
\end{eqnarray}
Using the expression \eqref{ED_V_1} for the energy denominator $(ED_V)$, it is then clear that the diagram $3$ is proportional to a vanishing integral of the type \eqref{trivial_Aj_int}, so that
\begin{eqnarray}
\Psi_{\gamma_T^{*}\rightarrow q_0+\bar{q}_1}^{3}=0
\, .
\end{eqnarray}



\subsection{Diagram 1 for transverse photon\label{sec:diag_1_T}}

The contribution of the diagram 1 from fig. \ref{Fig:gammaT_NLO} is written
\begin{eqnarray}
\Psi_{\gamma_T^{*}\rightarrow q_0+\bar{q}_1}^{1}
&=&  \sum_{q_{0'} \bar{q}_{1'} \textrm{ states}}
\sum_{g_{2} \textrm{ states}}
\frac{
\langle 0| b_0\, V_I(0)\, b_{0'}^{\dag} a_{2}^{\dag} |0 \rangle\;
\langle 0| a_{2} d_{1} \, V_I(0)\,  d_{1'}^{\dag} |0 \rangle\;
\langle 0| d_{1'}  b_{0'}\, V_I(0)\, a_{\gamma}^{\dag} |0 \rangle
}{(ED_{LO})\; (ED_A)\; (ED_V)}
\nonumber\\
&=&
\frac{e\, e_f}{(ED_{LO})}\, (\mu^3)^{2-\frac{D}{2}}\, (-g^2)C_F\, \delta_{\alpha_{0},\, \alpha_{1}}\;
 \int \frac{d^{D-1} \underline{k_{2}}}{(2\pi)^{D-1} }\; \frac{\theta(k_{2}^+)}{(2 k_{2}^+)}\;
 \int \frac{d^{D-1} \underline{k_{0'}}}{(2\pi)^{D-1} }\; \frac{\theta(k_{0'}^+)}{(2 k_{0'}^+)}\;
 \nonumber\\
&& \times\;
 \int \frac{d^{D-1} \underline{k_{1'}}}{(2\pi)^{D-1} }\; \frac{\theta(k_{1'}^+)}{(2 k_{1'}^+)}\;
  (2\pi)^{D-1}\delta^{(D-1)}(\underline{k_{0'}}+\underline{k_{2}}\!-\!
\underline{k_{0}})\;
(2\pi)^{D-1}\delta^{(D-1)}(\underline{k_{1'}}\!-\!
\underline{k_{2}}\!-\!\underline{k_1})\;
\nonumber\\
&& \times\;
(2\pi)^{D-1}\delta^{(D-1)}(\underline{k_{0'}}+\underline{k_{1'}}\!-\!
\underline{q})\;
\frac{\textrm{Num}_{1}^T}{(ED_A)\; (ED_V)}
\nonumber\\
&=&(2\pi)^{D-1}\delta^{(D-1)}(\underline{k_{0}}+\underline{k_{1}}\!-\!
\underline{q})\; \delta_{\alpha_{0},\, \alpha_{1}}\;
\frac{e\, e_f}{(ED_{LO})}\, (\mu^3)^{2-\frac{D}{2}}\, \frac{(-g^2)C_F}{2\pi}\,
\nonumber\\
&& \times\;
\int_{0}^{k_0^+}\!\!
\frac{d {k_2^+}}{2 k_{2}^+}\;
\frac{1}{2(k_{0}^+\!-\! k_{2}^+)}\;
\frac{1}{2(k_{1}^+\!+\! k_{2}^+)}\;
\int \frac{d^{D-2} \k_{2}}{(2\pi)^{D-2} }\;
\frac{\textrm{Num}_{1}^T}{(ED_A)\; (ED_V)}
\, ,
\label{WF_1_T_1}
\end{eqnarray}
where
\begin{eqnarray}
\textrm{Num}_{1}^T &=&
\sum_{\textrm{phys. pol. }\lambda_2} \;\;\; \sum_{h_{0'}, h_{1'}=\pm 1/2}  \overline{u}(0)\; \slashed{\epsilon}_{\lambda_2}\!(\underline{k_2})\; u(0')\;\;
 \overline{u}(0')\;  \slashed{\epsilon}_{\lambda}\!(\underline{q})\; v(1')\;\;
\overline{v}(1')\,  \slashed{\epsilon}_{\lambda_2}^*\!(\underline{k_2})\;  v(1)
\, .
\label{num_1_T_def}
\end{eqnarray}

The calculation of the numerator $\textrm{Num}_{1}^T$, rather tedious, is performed in appendix \ref{sec:num_1_T}, and gives
\begin{eqnarray}
\textrm{Num}_{1}^T &=&
\left(\k_2\!-\! \frac{k_2^+}{k_0^+}\, \k_0 \right)^2
\left(\k_2^j\!-\! \frac{k_2^+}{k_1^+}\, \k_1^j  \right)\;
\mathfrak{C}_{I}^j
+\left(\k_2\!-\! \frac{k_2^+}{k_1^+}\, \k_1 \right)^2
\left(\k_2^j\!-\! \frac{k_2^+}{k_0^+}\, \k_0^j  \right)\;
\mathfrak{C}_{II}^j
\nonumber\\
&&
+\, 2\Bigg\{\frac{k_0^+k_1^+}{(k_2^+)^2}\,
\left(\k_2\!-\! \frac{k_2^+}{k_0^+}\, \k_0 \right)^2\,
+\frac{k_0^+k_1^+}{(k_2^+)^2}\,
\left(\k_2\!-\! \frac{k_2^+}{k_1^+}\, \k_1 \right)^2\,
-\frac{(q^+)^2}{k_0^+k_1^+}\,
\left(\k_0\!-\! \frac{k_0^+}{q^+}\, \q \right)^2\,
\Bigg\}\;
\overline{u}(0)\,\slashed{\epsilon}_{\lambda}\!(\underline{q})\;v(1)
\nonumber\\
&&
+\, \frac{\varepsilon_{\lambda}^{i}}{k_2^+}\;
 \left[4 \left(\k_0^i \!-\!\frac{k_0^+}{q^+}\, \q^i\right)
   -(D\!-\!2)\, \left(\k_2^i \!-\!\frac{k_2^+}{q^+}\, \q^i\right)
\right]\;
\left(\k_{2}^j \!-\!\frac{k_{2}^+}{k_{0}^+}\,\k_{0}^j\right)
\left(\k_{2}^l \!-\!\frac{k_{2}^+}{k_{1}^+}\,\k_{1}^l\right)\,
\overline{u}(0)\, \gamma^+ \gamma^{j} \gamma^{l}\,  v(1)
\, ,
\label{num_1_T_result}
\end{eqnarray}
where
\begin{eqnarray}
\mathfrak{C}_{I}^j &\equiv & \frac{k_0^+ k_1^+}{(k_2^+)^2 (k_0^+\!-\!k_2^+)} \; \overline{u}(0)\, \gamma^+ \gamma^j \slashed{\epsilon}_{\lambda}\!(\underline{q})\;  v(1)
-\frac{k_0^+}{k_2^+}\,
\left[\frac{1}{k_2^+}+\frac{(D\!-\!4)}{2(k_0^+\!-\!k_2^+)}\right]\;
\overline{u}(0)\, \gamma^+ \slashed{\epsilon}_{\lambda}\!(\underline{q})\; \gamma^j v(1)
\label{def_coeff_CjI}
\\
\mathfrak{C}_{II}^j &\equiv & -\frac{k_0^+ k_1^+}{(k_2^+)^2 (k_1^+\!+\!k_2^+)} \; \overline{u}(0)\, \gamma^+ \slashed{\epsilon}_{\lambda}\!(\underline{q})\; \gamma^j  v(1)
+\frac{k_1^+}{k_2^+}\,
\left[\frac{1}{k_2^+}-\frac{(D\!-\!4)}{2(k_1^+\!+\!k_2^+)}\right]\;
\overline{u}(0)\, \gamma^+ \gamma^j \slashed{\epsilon}_{\lambda}\!(\underline{q})\; v(1)
\, .
\label{def_coeff_CjII}
\end{eqnarray}

The next step is to perform in this numerator the change of variable
\begin{eqnarray}
\k_2 \mapsto \K \equiv \k_2 -\frac{k_2^+}{k_0^+}\, \k_0
\label{cv_k2_to_K}
\end{eqnarray}
suggested by the form of the denominators. Using also the notation
\eqref{cv_k0_to_P}, one obtains
\begin{eqnarray}
\textrm{Num}_{1}^T &=& \textrm{Num}_{1}^T\bigg|_{I} +\textrm{Num}_{1}^T\bigg|_{II} + \textrm{Num}_{1}^T\bigg|_{III} +\textrm{Num}_{1}^T\bigg|_{IV}
\label{num_1_T_result_cv}
\\
\textrm{Num}_{1}^T\bigg|_{I} &=& \K^2\;
\left[\K^j + \frac{q^+ k_2^+}{k_0^+k_1^+}\, \P^j \right] \;
\mathfrak{C}_{I}^j
\label{num_1_T_result_cv_I}
\\
\textrm{Num}_{1}^T\bigg|_{II} &=&
\left[\K^2 + 2\, \frac{q^+ k_2^+}{k_0^+k_1^+}\, \P\!\cdot\!\K
+\left(\frac{q^+ k_2^+}{k_0^+k_1^+}\right)^2\, \P^2 \right] \;
\K^j\; \mathfrak{C}_{II}^j
\label{num_1_T_result_cv_II}
\\
\textrm{Num}_{1}^T\bigg|_{III} &=&
4\, \left[ \frac{k_0^+k_1^+}{(k_2^+)^2}\, \K^2
      + \frac{q^+}{k_2^+}\, \P\!\cdot\!\K \right]\;
\overline{u}(0)\,\slashed{\epsilon}_{\lambda}\!(\underline{q})\;v(1)
\label{num_1_T_result_cv_III}
\\
\textrm{Num}_{1}^T\bigg|_{IV} &=&
  \frac{{\varepsilon}_{\lambda}^i}{k_2^+}\;
 \Bigg\{-(D\!-\!2)
    \K^i
    + \left[4-\frac{(D\!-\!2)k_2^+}{k_0^+}\right]\;
    \P^i
 \Bigg\}\;
 \K^j\;
 \left[\K^l+ \frac{q^+ k_2^+}{k_0^+ k_1^+}\, \P^l  \right]\;
 \overline{u}(0)\, \gamma^+ \gamma^j \gamma^l v(1)
\label{num_1_T_result_cv_IV}
\, .
\end{eqnarray}
It is convenient to separate symmetric and antisymmetric pieces of the Dirac structure in the last term using the identity
\begin{eqnarray}
 \overline{u}(0)\, \gamma^+ \gamma^j \gamma^l v(1)
 &=& -\delta^{jl} \; \overline{u}(0)\, \gamma^+  v(1)
    +\frac{1}{2}\; \overline{u}(0)\, \gamma^+
    \left[\gamma^j, \gamma^l\right]\, v(1)\, .
    \label{sym_antisym_decomp_spinor}
\end{eqnarray}
The contribution $IV$ to the numerator then becomes
\begin{eqnarray}
\textrm{Num}_{1}^T\bigg|_{IV} &=&
  \frac{{\varepsilon}_{\lambda}^i}{k_2^+}\;
 \Bigg\{-(D\!-\!2)
    \K^i
    + \left[4-\frac{(D\!-\!2)k_2^+}{k_0^+}\right]\;
    \P^i
 \Bigg\}\;\nonumber\\
 && \times
 \Bigg\{\frac{q^+ k_2^+}{2k_0^+ k_1^+}\;\K^j\, \P^l\;
 \overline{u}(0)\, \gamma^+ \left[\gamma^j, \gamma^l\right]\, v(1)
 - \bigg[ \K^2+ \frac{q^+ k_2^+}{k_0^+ k_1^+}\, \P\!\cdot\!\K\bigg]\;
  \overline{u}(0)\, \gamma^+  v(1)\Bigg\}
\label{num_1_T_result_cv_IV_bis}
\, .
\end{eqnarray}

At this stage, the wave-function contribution corresponding to the diagram 1 is written
\begin{eqnarray}
\Psi_{\gamma_T^{*}\rightarrow q_0+\bar{q}_1}^{1}
&=&(2\pi)^{D-1}\delta^{(D-1)}(\underline{k_{0}}+\underline{k_{1}}\!-\!
\underline{q})\; \delta_{\alpha_{0},\, \alpha_{1}}\;
\frac{e\, e_f}{(ED_{LO})}\, \mu^{2-\frac{D}{2}}\,
\left[\frac{\alpha_s\, C_F}{2\pi}\right]\,
\frac{(-1)}{2q^+}\;
\int_{0}^{k_0^+}\!\!
d{k_2^+}\;
\frac{(k_{0}^+\!-\! k_{2}^+)}{k_{0}^+}\;
\mathcal{I}_1^{T}
\, ,\label{WF_1_T_2}
\end{eqnarray}
where
\begin{eqnarray}
\mathcal{I}_1^{T}&=& 4\pi\, (\mu^2)^{2-\frac{D}{2}}\,
\int \frac{d^{D-2} \K}{(2\pi)^{D-2} }\;
\frac{\textrm{Num}_{1}^T}{\Big[\K^2 +\Delta_1\Big]\,
\Big[(\K\!+\!\L)^2 +\Delta_2\Big]}
\, ,
\label{int_perp_1_T}
\end{eqnarray}
using the notations introduced in eqs. \eqref{def_Delta1}, \eqref{def_Delta2} and \eqref{def_L}.

In order to calculate the integral $\mathcal{I}_1^{T}$, the most efficient way is to rewrite it as a linear combination of a few simple and independent scalar integrals. As a first step towards that goal, one should substitute the $\K^2$ and $\P\!\cdot\!\K$
appearing in the numerators \eqref{num_1_T_result_cv_I}, \eqref{num_1_T_result_cv_II}, \eqref{num_1_T_result_cv_III} and \eqref{num_1_T_result_cv_IV_bis} by the expressions
\begin{eqnarray}
\K^2 &=& \Big[\K^2+\Delta_1\Big] - \Delta_1
\label{K2_replace}\\
\P\!\cdot\!\K &=& - \frac{k_0^+}{2(k_0^+\!-\! k_2^+)}\,
\bigg\{ \Big[(\K+\L)^2+\Delta_2\Big] - \Big[\K^2+\Delta_1\Big]
    -\L^2 -\Delta_2 +\Delta_1
\bigg\}
\label{PdotK_replace}
\, ,
\end{eqnarray}
and make the obvious simplifications with the denominators. Then, after some algebra, each of the four contributions to $\mathcal{I}_1^{T}$, associated with the decomposition \eqref{num_1_T_result_cv} of the numerator of the diagram 1, can be written as a linear combination of the integrals $\mathcal{A}_0(\Delta_1)$, $\mathcal{A}_0(\Delta_2)$, $\mathcal{B}_0(\Delta_1,\Delta_2,\L)$, $\mathcal{B}^i(\Delta_1,\Delta_2,\L)$ and $\mathcal{B}^{ij}(\Delta_1,\Delta_2,\L)$ defined in appendix \eqref{sec:PassVelt} (plus some integrals which vanish identically due to rotational symmetry, like in eq. \eqref{trivial_Aj_int}), as
\begin{eqnarray}
\mathcal{I}_1^{T}\bigg|_{I} &=&
\bigg\{-\Delta_1\; \mathcal{B}^j
- \Delta_1\; \frac{q^+ k_2^+}{k_0^+ k_1^+}\; \P^j\, \mathcal{B}_0
+ \frac{(k_1^+\!+\!k_2^+)}{k_1^+}\, \P^j\, \mathcal{A}_0(\Delta_2)  \bigg\} \;
\mathfrak{C}_{I}^j
\label{I_1_T_I_result}
\\
\mathcal{I}_1^{T}\bigg|_{II} &=&
\frac{(k_1^+\!+\!k_2^+)}{k_1^+}\, \P^j\, \mathcal{A}_0(\Delta_2)  \;
\mathfrak{C}_{II}^j
\label{I_1_T_II_result}
\\
\mathcal{I}_1^{T}\bigg|_{III} &=&
\bigg\{ \frac{4 k_0^+ k_1^+}{(k_2^+)^2}\, \mathcal{A}_0(\Delta_2)
- \frac{2 q^+ k_0^+}{k_2^+(k_0^+\!-\!k_2^+)}\,
\Big[\mathcal{A}_0(\Delta_1)-\mathcal{A}_0(\Delta_2)\Big]\nonumber\\
&& \hspace{0.5cm}
- \frac{2 q^+}{k_2^+}\,
\left[\frac{(k_1^+\!+\!k_2^+)}{k_1^+}\, \P^2
  + \frac{(k_0^+\!-\!k_2^+)}{k_0^+}\, \overline{Q}^2 \right]
\mathcal{B}_0
 \bigg\} \,
\overline{u}(0)\,\slashed{\epsilon}_{\lambda}\!(\underline{q})\;v(1)
\label{I_1_T_III_result}
\\
\mathcal{I}_1^{T}\bigg|_{IV} &=&
   \frac{q^+}{2k_0^+ k_1^+}\,
\Bigg\{-(D\!-\!2)\, \mathcal{B}^{ij}
    + \left(4-\frac{(D\!-\!2)k_2^+}{k_0^+}\right)\;
    \P^i\,  \mathcal{B}^{j}
 \Bigg\}\; {\varepsilon}_{\lambda}^i\;
\P^l\; \overline{u}(0)\, \gamma^+ \left[\gamma^j, \gamma^l\right] v(1)
 \nonumber\\
&& \hspace{-1cm}
+ \Bigg\{
   \bigg[\frac{(D\!-\!6)}{k_2^+}+ \frac{(D\!-\!2)q^+}{2 k_0^+ k_1^+}\bigg]\,
       \P^i\, \mathcal{A}_0(\Delta_2)
   +\frac{q^+}{2 k_1^+(k_0^+\!-\!k_2^+)}\,
     \left(4-\frac{(D\!-\!2)k_2^+}{k_0^+}\right)\,
     \P^i\, \Big[\mathcal{A}_0(\Delta_1)\!-\!\mathcal{A}_0(\Delta_2)\Big]
\nonumber\\
&& \hspace{-0.5cm}
    +\frac{q^+}{2 k_0^+ k_1^+}
     \left[\frac{(k_1^+\!+\!k_2^+)}{k_1^+}\, \P^2
       +\frac{(k_0^+\!-\!k_2^+)}{k_0^+}\, \overline{Q}^2\right]\;
   \left[-(D\!-\!2)\,\mathcal{B}^{i}
    +\left(4-\frac{(D\!-\!2)k_2^+}{k_0^+}\right)\,\P^i\, \mathcal{B}_0\right]
 \Bigg\}\;
 {\varepsilon}_{\lambda}^i\; \overline{u}(0)\, \gamma^+ v(1)
 \label{I_1_T_IV_result}
 \, ,
\end{eqnarray}
keeping the arguments of the integrals $\mathcal{B}_0$, $\mathcal{B}^i$ and $\mathcal{B}^{ij}$ implicit, since they are always the same.

It remains now to eliminate the vector and tensor integrals $\mathcal{B}^i$ and $\mathcal{B}^{ij}$ in favor of the scalar
integrals $\mathcal{A}_0(\Delta_1)$, $\mathcal{A}_0(\Delta_2)$ and $\mathcal{B}_0$ thanks to the Passarino-Veltman reduction technique \cite{Passarino:1978jh}, described in more detail in appendix \ref{sec:PassVelt}. Due to rotational invariance in the $(D\!-\!2)$ transverse space, $\mathcal{B}^i$ and $\mathcal{B}^{ij}$ are of the form
\begin{eqnarray}
\mathcal{B}^i(\Delta_1,\Delta_2,\L)&= & \L^i\; \mathcal{B}_1(\Delta_1,\Delta_2,\L^2)
\label{Bi_PV}
\\
\mathcal{B}^{ij}(\Delta_1,\Delta_2,\L)&= & \L^i \L^j\; \mathcal{B}_{21}(\Delta_1,\Delta_2,\L^2)
+ \delta^{ij}\; \mathcal{B}_{22}(\Delta_1,\Delta_2,\L^2)
\label{Bij_PV}
\,
\end{eqnarray}
where  $\mathcal{B}_1$, $\mathcal{B}_{21}$ and $\mathcal{B}_{22}$ are scalar integrals which can be written as linear combinations of
$\mathcal{A}_0(\Delta_1)$, $\mathcal{A}_0(\Delta_2)$ and $\mathcal{B}_0$, see appendix \ref{sec:PassVelt}. However, before using these relations between the scalar integrals, it is convenient to use the particular form \eqref{Bi_PV} and \eqref{Bij_PV} of the vector and scalar integrals in order to simplify the Dirac algebra.

The relation \eqref{Bi_PV} implies that, in the expression \eqref{I_1_T_I_result} for the contribution $I$, all the terms in the bracket are proportional to $\P^j$, so that the contribution $I$ to $\mathcal{I}_1^{T}$ is proportional $\P^j \mathfrak{C}_{I}^j$. Then, using the relations
\eqref{quark_current_plus_epsilon j} and \eqref{quark_current_plus j_epsilon}, one finds
\begin{eqnarray}
\mathcal{I}_1^{T}\bigg|_{I} &=& -\frac{k_0^+ k_1^+}{k_2^+(k_0^+\!-\!k_2^+)}\,
\bigg\{\Delta_1\,\left[\frac{(k_0^+\!-\!k_2^+)}{k_0^+}\, \mathcal{B}_1
                  -\frac{q^+ k_2^+}{k_0^+ k_1^+}\, \mathcal{B}_0\right]
        + \frac{(k_1^+\!+\!k_2^+)}{k_1^+}\,  \mathcal{A}_0(\Delta_2)\bigg\}\;
\nonumber\\
&& \times
\bigg\{\left[\frac{2k_0^+}{k_2^+} + \frac{(D\!-\!6)k_0^+}{q^+}\right]\,
       \overline{u}(0)\,\slashed{\epsilon}_{\lambda}\!(\underline{q})\;v(1)
      + (D\!-\!6)\; \frac{(\P\!\cdot\! \varepsilon_{\lambda})}{q^+}\;
              \overline{u}(0)\, \gamma^+\,  v(1)
\bigg\}
\label{I_1_T_I_result_spin_reduced}
 \, .
\end{eqnarray}
Similarly, the contribution $II$ can be expanded as
\begin{eqnarray}
\mathcal{I}_1^{T}\bigg|_{II} &=&
-\frac{k_0^+}{k_2^+}\,
\bigg\{\left[\frac{2k_1^+}{k_2^+} - \frac{(D\!-\!6)k_1^+}{q^+}\right]\,
       \overline{u}(0)\,\slashed{\epsilon}_{\lambda}\!(\underline{q})\;v(1)
      + (D\!-\!6)\; \frac{(\P\!\cdot\! \varepsilon_{\lambda})}{q^+}\;
              \overline{u}(0)\, \gamma^+\,  v(1)
\bigg\}\;\mathcal{A}_0(\Delta_2)
\label{I_1_T_II_result_spin_reduced}
 \, .
\end{eqnarray}

The relations \eqref{Bi_PV} and \eqref{Bij_PV} imply that only the $\mathcal{B}_{22}$ integral survives in the the first line in the contribution $IV$ in the expression \eqref{I_1_T_IV_result}. Then, using the identity \eqref{quark_current_times_epsilon_mom_cons}, one obtains
\begin{eqnarray}
\mathcal{I}_1^{T}\bigg|_{IV} &=& 2(D\!-\!2)\, \mathcal{B}_{22}\;
    \overline{u}(0)\,\slashed{\epsilon}_{\lambda}\!(\underline{q})\;v(1)
    +\Bigg\{-\frac{(D\!-\!2)(k_0^+\!-\!k_1^+)}{k_0^+ k_1^+}\, \mathcal{B}_{22}
    + \bigg[\frac{(D\!-\!6)}{k_2^+}+ \frac{(D\!-\!2)q^+}{2 k_0^+k_1^+}\bigg]\,
        \mathcal{A}_0(\Delta_2)
\nonumber\\
&&
+\frac{q^+}{2 k_0^+ k_1^+}
     \left[\frac{(k_1^+\!+\!k_2^+)}{k_1^+}\, \P^2
       +\frac{(k_0^+\!-\!k_2^+)}{k_0^+}\, \overline{Q}^2\right]\;
   \left[(D\!-\!2)\, \frac{(k_0^+\!-\!k_2^+)}{k_0^+}\, \mathcal{B}_1
    +\left(4-\frac{(D\!-\!2)k_2^+}{k_0^+}\right)\, \mathcal{B}_0\right]
\nonumber\\
&& 
+\frac{q^+}{2 k_1^+(k_0^+\!-\!k_2^+)}\,
     \left(4-\frac{(D\!-\!2)k_2^+}{k_0^+}\right)\,
      \Big[\mathcal{A}_0(\Delta_1)\!-\!\mathcal{A}_0(\Delta_2)\Big]
\Bigg\}\;
(\P\!\cdot\! \varepsilon_{\lambda})\; \overline{u}(0)\, \gamma^+ v(1)
\label{I_1_T_IV_result_spin_reduced}
 \, .
\end{eqnarray}

At this stage, note that only two different Dirac structures are left in the expression for the integral $\mathcal{I}_1^{T}$. Hence, the diagram $1$ for transverse photon can be written as a sum of one term proportional to $\overline{u}(0)\,\slashed{\epsilon}_{\lambda}\!(\underline{q})\;v(1)$ and thus proportional to the LO wave function \eqref{WF_T_LO}, and one term proportional to the new structure $(\P\!\cdot \varepsilon_{\lambda})\; \overline{u}(0)\, \gamma^+ v(1)$. This is reminiscent of the NLO corrections to the electron-photon vertex in QED, which can be written as a sum of two contributions with different Dirac structures, multiplied by the Dirac and Pauli form factors respectively.

Now, it is time to perform the last step in the Passarino-Veltman tensor reduction method \cite{Passarino:1978jh}, which is to use the algebraic relations \eqref{B1 result}, \eqref{B21 result} and \eqref{B22 result} in order to eliminate the scalar integrals $\mathcal{B}_{1}$, $\mathcal{B}_{21}$ and $\mathcal{B}_{22}$ in favor of $\mathcal{A}_0(\Delta_1)$, $\mathcal{A}_0(\Delta_2)$ and $\mathcal{B}_{0}$ in the expression for $\mathcal{I}_1^{T}$.
Using the definitions \eqref{def_Delta1}, \eqref{def_Delta2} and \eqref{def_L} relevant for the diagram 1, the algebraic relation \eqref{B1 result} is written
\begin{eqnarray}
 \mathcal{B}_1(\Delta_1,\Delta_2,\L^2)&=&
\frac{1}{2\, \P^2}\, \left(\frac{k_0^+}{k_0^+\!-\!k_2^+}\right)^2\,
\Bigg\{ \mathcal{A}_0(\Delta_1) - \mathcal{A}_0(\Delta_2)
    + \left(\frac{k_0^+\!-\!k_2^+}{k_0^+}\right)\,
    \bigg[\frac{q^+ k_2^+}{k_0^+ k_1^+}\, \left(\P^2+\overline{Q}^2\right)
\nonumber\\
&&
\hspace{3.5cm}
 -\left(\frac{k_1^+\!+\!k_2^+}{k_1^+}\right)\, \overline{Q}^2
    -\left(\frac{k_0^+\!-\!k_2^+}{k_0^+}\right)\, \P^2 \bigg]\, \mathcal{B}_0(\Delta_1,\Delta_2,\L)
\Bigg\}
\label{B1 result_applied}\, .
\end{eqnarray}
Inserting this relation in the expression \eqref{I_1_T_I_result_spin_reduced},
one finds
\begin{eqnarray}
\mathcal{I}_1^{T}\bigg|_{I} &=& -\frac{k_0^+ k_1^+}{k_2^+(k_0^+\!-\!k_2^+)}\,
\bigg\{\left[\frac{2k_0^+}{k_2^+} + \frac{(D\!-\!6)k_0^+}{q^+}\right]\,
       \overline{u}(0)\,\slashed{\epsilon}_{\lambda}\!(\underline{q})\;v(1)
      + (D\!-\!6)\; \frac{(\P\!\cdot\! \varepsilon_{\lambda})}{q^+}\;
              \overline{u}(0)\, \gamma^+\,  v(1)
\bigg\}\;
\Bigg\{\frac{(k_1^+\!+\!k_2^+)}{k_1^+}\,  \mathcal{A}_0(\Delta_2)
\nonumber\\
&& +\frac{q^+ k_2^+}{k_0^+ k_1^+}\,
\left(\frac{\P^2+\overline{Q}^2}{2\, \P^2}\right)\,
\bigg[\mathcal{A}_0(\Delta_1)-\mathcal{A}_0(\Delta_2)
-\left(\frac{k_0^+\!-\!k_2^+}{k_0^+}\right)\,
\left[\frac{(k_1^+\!+\!k_2^+)}{k_1^+}\, \P^2
  + \frac{(k_0^+\!-\!k_2^+)}{k_0^+}\, \overline{Q}^2 \right]
\mathcal{B}_0\bigg]
\Bigg\}
\label{I_1_T_I_result_tensor_reduced}
 \, .
\end{eqnarray}
For the contributions $II$ and $III$, the expressions \eqref{I_1_T_II_result_spin_reduced} and \eqref{I_1_T_III_result} are already expressed in terms of the scalar integrals $\mathcal{A}_0(\Delta_1)$, $\mathcal{A}_0(\Delta_2)$ and $\mathcal{B}_0$ only. Examining the contributions  \eqref{I_1_T_I_result_tensor_reduced}, \eqref{I_1_T_II_result_spin_reduced} and \eqref{I_1_T_III_result} to the integral $\mathcal{I}_1^{T}$, as well as the
expansions \eqref{A0D1_result},  \eqref{A0D2_result} and \eqref{B0_result} of the scalar integrals $\mathcal{A}_0(\Delta_1)$, $\mathcal{A}_0(\Delta_2)$ and $\mathcal{B}_0$ for $D\rightarrow 4$, one can see that a better basis of scalar integrals consists in $\mathcal{A}_0(\Delta_2)$, $\mathcal{I}_+$ and $\mathcal{I}_-$, with the definitions
\begin{eqnarray}
\mathcal{I}_{\pm} &\equiv & \frac{1}{2}\,
\bigg\{\mathcal{A}_0(\Delta_1)-\mathcal{A}_0(\Delta_2)
\pm\left(\frac{k_0^+\!-\!k_2^+}{k_0^+}\right)\,
\left[\left(\frac{k_1^+\!+\!k_2^+}{k_1^+}\right)\, \P^2
  + \left(\frac{k_0^+\!-\!k_2^+}{k_0^+}\right)\, \overline{Q}^2 \right]
\mathcal{B}_0\bigg\}
\label{I_plus_minus_def}
 \, .
\end{eqnarray}
From the expansions \eqref{A0D1_result},  \eqref{A0D2_result} and \eqref{B0_result}, one finds
\begin{eqnarray}
\mathcal{I}_+
&=& - \log\left(\frac{k_2^+}{k_0^+}\right)
+ \log\left(\frac{k_1^+\!+\!k_2^+}{k_1^+}\right)
- \log\left(\frac{q^+}{k_1^+}\right)
+ O\left(D\!-\!4\right)
\label{I_plus_result}
\\
\mathcal{I}_-
&=&
- \log\left(\frac{\P^2+\overline{Q}^2}{\overline{Q}^2}\right)
+ O\left(D\!-\!4\right)\, .
\label{I_minus_result}
\end{eqnarray}
From the expressions \eqref{A0D2_result}, \eqref{I_plus_result} and \eqref{I_minus_result}, it is clear that the three scalar integrals $\mathcal{A}_0(\Delta_2)$, $\mathcal{I}_+$ and $\mathcal{I}_-$ are linearly independent, and thus form a basis.
That basis of scalar integrals has various appealing properties. For $D\rightarrow 4$, the $\P$ dependence is entirely contained in $\mathcal{I}_-$, and separated from the dependence on $k^+_2$, $k^+_1$ and $k^+_0$, contained in $\mathcal{A}_0(\Delta_2)$ and $\mathcal{I}_+$. Only $\mathcal{A}_0(\Delta_2)$ is UV divergent. And finally, only $\mathcal{I}_+$ contains the term
$-\log(k_2^+/k_0^+)$, which might lead to an unphysical double logarithmic low $x$ divergence after integration over $k_2^+$.
Written with that basis of scalar integrals, the contributions $I$ and $III$ to $\mathcal{I}_1^{T}$ simplify as
\begin{eqnarray}
\mathcal{I}_1^{T}\bigg|_{I} &=& -\left(\frac{k_0^+}{k_0^+\!-\!k_2^+}\right)\,
\Bigg\{\left(\frac{k_1^+\!+\!k_2^+}{k_2^+}\right)\,  \mathcal{A}_0(\Delta_2)
+\frac{q^+}{k_0^+}\,
\left(\frac{\P^2+\overline{Q}^2}{\P^2}\right)\, \mathcal{I}_-
\Bigg\}\;
\nonumber\\
&& \times
\bigg\{\left[\frac{2k_0^+}{k_2^+} + \frac{(D\!-\!6)k_0^+}{q^+}\right]\,
       \overline{u}(0)\,\slashed{\epsilon}_{\lambda}\!(\underline{q})\;v(1)
      +(D\!-\!6)\; \frac{(\P\!\cdot\! \varepsilon_{\lambda})}{q^+}\;
              \overline{u}(0)\, \gamma^+\,  v(1)
\bigg\}
\label{I_1_T_I_result_final}\\
\mathcal{I}_1^{T}\bigg|_{III} &=& -4 \left(\frac{k_0^+}{k_0^+\!-\!k_2^+}\right)\,
\bigg\{-\frac{k_1^+(k_0^+\!-\!k_2^+)}{(k_2^+)^2}\, \mathcal{A}_0(\Delta_2)
+ \frac{q^+}{k_2^+}\, \mathcal{I}_+
 \bigg\} \,
\overline{u}(0)\,\slashed{\epsilon}_{\lambda}\!(\underline{q})\;v(1)
\label{I_1_T_III_result_final}
 \, ,
\end{eqnarray}
whereas the expression \eqref{I_1_T_II_result_spin_reduced} for the contribution $II$ involves only the scalar integral $\mathcal{A}_0(\Delta_2)$ already.

The expression \eqref{I_1_T_IV_result_spin_reduced} for the contribution $IV$ involves the integral $\mathcal{B}_{22}$. Translating the relation \eqref{B22 result} to the new base, one gets
\begin{eqnarray}
 (D\!-\!3)\, \mathcal{B}_{22}
&=&
\frac{1}{2}\, \mathcal{A}_0(\Delta_2)
- \left(\frac{k_0^+}{k_0^+\!-\!k_2^+}\right)\,
    \frac{q^+k_2^+}{2 k_0^+k_1^+}\,  \mathcal{I}_+
+ \left(\frac{\P^2+\overline{Q}^2}{2 \P^2}\right)\, \mathcal{I}_-
\label{B22 result_new_base}\, .
\end{eqnarray}
Using the relation \eqref{B22 result_new_base} to completely eliminate $\mathcal{B}_{22}$ from  the contribution $IV$ in \eqref{I_1_T_IV_result_spin_reduced}, one would obtain terms with a pole at $D=3$. That pole is not associated with a UV or IR divergence. Instead, it is just a reminder that the method used to calculate the tensor integral $\mathcal{B}^{ij}$ is not valid at $D=3$ because the system of equations
\eqref{Bij rel_1} and \eqref{Bij rel_2} becomes degenerate in that case.
However, this is not a problem, since only the vicinity of $D=4$ is of interest for the present calculation. Instead of completely eliminating $\mathcal{B}_{22}$, it turns out more convenient to make the replacement
\begin{eqnarray}
\mathcal{B}_{22}
&=&
\frac{1}{2}\, \mathcal{A}_0(\Delta_2)
- \left(\frac{k_0^+}{k_0^+\!-\!k_2^+}\right)\,
    \frac{q^+k_2^+}{2k_0^+k_1^+}\,  \mathcal{I}_+
+ \left(\frac{\P^2+\overline{Q}^2}{2\P^2}\right)\, \mathcal{I}_-
- (D\!-\!4)\, \mathcal{B}_{22}
\label{B22 result_with_rational_term}\, .
\end{eqnarray}
Thanks to the expansions \eqref{A0D2_result}, \eqref{I_plus_result} and \eqref{I_minus_result} of the scalar integrals $\mathcal{A}_0(\Delta_2)$, $\mathcal{I}_+$ and $\mathcal{I}_-$, and the relation \eqref{B22 result_with_rational_term} itself, one finds
\begin{eqnarray}
- (D\!-\!4)\, \mathcal{B}_{22}
= - \frac{(D\!-\!4)}{2}\,  \mathcal{A}_0(\Delta_2)
     + O\left(D\!-\!4\right)
=  1
+ O\left(D\!-\!4\right)
\label{rational_term_expansion}\, .
\end{eqnarray}
This term, arising in the $D\rightarrow 4$ expansion with a coefficient $(D\!-\!4)/(D\!-\!4)$ is a rational term, as defined in the context of calculations of scattering amplitudes with loops in covariant perturbation theory. However, it corresponds to a different type of rational term than the one already encountered and discussed in section \ref{sec:self_energy}. The rational term \eqref{rational_term_expansion} is induced by the tensor reduction of the integral $\mathcal{B}^{ij}$ and does not depend on which variant of dimensional regularization is used. Such terms are typically missed by naive UV regularizations which do not preserve invariance of the integrals by shift of the integration variable (such as cutoff regularization) and represent one of the main challenges for attempts at formulating consistent UV regularizations directly in $4$ dimensions.

Using the relations \eqref{B1 result_applied}, \eqref{I_plus_minus_def} and \eqref{B22 result_with_rational_term}, one can now rewrite \eqref{I_1_T_IV_result_spin_reduced}, after some algebra, as
\begin{eqnarray}
\mathcal{I}_1^{T}\bigg|_{IV} &=& (D\!-\!2)\,\Bigg\{ \mathcal{A}_0(\Delta_2)
- \left(\frac{k_0^+}{k_0^+\!-\!k_2^+}\right)\,
    \frac{q^+k_2^+}{k_0^+k_1^+}\,  \mathcal{I}_+
+ \left(\frac{\P^2+\overline{Q}^2}{\P^2}\right)\, \mathcal{I}_-
-2 (D\!-\!4)\, \mathcal{B}_{22}
\Bigg\}\;
\overline{u}(0)\,\slashed{\epsilon}_{\lambda}\!(\underline{q})\;v(1)
\nonumber\\
&&
+\Bigg\{
\bigg[\frac{(D\!-\!6)}{k_2^+}+ \frac{(D\!-\!2)}{k_0^+}\bigg]\,
        \mathcal{A}_0(\Delta_2)
+ \left(\frac{k_0^+}{k_0^+\!-\!k_2^+}\right)\,
    \frac{q^+}{k_0^+k_1^+}\, \left[4+(D\!-\!2)\frac{k_2^+(k_0^+\!-\!k_1^+)}{k_0^+k_1^+}\right] \mathcal{I}_+
\nonumber\\
&& \hspace{0.5cm}
+\frac{(D\!-\!2)}{k_0^+}\,
    \left(\frac{\P^2+\overline{Q}^2}{\P^2}\right)\, \mathcal{I}_-
+ \frac{(k_0^+\!-\!k_1^+)}{k_0^+ k_1^+}\, (D\!-\!2)(D\!-\!4) \mathcal{B}_{22}
\Bigg\}\;
(\P\!\cdot\! \varepsilon_{\lambda})\; \overline{u}(0)\, \gamma^+ v(1)
\label{I_1_T_IV_result_final}
 \, .
\end{eqnarray}
At this stage, the integral $\mathcal{I}_1^{T}$ has been fully decomposed on a basis of two independent Dirac structures and on a basis of three scalar integrals, up to a rational term $(D\!-\!4) \mathcal{B}_{22}$, and the result is the sum of the expressions \eqref{I_1_T_I_result_final}, \eqref{I_1_T_II_result_spin_reduced}, \eqref{I_1_T_III_result_final} and \eqref{I_1_T_IV_result_final}. A few comments are in order. In several terms, there is one factor of $(k_0^+\!-\!k_2^+)$ in the denominator. This is not a problem, since $\mathcal{I}_1^{T}$ is multiplied by $(k_0^+\!-\!k_2^+)$ before taking the integral over $k_2^+$, see eq. \eqref{WF_1_T_2}. Hence, divergences in the $k_2^+$ integral can come only from the $k_2^+\rightarrow 0$ regime.
In the contributions $I$, $II$ and $III$, the coefficient of $\mathcal{A}_0(\Delta_2)\, \overline{u}(0)\,\slashed{\epsilon}_{\lambda}\!(\underline{q})\;v(1)$ is of order $1/(k_2^+)^2$ for $k_2^+\rightarrow 0$, which would produce an unphysical power divergence. All the other coefficients are of order  $1/k_2^+$ at most, and thus leading to logarithmic divergences at most, except for the $\mathcal{I}_+\, \overline{u}(0)\,\slashed{\epsilon}_{\lambda}\!(\underline{q})\;v(1)$ term in the contribution $III$ in eq. \eqref{I_1_T_III_result_final}, which gives a double log divergence at small $k_2^+$.

Defining the two form factors ${\cal V}_{1}^{T}$ and ${\cal N}_{1}^{T}$ through the relation
\begin{eqnarray}
-\frac{1}{2q^+}\;
\int_{0}^{k_0^+}\!\!
d{k_2^+}\;
\frac{(k_{0}^+\!-\! k_{2}^+)}{k_{0}^+}\;
\mathcal{I}_1^{T}
&=&  {\cal V}_{1}^{T}\; \overline{u}(0)\,\slashed{\epsilon}_{\lambda}\!(\underline{q})\;v(1)
+ {\cal N}_{1}^{T}\; \frac{(\P\!\cdot\! \varepsilon_{\lambda})}{q^+}\;
              \overline{u}(0)\, \gamma^+\,  v(1)
\, ,
\label{def_form_fact}
\end{eqnarray}
the diagram 1 contribution to the wave-function is written
\begin{eqnarray}
\Psi_{\gamma_T^{*}\rightarrow q_0+\bar{q}_1}^{1}
&=&(2\pi)^{D-1}\delta^{(D-1)}(\underline{k_{0}}+\underline{k_{1}}\!-\!
\underline{q})\; \delta_{\alpha_{0},\, \alpha_{1}}\;
\frac{e\, e_f}{(ED_{LO})}\, \mu^{2-\frac{D}{2}}\,
\left[\frac{\alpha_s\, C_F}{2\pi}\right]\,
\nonumber\\
&& \times\;
\bigg\{
{\cal V}_{1}^{T}\; \overline{u}(0)\,\slashed{\epsilon}_{\lambda}\!(\underline{q})\;v(1)
+ {\cal N}_{1}^{T}\; \frac{(\P\!\cdot\! \varepsilon_{\lambda})}{q^+}\;
              \overline{u}(0)\, \gamma^+\,  v(1)
\bigg\}
\, .\label{WF_1_T_3}
\end{eqnarray}
Collecting contributions from the results \eqref{I_1_T_I_result_final}, \eqref{I_1_T_II_result_spin_reduced}, \eqref{I_1_T_III_result_final} and \eqref{I_1_T_IV_result_final}, the two form factors are found to be
\begin{eqnarray}
{\cal V}_{1}^{T}&=& \int_{0}^{k_0^+}\!\! d{k_2^+}\;
\Bigg\{ \bigg[\frac{1}{k_2^+}-\frac{2}{q^+}
      +(D\!-\!2)\frac{k_2^+}{2q^+k_0^+}\bigg]\,
\bigg[\mathcal{A}_0(\Delta_2)
      +\left(\frac{\P^2+\overline{Q}^2}{\P^2}\right)\, \mathcal{I}_-\bigg]
\nonumber\\
&& \hspace{2cm}
+ \bigg[\frac{2}{k_2^+}+(D\!-\!2)\frac{k_2^+}{2k_0^+k_1^+}\bigg]\,
           \mathcal{I}_+
+ \left(\frac{k_0^+\!-\!k_2^+}{k_0^+}\right)\, \frac{(D\!-\!2)}{q^+}\,(D\!-\!4)\,\mathcal{B}_{22}
\Bigg\}
\label{V1T_decomp}
\\
{\cal N}_{1}^{T}&=& \int_{0}^{k_0^+}\!\! \frac{d{k_2^+}}{k_0^+}\;
\Bigg\{ \bigg[-2 +(D\!-\!2)\frac{k_2^+}{2k_0^+}\bigg]\,
\bigg[\mathcal{A}_0(\Delta_2)
      +\left(\frac{\P^2+\overline{Q}^2}{\P^2}\right)\, \mathcal{I}_-\bigg]
\nonumber\\
&& \hspace{2cm}
-\frac{q^+}{k_1^+}\, \bigg[2 +(D\!-\!2)\frac{k_2^+ (k_0^+\!-\!k_1^+)}{2k_0^+k_1^+}\bigg]\,
           \mathcal{I}_+
- \left(\frac{k_0^+\!-\!k_2^+}{k_0^+}\right)\,
\left(\frac{k_0^+\!-\!k_1^+}{k_1^+}\right)\, \frac{(D\!-\!2)}{2}\,(D\!-\!4)\,\mathcal{B}_{22}
\Bigg\}
\label{N1T_decomp}
\, .
\end{eqnarray}
Note that the terms of order $1/(k_2^+)^2$ for $k_2^+\rightarrow 0$ have canceled each other, so that the $k_2^+$ integral in ${\cal V}_{1}^{T}$ has only single and double logarithmic divergences at small $k_2^+$.
Moreover, in ${\cal N}_{1}^{T}$, all the terms of order $1/k_2^+$ have canceled each other as well, so that $k_2^+$ integral in ${\cal N}_{1}^{T}$ is convergent.

The small $k_2^+$ divergences are chosen to be regulated by a cutoff $k_2^+>k^+_{\min}$, like in section \ref{sec:self_energy} and not by the dimensional regularization itself, only used to regulate UV divergences. Hence, the evaluation of ${\cal V}_{1}^{T}$ and ${\cal N}_{1}^{T}$ is best done by first expanding around $D=4$, and then integrating over $k_2^+$.
For ${\cal N}_{1}^{T}$, there is no small $k_2^+$ divergence, as already said, but that method still simplifies the calculation. Remembering that $\mathcal{A}_0(\Delta_2)$ and $\mathcal{B}_{22}$ are UV divergent in four dimensions whereas $\mathcal{I}_+$ and $\mathcal{I}_-$ are UV finite, one can first isolate all the rational terms in ${\cal N}_{1}^{T}$ and then use the expansions \eqref{A0D2_result}, \eqref{I_plus_result}, \eqref{I_minus_result} and \eqref{rational_term_expansion} of the integrals as
\begin{eqnarray}
{\cal N}_{1}^{T}&=& \int_{0}^{k_0^+}\!\! \frac{d{k_2^+}}{k_0^+}\;
\Bigg\{ \bigg[-2 +\frac{k_2^+}{k_0^+}\bigg]\,
\bigg[\mathcal{A}_0(\Delta_2)
      +\left(\frac{\P^2+\overline{Q}^2}{\P^2}\right)\, \mathcal{I}_-\bigg]
-\frac{q^+}{k_1^+}\, \bigg[2+\frac{k_2^+(k_0^+\!-\!k_1^+)}{k_0^+k_1^+}\bigg]\,
    \mathcal{I}_+
\nonumber\\
&& \hspace{1cm}
   - \frac{k_2^+}{k_0^+}\; \left(2\!-\!\frac{D}{2}\right)\, \mathcal{A}_0(\Delta_2)
- \left(\frac{k_0^+\!-\!k_2^+}{k_0^+}\right)\,
\left(\frac{k_0^+\!-\!k_1^+}{k_1^+}\right)\, (D\!-\!4)\,\mathcal{B}_{22}
\Bigg\} + O\left(D\!-\!4\right)
\nonumber\\
&=& \int_{0}^{k_0^+}\!\! \frac{d{k_2^+}}{k_0^+}\;
\Bigg\{ \bigg[-2 +\frac{k_2^+}{k_0^+}\bigg]\,
\bigg[
\Gamma\!\left(2\!-\! \frac{D}{2}\right)\,
\left(\frac{\overline{Q}^2}{4\pi\, \mu^2}\right)^{\frac{D}{2}-2}
\!\!\!\!\! -\frac{(\P^2+\overline{Q}^2)}{\P^2}\; \log\left(\frac{\P^2+\overline{Q}^2}{\overline{Q}^2}\right)
- \log\left(\frac{k_0^+\!-\!k_2^+}{k_0^+}\right)
-\log\left(\frac{k_1^+\!+\!k_2^+}{k_1^+}\right)
\bigg]
\nonumber\\
&& \hspace{2cm}
-\frac{q^+}{k_1^+}\, \bigg[2+\frac{k_2^+(k_0^+\!-\!k_1^+)}{k_0^+k_1^+}\bigg]\,
   \bigg[-\log\left(\frac{k_2^+}{k_0^+}\right)
         +\log\left(\frac{k_1^+\!+\!k_2^+}{k_1^+}\right)
         -\log\left(\frac{q^+}{k_1^+}\right)
   \bigg]
\nonumber\\
&& \hspace{2cm}
   - \frac{k_2^+}{k_0^+}
+ \left(\frac{k_0^+\!-\!k_2^+}{k_0^+}\right)\,
\left(\frac{k_0^+\!-\!k_1^+}{k_1^+}\right)
\Bigg\} + O\left(D\!-\!4\right)
\nonumber\\
&=& -\frac{3}{2}\;
\Gamma\!\left(2\!-\! \frac{D}{2}\right)\,
\left(\frac{\overline{Q}^2}{4\pi\, \mu^2}\right)^{\frac{D}{2}-2}
+\frac{3(\P^2+\overline{Q}^2)}{2\P^2}\; \log\left(\frac{\P^2+\overline{Q}^2}{\overline{Q}^2}\right)
-3-\frac{k_0^+}{2k_1^+}
\nonumber\\
&& \hspace{2cm}
  -\frac{1}{2} +  \frac{(k_0^+\!-\!k_1^+)}{2k_1^+} + O\left(D\!-\!4\right)
\label{N1T_eval_1}
\, .
\end{eqnarray}
In the last step, the integration over $k_2^+$ has been performed using the identities given in appendix \ref{sec:integrals}, and obvious algebraic simplifications have been performed. In particular, all of the terms in $\log(q^+/k_1^+)$ are found to cancel each other in a nontrivial way.
Nevertheless, for illustrative purposes, the contribution of the rational terms is kept separate, in the last line of eq. \eqref{N1T_eval_1}. The $-1/2$ term is the rational term induced by the $D$ dependence of the coefficient of $\mathcal{A}_0(\Delta_2)$, whereas the last term is the rational term produced in the Passarino-Veltman tensor reduction of the integral $\mathcal{B}^{ij}$. Obviously, the final result for ${\cal N}_{1}^{T}$ is
\begin{eqnarray}
{\cal N}_{1}^{T}
&=& -\frac{3}{2}\;
\Gamma\!\left(2\!-\! \frac{D}{2}\right)\,
\left(\frac{\overline{Q}^2}{4\pi\, \mu^2}\right)^{\frac{D}{2}-2}
+\frac{3(\P^2+\overline{Q}^2)}{2\P^2}\; \log\left(\frac{\P^2+\overline{Q}^2}{\overline{Q}^2}\right)
-4 + O\left(D\!-\!4\right)
\label{N1T_eval_2}
\, .
\end{eqnarray}

The calculation of ${\cal V}_{1}^{T}$ is done in the same way starting from eq. \eqref{V1T_decomp}, except that the cutoff $k_2^+>k^+_{\min}$ has to be imposed when needed. One finds
\begin{eqnarray}
{\cal V}_{1}^{T}&=& \int_{0}^{k_0^+}\!\! \frac{d{k_2^+}}{k_0^+}\;
\Bigg\{ \bigg[\frac{k_0^+}{k_2^+}-\frac{2k_0^+}{q^+}
      +\frac{k_2^+}{q^+}\bigg]\,
\bigg[\mathcal{A}_0(\Delta_2)
      +\left(\frac{\P^2+\overline{Q}^2}{\P^2}\right)\, \mathcal{I}_-\bigg]
+ \bigg[\frac{2k_0^+}{k_2^+}+\frac{k_2^+}{k_1^+}\bigg]\,
           \mathcal{I}_+
\nonumber\\
&& \hspace{2cm}
-\frac{k_2^+}{q^+}\; \left(2\!-\!\frac{D}{2}\right)\, \mathcal{A}_0(\Delta_2)
+ 2\left(\frac{k_0^+\!-\!k_2^+}{q^+}\right)\, \,(D\!-\!4)\,\mathcal{B}_{22}
\Bigg\}
+ O\left(D\!-\!4\right)
\nonumber\\
&=& \int_{0}^{k_0^+}\!\! \frac{d{k_2^+}}{k_0^+}\;
\Bigg\{ \bigg[\frac{k_0^+}{k_2^+}-\frac{2k_0^+}{q^+}
      +\frac{k_2^+}{q^+}\bigg]\,
\bigg[
\Gamma\!\left(2\!-\! \frac{D}{2}\right)\,
\left(\frac{\overline{Q}^2}{4\pi\, \mu^2}\right)^{\frac{D}{2}-2}
\!\!\!\!\! -\frac{(\P^2+\overline{Q}^2)}{\P^2}\; \log\left(\frac{\P^2+\overline{Q}^2}{\overline{Q}^2}\right)
\nonumber\\
&& \hspace{1cm}
- \log\left(\frac{k_0^+\!-\!k_2^+}{k_0^+}\right)
-\log\left(\frac{k_1^+\!+\!k_2^+}{k_1^+}\right)
\bigg]
+ \bigg[\frac{2k_0^+}{k_2^+}+\frac{k_2^+}{k_1^+}\bigg]\,
   \bigg[-\log\left(\frac{k_2^+}{k_0^+}\right)
         +\log\left(\frac{k_1^+\!+\!k_2^+}{k_1^+}\right)
         -\log\left(\frac{q^+}{k_1^+}\right)
   \bigg]
\nonumber\\
&& \hspace{2cm}
   -\frac{k_2^+}{q^+}
-2\left(\frac{k_0^+\!-\!k_2^+}{q^+}\right)
\Bigg\} + O\left(D\!-\!4\right)
\nonumber\\
&=&
- \left[\log\left(\frac{k^+_{\min}}{k_0^+}\right) +\frac{3k_0^+}{2q^+} \right]\,
\left[
\Gamma\!\left(2\!-\! \frac{D}{2}\right)\,
\left(\frac{\overline{Q}^2}{4\pi\, \mu^2}\right)^{\frac{D}{2}-2}
-\frac{(\P^2+\overline{Q}^2)}{\P^2}\; \log\left(\frac{\P^2+\overline{Q}^2}{\overline{Q}^2}\right)
\right]
\nonumber\\
&&
+\left[\log\left(\frac{k^+_{\min}}{k_0^+}\right)\right]^2
 + \left[2\log\left(\frac{k^+_{\min}}{k_0^+}\right) +\frac{3}{2} \right]
  \log\left(\frac{q^+}{k_1^+}\right)
 +\frac{\pi^2}{6} - \textrm{Li}_2\left(-\frac{k_0^+}{k_1^+}\right)
 -\frac{5k_0^+}{2q^+}
\nonumber\\
&&
 -\frac{k_0^+}{2q^+} -\frac{k_0^+}{q^+}
 + O\left(D\!-\!4\right)
\label{V1T_eval_1}
\, ,
\end{eqnarray}
where $\textrm{Li}_2$ is the dilogarithm function, defined as
\begin{eqnarray}
\textrm{Li}_2(z)&\equiv & -\int_{0}^{z} \frac{d\xi}{\xi}\; \log(1\!-\!\xi)
\, .
\label{def_Li2}
\end{eqnarray}
In the result \eqref{V1T_eval_1}, the last two terms correspond to the two types of rational terms, due, respectively, to the $D$ dependence of the coefficient of $\mathcal{A}_0(\Delta_2)$ and to the tensor reduction of the integral $\mathcal{B}^{ij}$.


\subsection{Summing the graphs}

The calculations performed so far in this section show that, at NLO accuracy (which means including terms of order $e\, \alpha_s$),
the LFWF for the splitting of a transverse photon into a quark-antiquark pair can be written in momentum space as
\begin{eqnarray}
\Psi_{\gamma_T^{*}\rightarrow q_0+\bar{q}_1}^{NLO}
&=&(2\pi)^{D-1}\delta^{(D-1)}(\underline{k_{0}}+\underline{k_{1}}\!-\!
\underline{q})\; \delta_{\alpha_{0},\, \alpha_{1}}\;
\frac{e\, e_f}{(ED_{LO})}\, \mu^{2-\frac{D}{2}}\,
\nonumber\\
&& \times\;
\Bigg\{\left[1+\left(\frac{\alpha_s\, C_F}{2\pi}\right)\; {\cal V}^{T}\;\right]\,
 \overline{u}(0)\,\slashed{\epsilon}_{\lambda}\!(\underline{q})\;v(1)
+ \left(\frac{\alpha_s\, C_F}{2\pi}\right)\,{\cal N}^{T}\; \frac{(\P\!\cdot\! \varepsilon_{\lambda})}{q^+}\;
              \overline{u}(0)\, \gamma^+\,  v(1)
\Bigg\}
\, ,\label{WF_NLO_T_1}
\end{eqnarray}
with the form factors
\begin{eqnarray}
{\cal V}^{T} &\equiv& {\cal V}_{1}^{T} +{\cal V}_{2}^{T} +{\cal V}_{A}^{T} +{\cal V}_{B}^{T}
\nonumber\\
{\cal N}^{T} &\equiv& {\cal N}_{1}^{T} +{\cal N}_{2}^{T}
\, ,
\end{eqnarray}
with the contributions ${\cal V}_{A}^{T}$, ${\cal V}_{1}^{T}$ and ${\cal N}_{1}^{T}$ given in eqs. \eqref{VAT_eval_1}, \eqref{V1T_eval_1} and \eqref{N1T_eval_2} respectively, whereas the others correspond to the symmetric diagrams obtained by exchange of the quark and antiquark.

It is instructive to first add together ${\cal V}_{A}^{T}$ and ${\cal V}_{1}^{T}$. One gets
\begin{eqnarray}
{\cal V}_{1}^{T}+{\cal V}_{A}^{T}
&=&
\left[\log\left(\frac{k^+_{\min}}{k_0^+}\right) +\frac{3k_1^+}{2q^+} \right]\,
\Gamma\!\left(2\!-\! \frac{D}{2}\right)\,
\left(\frac{\overline{Q}^2}{4\pi\, \mu^2}\right)^{\frac{D}{2}-2}
+\left[\log\left(\frac{k^+_{\min}}{k_0^+}\right) +\frac{3k_0^+}{2q^+}\right]\,
  \frac{(\P^2+\overline{Q}^2)}{\P^2}\; \log\left(\frac{\P^2+\overline{Q}^2}{\overline{Q}^2}\right)
\nonumber\\
&&
-\left[2\log\left(\frac{k^+_{\min}}{k_0^+}\right) +\frac{3}{2}\right]\,
   \log\left(\frac{\P^2+\overline{Q}^2}{\overline{Q}^2}\right)
-\frac{\pi^2}{6} - \textrm{Li}_2\left(-\frac{k_0^+}{k_1^+}\right)
 +\frac{5k_1^+}{2q^+}+\frac{1}{2}
\nonumber\\
&&
 +\frac{k_1^+}{2q^+} -\frac{k_0^+}{q^+}
 + O\left(D\!-\!4\right)
\label{V1T_plus_VAT_1}
\, ,
\end{eqnarray}
isolating again the two types of rational terms. Note that the unphysical double logarithmic divergences at small $k_2^+$ have canceled between the graphs $1$ and $A$, as well as all the terms proportional to $\log(q^+/k_1^+)$.

In terms of kinematics, the diagrams $2$ and $B$ are the images of the diagrams $1$ and $A$ by interchange on the quark with momentum
$\underline{k_{0}}$ and of the antiquark with momentum $\underline{k_{1}}$.
This implies that the contributions ${\cal V}_{B}^{T}$, ${\cal V}_{2}^{T}$ and ${\cal N}_{2}^{T}$ to the form factors can be obtained from ${\cal V}_{A}^{T}$, ${\cal V}_{1}^{T}$ and ${\cal N}_{1}^{T}$ respectively, by interchange of $\underline{k_{0}}$ and $\underline{k_{1}}$.
Under this interchange, the relative momentum $\P$ defined in eq. \eqref{cv_k0_to_P} flips sign.
As ${\cal V}_{A}^{T}$ and ${\cal V}_{1}^{T}$ depend on the transverse momenta $\k_0$ and $\k_1$ only through $\P^2$, one finds
\begin{eqnarray}
{\cal V}^{T}
&=& \bigg[ {\cal V}_{1}^{T}  +{\cal V}_{A}^{T}\bigg]
+ \Big(k_0^+ \leftrightarrow k_1^+\Big)
\nonumber\\
&=&
\left[\log\left(\frac{k^+_{\min}}{k_0^+}\right) +\log\left(\frac{k^+_{\min}}{k_1^+}\right)
+\frac{3}{2} \right]\,
\left[
\Gamma\!\left(2\!-\! \frac{D}{2}\right)\,
\left(\frac{\overline{Q}^2}{4\pi\, \mu^2}\right)^{\frac{D}{2}-2}
+\frac{\left(\overline{Q}^2\!-\!\P^2\right)}{\P^2}\; \log\left(\frac{\P^2+\overline{Q}^2}{\overline{Q}^2}\right)
\right]
\nonumber\\
&&
- \textrm{Li}_2\left(-\frac{k_0^+}{k_1^+}\right)
- \textrm{Li}_2\left(-\frac{k_1^+}{k_0^+}\right)
-\frac{\pi^2}{3}
 +\frac{7}{2}
 +\frac{1}{2} -1
 + O\left(D\!-\!4\right)
\label{VT_result_1}
\, .
\end{eqnarray}
As previously, the last two terms in eq. \eqref{VT_result_1} collect the contributions from the two types of rational terms.
Making use of the identity (see ref. \cite{gradshteyn2000table})
\begin{eqnarray}
\textrm{Li}_2\left(-z\right)+\textrm{Li}_2\left(-\frac{1}{z}\right)
&=& -\frac{\pi^2}{6} -\frac{1}{2}\, \Big[\log(z)\Big]^2\, ,
\label{prop_Li2}
\end{eqnarray}
valid in particular for $z>0$, the final result for the form factor ${\cal V}^{T}$  can be written as
\begin{eqnarray}
{\cal V}^{T}
&=&
\left[\log\left(\frac{k^+_{\min}}{k_0^+}\right) +\log\left(\frac{k^+_{\min}}{k_1^+}\right)
+\frac{3}{2} \right]\,
\left[
\Gamma\!\left(2\!-\! \frac{D}{2}\right)\,
\left(\frac{\overline{Q}^2}{4\pi\, \mu^2}\right)^{\frac{D}{2}-2}
+\frac{\left(\overline{Q}^2\!-\!\P^2\right)}{\P^2}\; \log\left(\frac{\P^2+\overline{Q}^2}{\overline{Q}^2}\right)
\right]
\nonumber\\
&&
+\frac{1}{2}\; \left[\log\left(\frac{k_0^+}{k_1^+}\right)\right]^2
-\frac{\pi^2}{6}
 +3
 + O\left(D\!-\!4\right)
\label{VT_result_2}
\, .
\end{eqnarray}

For the other form factor ${\cal N}^{T}$, one sees from eq. \eqref{N1T_eval_2} that also ${\cal N}_{1}^{T}$ depends on the transverse momenta $\k_0$ and $\k_1$ only through $\P^2$. However, there is a factor $(\P\!\cdot\! \varepsilon_{\lambda})$ in front of ${\cal N}^{T}$ in eq. \eqref{WF_NLO_T_1}, which flips sign under the interchange of $\underline{k_{0}}$ and $\underline{k_{1}}$. Hence, one has
\begin{eqnarray}
{\cal N}^{T}
=  {\cal N}_{1}^{T} + {\cal N}_{2}^{T}
= {\cal N}_{1}^{T}
- \Big(k_0^+ \leftrightarrow k_1^+\Big)
= 0
\label{NT_result}
\, .
\end{eqnarray}
Since ${\cal N}^{T}$ would have been a new type of form factor, not present in the LO wave function, one should have expected a cancellation of all the divergences in ${\cal N}^{T}$, both the UV ones and the small $k_2^+$ ones. One obtains actually a full cancellation of ${\cal N}^{T}$. This could be the result of a symmetry, but the use of light-front perturbation theory makes it complicated to investigate further the reason for that cancellation.

It is important to note that the cancellation of ${\cal N}^{T}$ could not have been obtained in a calculation performed in $4$ dimensions with a naive UV regulator, thus missing the rational terms. It is clear from the expression \eqref{N1T_eval_1} that the rational term associated with the Passarino-Veltman decomposition of the tensor integral $\mathcal{B}^{ij}$ is crucial to make ${\cal N}_{1}^{T}$ invariant under the exchange of $\underline{k_{0}}$ and $\underline{k_{1}}$, leading to the vanishing of ${\cal N}^{T}$.
By contrast, the other rational term (coming from the $D$ dependence of the coefficient of the UV-divergent integral $\mathcal{A}_0(\Delta_2)$) gives a symmetric contribution to ${\cal N}_{1}^{T}$ which drops identically from \eqref{NT_result}. That type of rational term is the most scheme-dependent part of our calculation. Indeed, it would be most likely the only piece changing if that calculation was done in another variant of dimensional regularization, like in the dimensional reduction scheme \cite{Siegel:1979wq} or in the four-dimensional helicity scheme \cite{Bern:1991aq,Bern:2002zk}, instead of the conventional dimensional regularization \cite{'tHooft:1972fi} used in the present calculation.


\section{Longitudinal photon: quark-antiquark Fock component at one loop\label{sec:NLO_WF_L}}

In the case of a longitudinal photon, the graphs contributing to the $\gamma^*_{L}\rightarrow q\bar{q}$ LFWF at one loop are the same as the graphs $A$, $B$, $1$, $2$ and $3$ from fig. \ref{Fig:gammaT_NLO}, up to a replacement of the $\gamma\rightarrow q\bar{q}$ vertex by the effective vertex \eqref{gammaL_to_qqbar_vertex} for splitting of a longitudinal photon into $q\bar{q}$.
Note that in the longitudinal photon case, there is no graph analog to $A'$, $B'$, $1'$ or $2'$, since the longitudinal photon is already an internal piece of an instantaneous Coulomb interaction vertex.


\subsection{Diagram A for longitudinal photon\label{sec:diag_A_L}}

According to the results of the section \ref{sec:self_energy} the contribution of the diagram $A$ to the longitudinal photon LFWF factorizes as
\begin{eqnarray}
\Psi_{\gamma_L^{*}\rightarrow q_0+\bar{q}_1}^{A}
&=&
\Psi_{\gamma_L^{*}\rightarrow q_0+\bar{q}_1}^{LO}\;
\times\, \left[\frac{\alpha_s\, C_F}{2\pi}\right]\,
{\cal V}_{A}^{L}
\, ,\label{WF_A_L}
\end{eqnarray}
in the same way as in the transverse photon case, and with the same factor given in eq. \eqref{VAT_eval_1}
\begin{eqnarray}
{\cal V}_{A}^{L}
&=&
{\cal V}_{A}^{T}
\nonumber\\
&=&
\left[2\log\left(\frac{k^+_{\min}}{k_0^+}\right) +\frac{3}{2} \right]\,
\bigg[
\frac{\Gamma\left(3\!-\! \frac{D}{2}\right)}{\left(2\!-\!\frac{D}{2}\right)}\;
\left(\frac{\overline{Q}^2}{4\pi\, \mu^2}\right)^{\frac{D}{2}-2}
-\log\left(\frac{\P^2+\overline{Q}^2}{\overline{Q}^2}\right)
-\log\left(\frac{q^+}{k_1^+}\right)
\bigg]
\nonumber\\
&&
 -\left[\log\left(\frac{k^+_{\min}}{k_0^+}\right)\right]^2
 -\frac{\pi^2}{3} + 3+\frac{1}{2}
 + O\left(D\!-\!4\right)
\label{VAL_eval_1}
\, .
\end{eqnarray}


\subsection{Diagram 1 for longitudinal photon\label{sec:diag_1_L}}

In the longitudinal photon case, the contribution of the diagram $1$ to the  $\gamma^*_{L}\rightarrow q\bar{q}$ LFWF is written
\begin{eqnarray}
\Psi_{\gamma_L^{*}\rightarrow q_0+\bar{q}_1}^{1}
&=&  \sum_{q_{0'} \bar{q}_{1'} \textrm{ states}}
\sum_{g_{2} \textrm{ states}}
\frac{
\langle 0| b_0\, V_I(0)\, b_{0'}^{\dag} a_{2}^{\dag} |0 \rangle\;
\langle 0| a_{2} d_{1} \, V_I(0)\,  d_{1'}^{\dag} |0 \rangle\;\;
V_{\gamma_L \rightarrow q_{0'} \bar{q}_{1'}}
}{(ED_{LO})\; (ED_A)\; (ED_V)}
\nonumber\\
&=&(2\pi)^{D-1}\delta^{(D-1)}(\underline{k_{0}}+\underline{k_{1}}\!-\!
\underline{q})\; \delta_{\alpha_{0},\, \alpha_{1}}\;
\frac{e\, e_f}{(ED_{LO})}\; \frac{Q}{q^+}\; (\mu^3)^{2-\frac{D}{2}}\, \frac{(-g^2)C_F}{2\pi}\,
\nonumber\\
&& \times\;
\int_{0}^{k_0^+}\!\!
\frac{d {k_2^+}}{2 k_{2}^+}\;
\frac{1}{2(k_{0}^+\!-\! k_{2}^+)}\;
\frac{1}{2(k_{1}^+\!+\! k_{2}^+)}\;
\int \frac{d^{D-2} \k_{2}}{(2\pi)^{D-2} }\;
\frac{\textrm{Num}_{1}^L}{(ED_A)\; (ED_V)}
\, ,
\label{WF_1_L_1}
\end{eqnarray}
where
\begin{eqnarray}
\textrm{Num}_{1}^L &=&
\sum_{\textrm{phys. pol. }\lambda_2} \;\;\; \sum_{h_{0'}, h_{1'}=\pm 1/2}  \overline{u}(0)\; \slashed{\epsilon}_{\lambda_2}\!(\underline{k_2})\; u(0')\;\;
 \overline{u}(0')\,  \gamma^+\, v(1')\;\;
\overline{v}(1')\,  \slashed{\epsilon}_{\lambda_2}^*\!(\underline{k_2})\;  v(1)
\, .
\label{num_1_L_def}
\end{eqnarray}
The calculation of the numerator $\textrm{Num}_{1}^L$ is done in appendix \ref{sec:num_1_L}, and gives
\begin{eqnarray}
\textrm{Num}_{1}^L &=&
\, 2\Bigg\{\frac{k_0^+(k_1^+\!+\!k_2^+)}{(k_2^+)^2}\,
\left(\k_2\!-\! \frac{k_2^+}{k_0^+}\, \k_0 \right)^2\,
+\frac{(k_0^+\!-\!k_2^+)k_1^+}{(k_2^+)^2}\,
\left(\k_2\!-\! \frac{k_2^+}{k_1^+}\, \k_1 \right)^2\,
\Bigg\}\;
\overline{u}(0)\,\gamma^+\, v(1)
\nonumber\\
&&
+\Bigg\{
 -2\left[\k_1^i\!+\!\k_2^i \!-\!\frac{(k_1^+\!+\!k_2^+)}{k_0^+}\, \k_0^i\right]\,
 \left[\k_0^j\!-\!\k_2^j \!-\!\frac{(k_0^+\!-\!k_2^+)}{k_1^+}\, \k_1^j\right]
\nonumber\\
&& \hspace{2cm}
+(D\!-\!4)\, \left[\k_2^i\!-\!\frac{k_2^+}{k_0^+}\,\k_0^i\right]\,
             \left[\k_2^j\!-\!\frac{k_2^+}{k_1^+}\,\k_1^j\right]
\Bigg\}\;
\overline{u}(0)\, \gamma^+ \gamma^{i} \gamma^{j}\,  v(1)
\, .
\label{num_1_L_result}
\end{eqnarray}
Performing the change of variable \eqref{cv_k2_to_K} and using the notation \eqref{cv_k0_to_P}, as well as the decomposition \eqref{sym_antisym_decomp_spinor}, one finds after some algebra that the numerator $\textrm{Num}_{1}^L$ becomes
\begin{eqnarray}
\textrm{Num}_{1}^L &=&
 \frac{q^+}{k_0^+ k_1^+}\;
 \left[k_1^+\!-\!k_0^+\!+\!\frac{(D\!-\!2)}{2}\, k_2^+\right]
 \K^i\, \P^j\;
 \overline{u}(0)\, \gamma^+ \left[\gamma^i, \gamma^j\right]\, v(1)
 \nonumber\\
 && +\left[\frac{4 k_0^+ k_1^+}{(k_2^+)^2}
        \!+\!\frac{2(k_0^+\!-\!k_1^+)}{k_2^+}
        \!-\!(D\!-\!2)
 \right]
 \left[\K^2+ \frac{q^+ k_2^+}{k_0^+ k_1^+}\, \P\!\cdot\!\K\right]\; \overline{u}(0)\, \gamma^+  v(1)
 \label{num_1_L_result_2}
\end{eqnarray}

Like in the transverse photon case, it is then convenient to rewrite the wave-function contribution corresponding to the diagram 1 as
\begin{eqnarray}
\Psi_{\gamma_L^{*}\rightarrow q_0+\bar{q}_1}^{1}
&=&(2\pi)^{D-1}\delta^{(D-1)}(\underline{k_{0}}+\underline{k_{1}}\!-\!
\underline{q})\; \delta_{\alpha_{0},\, \alpha_{1}}\;
\frac{e\, e_f}{(ED_{LO})}\, \mu^{2-\frac{D}{2}}\,
\frac{Q}{q^+}\;
\left[\frac{\alpha_s\, C_F}{2\pi}\right]\,
\nonumber\\
&& \hspace{2cm} \times  \;
\frac{(-1)}{2q^+}\;
\int_{0}^{k_0^+}\!\!
d{k_2^+}\;
\frac{(k_{0}^+\!-\! k_{2}^+)}{k_{0}^+}\;
\mathcal{I}_1^{L}
\, ,\label{WF_1_L_2}
\end{eqnarray}
defining the integral
\begin{eqnarray}
\mathcal{I}_1^{L}&=& 4\pi\, (\mu^2)^{2-\frac{D}{2}}\,
\int \frac{d^{D-2} \K}{(2\pi)^{D-2} }\;
\frac{\textrm{Num}_{1}^L}{\Big[\K^2 +\Delta_1\Big]\,
\Big[(\K\!+\!\L)^2 +\Delta_2\Big]}
\, .
\label{int_perp_1_L}
\end{eqnarray}
Substituting $\K^2$ and $\P\cdot\K$ thanks to the identities \eqref{K2_replace} and \eqref{PdotK_replace} in the expression \eqref{num_1_L_result_2} for the numerator $\textrm{Num}_{1}^L$, one can read off the decomposition of $\mathcal{I}_1^{L}$ in term of the integrals $\mathcal{A}_0$, $\mathcal{B}_0$ and $\mathcal{B}^i$, and find
\begin{eqnarray}
\mathcal{I}_1^{L}&=&
\frac{q^+}{k_0^+ k_1^+}\;
 \left[k_1^+\!-\!k_0^+\!+\!\frac{(D\!-\!2)}{2}\, k_2^+\right]
 \mathcal{B}^i\, \P^j\;
 \overline{u}(0)\, \gamma^+ \left[\gamma^i, \gamma^j\right]\, v(1)
 \nonumber\\
 && +\left[\frac{4 k_0^+ k_1^+}{(k_2^+)^2}
        \!+\!\frac{2(k_0^+\!-\!k_1^+)}{k_2^+}
        \!-\!(D\!-\!2)
 \right]
 \bigg\{\mathcal{A}_0(\Delta_2)
 - \frac{q^+ k_2^+}{2(k_0^+\!-\!k_2^+) k_1^+}\,
 \Big[\mathcal{A}_0(\Delta_1)-\mathcal{A}_0(\Delta_2)\Big]
 \nonumber\\
&& \hspace{2cm}
   - \frac{q^+ k_2^+}{2k_0^+ k_1^+}\,
    \left[
        \left(\frac{k_1^+\!+\!k_2^+}{k_1^+}\right)\, \P^2
       +\left(\frac{k_0^+\!-\!k_2^+}{k_0^+}\right)\, \overline{Q}^2
    \right]\;
    \mathcal{B}_0
 \bigg\}\; \overline{u}(0)\, \gamma^+  v(1)
 \, .
\label{int_perp_1_L_decomp_1}
\end{eqnarray}
Due to rotational symmetry, the vector integral $\mathcal{B}^i$ is proportional to $\L^i$ and thus to $\P^i$, following the Passarino-Veltman method. Hence, the term in the first line of eq. \eqref{int_perp_1_L_decomp_1} vanishes, because it is proportional to $\P^i\P^j$ contracted with an antisymmetric tensor. Using the base $\{\mathcal{A}_0(\Delta_2),\, \mathcal{I}_+,\, \mathcal{I}_-\}$ of scalar integrals (see the definition in eq. \eqref{I_plus_minus_def}), one finally finds
\begin{eqnarray}
\mathcal{I}_1^{L}&=&
\left[\frac{4 k_0^+ k_1^+}{(k_2^+)^2}
        \!+\!\frac{2(k_0^+\!-\!k_1^+)}{k_2^+}
        \!-\!(D\!-\!2)
\right]\,
\bigg\{\mathcal{A}_0(\Delta_2)
 - \frac{q^+ k_2^+}{(k_0^+\!-\!k_2^+) k_1^+}\, \mathcal{I}_+
 \bigg\}\; \overline{u}(0)\, \gamma^+  v(1)
 \, .
\label{int_perp_1_L_decomp_2}
\end{eqnarray}
Hence, in the longitudinal photon case, the diagram $1$ can be expressed with a single form factor multiplying the LO wave function, as
\begin{eqnarray}
\Psi_{\gamma_L^{*}\rightarrow q_0+\bar{q}_1}^{1}
&=&(2\pi)^{D-1}\delta^{(D-1)}(\underline{k_{0}}+\underline{k_{1}}\!-\!
\underline{q})\; \delta_{\alpha_{0},\, \alpha_{1}}\;
\frac{e\, e_f}{(ED_{LO})}\, \mu^{2-\frac{D}{2}}\,
\frac{Q}{q^+}\;
\left[\frac{\alpha_s\, C_F}{2\pi}\right]\,
{\cal V}_{1}^{L}\; \overline{u}(0)\, \gamma^+\,  v(1)\nonumber\\
&=&
\Psi_{\gamma_L^{*}\rightarrow q_0+\bar{q}_1}\Bigg|_{LO}\;
\times\, \left[\frac{\alpha_s\, C_F}{2\pi}\right]\,
{\cal V}_{1}^{L}
\, ,\label{WF_1_L_3}
\end{eqnarray}
by contrast to the transverse photon case, see eq. \eqref{WF_1_T_3}.
The form factor ${\cal V}_{1}^{L}$ is then found to be
\begin{eqnarray}
{\cal V}_{1}^{L}&=& \int_{0}^{k_0^+}\!\! \frac{d{k_2^+}}{q^+}\;
\Bigg\{-\frac{2 k_0^+k_1^+}{(k_2^+)^2}
       +\frac{(3k_1^+\!-\!k_0^+)}{k_2^+}
       +\frac{(k_0^+\!-\!k_1^+)}{k_0^+}
      +\frac{(D\!-\!2)}{2}\, \left(\frac{k_0^+\!-\!k_2^+}{k_0^+}\right)
\Bigg\}\; \mathcal{A}_0(\Delta_2)
\nonumber\\
&& \hspace{2cm}
+\int_{0}^{k_0^+}\!\! d{k_2^+}\;
\Bigg\{\frac{2}{k_2^+} +\frac{(k_0^+\!-\!k_1^+)}{k_0^+k_1^+}
        -\frac{(D\!-\!2)k_2^+}{2k_0^+k_1^+}
\Bigg\}\; \mathcal{I}_+
\label{V1L_decomp}
\, .
\end{eqnarray}
Note that the first term produces a power divergence at small $k_2^+$, which is of course unphysical and which was not present in the diagram $1$ for transverse photon, see eqs. \eqref{V1T_decomp} and \eqref{V1T_eval_1}. Since  the diagram $A$ (and by symmetry also the diagram $B$) only gives single and double logarithmic divergences at small $k_2^+$, the power divergence encountered here has to cancel in the sum of the diagrams $1$, $2$ and $3$ in the longitudinal photon case. One could go on and evaluate more explicitly the form factor ${\cal V}_{1}^{L}$ but it turns out more convenient to first add a contribution coming from the diagram $3$.


\subsection{Diagram 3 for longitudinal photon}

In the longitudinal photon case, the contribution of the diagram $3$ to the  $q\bar{q}$ LFWF is written
\begin{eqnarray}
\Psi_{\gamma_L^{*}\rightarrow q_0+\bar{q}_1}^{3}
&=&  \sum_{q_{0'} \bar{q}_{1'} \textrm{ states}}
\frac{
\langle 0| d_{1} \, b_0 V_I(0) b_{0'}^{\dag} d_{1'}^{\dag} |0 \rangle\;
V_{\gamma_L \rightarrow q_{0'} \bar{q}_{1'}}
}{(ED_{LO})\; (ED_V)}
\, ,
\label{WF_3_L_1}
\end{eqnarray}
with the two-to-two quark-antiquark vertex given in eq. \eqref{qqbar_to_qqbar_vertex}, and the effective vertex for longitudinal photon splitting in eq. \eqref{gammaL_to_qqbar_vertex}. Then, one gets
\begin{eqnarray}
\Psi_{\gamma_L^{*}\rightarrow q_0+\bar{q}_1}^{3}&=&
(2\pi)^{D-1}\delta^{(D-1)}(\underline{k_1}+\underline{k_{0}}\!-\!\underline{q})\;
\delta_{\alpha_{0} ,\,\alpha_{1}}\;
 \frac{\mu^{2-\frac{D}{2}}\, e\, e_f}{(ED_{LO})}\; \frac{Q}{q^+}\;
(\mu^2)^{2-\frac{D}{2}}\, g^2\, C_F\;\;
 \int \frac{d^{D-1} \underline{k_{0'}}}{(2\pi)^{D-1} }\; \frac{\theta(k_{0'}^+)}{(2 k_{0'}^+)}\;
 \nonumber\\
 &&\;\; \times\;\;
   \int \frac{d^{D-1} \underline{k_{1'}}}{(2\pi)^{D-1} }\; \frac{\theta(k_{1'}^+)}{(2 k_{1'}^+)}\;
(2\pi)^{D-1}\delta^{(D-1)}(\underline{k_{1'}}+\underline{k_{0'}}\!-\!\underline{q})\;
  \frac{(-1)}{(k_{0}^+\!-\!k_{0'}^+)^2}\;\;
 \frac{\textrm{Num}_{3}^T}{(ED_V)}
\, ,
\label{WF_3_L_2}
\end{eqnarray}
with the numerator
\begin{eqnarray}
\textrm{Num}_{3}^L &=&
 \sum_{h_{0'}, h_{1'}=\pm 1/2}  \overline{u}(0)\,  \gamma^+\,  u(0')\;\;
 \overline{u}(0')\,  \gamma^+\, v(1')\;\;
\overline{v}(1')\,  \gamma^+\,   v(1)\label{num_3_L_def}\\
&=& \overline{u}(0)\,  \gamma^+\,  \slashed{k}_{0'}\,  \gamma^+\, \slashed{k}_{1'}\,  \gamma^+\,   v(1)\nonumber\\
&=& 4 k_{0'}^+ k_{1'}^+\;\; \overline{u}(0)\, \gamma^+\, v(1)
\, .
\label{num_3_L_result}
\end{eqnarray}
Hence, one finds that, for this graph also, there is factorization between
the LO wave-function and a form factor ${\cal V}_{3}^{L}$, as
\begin{eqnarray}
\Psi_{\gamma_L^{*}\rightarrow q_0+\bar{q}_1}^{3}
&=& \Psi_{\gamma_L^{*}\rightarrow q_0+\bar{q}_1}^{LO}\;\;
     \times\;\;  \left[\frac{\alpha_s\, C_F}{2\pi}\right]\;\;  {\cal V}_{3}^{L}
\, ,
\label{WF_3_L_3}
\end{eqnarray}
where
\begin{eqnarray}
{\cal V}_{3}^{L}
&=&
(4\pi)(2\pi)\,  (\mu^2)^{2-\frac{D}{2}}\,
 \int \frac{d^{D-1} \underline{k_{0'}}}{(2\pi)^{D-1} }\; \theta(k_{0'}^+)\;
    \theta(q^+\!-\!k_{0'}^+)\;
  \frac{(-1)}{(k_{0}^+\!-\!k_{0'}^+)^2}\;\;
 \frac{1}{(ED_V)}
 \, .
 \label{V3L_1}
\end{eqnarray}
Using the expression \eqref{ED_V_1} for the energy denominator, one finds 
\begin{eqnarray}
{\cal V}_{3}^{L}
&=& \frac{2}{q^+}\,
 \int_{0}^{q^+} \!\!\!\! d k_{0'}^+ \;\;
  \frac{k_{0'}^+ (q^+\!-\!k_{0'}^+)}{(k_{0}^+\!-\!k_{0'}^+)^2}\;\;
 \mathcal{A}_0\left(\frac{k_{0'}^+ (q^+\!-\!k_{0'}^+)}{k_{0}^+ k_{1}^+}\; \overline{Q}^2 \right)
 \, .
 \label{V3L_2}
\end{eqnarray}

As discussed in the introduction of the section \ref{sec:NLO_WF_T}, the diagram $3$ can be split into two contributions, according to the direction of the $k^+$ flow along the instantaneous gluon line. One can see that for $k_{0'}^+<k_{0}^+$, the light-cone momentum of the gluon line is flowing upwards, into the quark line, whereas for  $k_{0'}^+>k_{0}^+$, the light-cone momentum of the gluon line is flowing downwards, into the antiquark line. Hence, it is natural to split the form factor associated with the diagram $3$ as
\begin{eqnarray}
{\cal V}_{3}^{L} &=& {\cal V}_{3a}^{L}+{\cal V}_{3b}^{L}
\label{V3L_3}\\
{\cal V}_{3a}^{L}
&\equiv & \frac{2}{q^+}\,
 \int_{0}^{k_{0}^+} \!\!\!\! d k_{0'}^+ \;\;
  \frac{k_{0'}^+ (q^+\!-\!k_{0'}^+)}{(k_{0}^+\!-\!k_{0'}^+)^2}\;\;
 \mathcal{A}_0\left(\frac{k_{0'}^+ (q^+\!-\!k_{0'}^+)}{k_{0}^+ k_{1}^+}\; \overline{Q}^2 \right)
 \label{V3aL_1}\\
 {\cal V}_{3b}^{L}
&\equiv & \frac{2}{q^+}\,
 \int_{k_{0}^+}^{q^+} \!\!\!\! d k_{0'}^+ \;\;
  \frac{k_{0'}^+ (q^+\!-\!k_{0'}^+)}{(k_{0}^+\!-\!k_{0'}^+)^2}\;\;
 \mathcal{A}_0\left(\frac{k_{0'}^+ (q^+\!-\!k_{0'}^+)}{k_{0}^+ k_{1}^+}\; \overline{Q}^2 \right)
 \label{V3bL_1}
 \, .
\end{eqnarray}
In ${\cal V}_{3a}^{L}$, it is then convenient to make a change of variable
\begin{eqnarray}
k_{0'}^+\rightarrow k_2^+= k_{0}^+\!-\!k_{0'}^+
\label{cv_for_3aL}
\end{eqnarray}
in order to make the gluon light-cone momentum explicit,
whereas for
${\cal V}_{3b}^{L}$, the relevant change of variable is
\begin{eqnarray}
k_{0'}^+\rightarrow k_2^+= k_{0'}^+\!-\!k_{0}^+
\label{cv_for_3bL}
\, .
\end{eqnarray}
Let us focus on the contribution ${\cal V}_{3a}^{L}$. Using the change of variable \eqref{cv_for_3aL}, one finds
\begin{eqnarray}
{\cal V}_{3a}^{L}
&=& \frac{2}{q^+}\,
 \int_{0}^{k_0^+} d k_2^+ \;\;
  \frac{(k_0^+\!-\!k_2^+) (k_1^+\!+\!k_2^+)}{(k_2^+)^2}\;\;
 \mathcal{A}_0\left(\frac{(k_0^+\!-\!k_2^+) (k_1^+\!+\!k_2^+)}{k_{0}^+ k_{1}^+}\; \overline{Q}^2 \right)
 \nonumber\\
 &=&
 \int_{0}^{k_0^+} \frac{d k_2^+}{q^+} \;\; \bigg[
   \frac{2k_0^+k_1^+}{(k_2^+)^2}
   +\frac{2(k_0^+\!-\!k_1^+)}{k_2^+}
   -2
  \bigg]\;\;
 \mathcal{A}_0\left(\Delta_2\right)
 \, .
 \label{V3aL_2}
\end{eqnarray}
Once again, one obtains a power divergence at small $k_2^+$, due to the first term in the bracket in eq. \eqref{V3aL_2}.


\subsection{Summing the graphs and integrating over $k_2^+$}

For the longitudinal photon case, the contribution from each NLO graph to the wave function in momentum space has been found to factorize into the LO wave function times a form factor. Hence, one has
\begin{eqnarray}
\Psi_{\gamma_L^{*}\rightarrow q_0+\bar{q}_1}^{NLO}
&=& \Psi_{\gamma_L^{*}\rightarrow q_0+\bar{q}_1}^{LO}\;\;
\times\;\;
\left[1+\left(\frac{\alpha_s\, C_F}{2\pi}\right)\; {\cal V}^{L}\;\right]
\, ,\label{WF_NLO_L_1}
\end{eqnarray}
where
\begin{eqnarray}
{\cal V}^{L} &\equiv& {\cal V}_{A}^{L} +{\cal V}_{B}^{L}+{\cal V}_{1}^{L} +{\cal V}_{2}^{L} +{\cal V}_{3a}^{L} +{\cal V}_{3b}^{L}
\, ,
\end{eqnarray}
with ${\cal V}_{A}^{L}$, ${\cal V}_{1}^{L}$ and ${\cal V}_{3a}^{L}$ given by the expressions \eqref{VAL_eval_1}, \eqref{V1L_decomp} and \eqref{V3aL_2} respectively.

Adding together the contributions ${\cal V}_{1}^{L}$ and ${\cal V}_{3a}^{L}$, one finds a cancellation of the power divergences at small $k_2^+$, and the result is simpler than the expression \eqref{V1L_decomp} for ${\cal V}_{1}^{L}$, and reads
\begin{eqnarray}
{\cal V}_{1}^{L}+ {\cal V}_{3a}^{L}&=& \int_{0}^{k_0^+}\!\! \frac{d{k_2^+}}{k_0^+}\;
\Bigg\{\frac{k_0^+}{k_2^+}
       -1
      +\frac{(D\!-\!2)}{2}\, \left(\frac{k_0^+\!-\!k_2^+}{q^+}\right)
\Bigg\}\; \mathcal{A}_0(\Delta_2)
\nonumber\\
&& 
+\int_{0}^{k_0^+}\!\! \frac{d{k_2^+}}{k_0^+}\;
\Bigg\{\frac{2k_0^+}{k_2^+} +\left(\frac{k_0^+\!-\!k_1^+}{k_1^+}\right)
        -\frac{(D\!-\!2)k_2^+}{2k_1^+}
\Bigg\}\; \mathcal{I}_+
\label{V1L_plus_V3aL_1}
\, .
\end{eqnarray}
The $D$ dependence of the coefficient of the UV-divergent integral $\mathcal{A}_0(\Delta_2)$ generates as usual a rational term when performing the $D\rightarrow 4$ expansion of \eqref{V1L_plus_V3aL_1} thanks to eqs. \eqref{A0D2_result} and \eqref{I_plus_result}, and one gets
\begin{eqnarray}
{\cal V}_{1}^{L}+ {\cal V}_{3a}^{L}&=& \int_{0}^{k_0^+}\!\! \frac{d{k_2^+}}{k_0^+}\;
\Bigg\{\frac{k_0^+}{k_2^+}
       -\frac{k_1^+}{q^+}
      -\frac{k_2^+}{q^+}
\Bigg\}\;
\Bigg\{
\Gamma\!\left(2\!-\! \frac{D}{2}\right)\,
\left(\frac{\overline{Q}^2}{4\pi\, \mu^2}\right)^{\frac{D}{2}-2}
- \log\left(\frac{k_1^+\!+\!k_2^+}{k_1^+}\right)
- \log\left(\frac{k_0^+\!-\!k_2^+}{k_0^+}\right)
\Bigg\}
\nonumber\\
&& 
+\int_{0}^{k_0^+}\!\! \frac{d{k_2^+}}{k_0^+}\;
\Bigg\{\frac{2k_0^+}{k_2^+} +\left(\frac{k_0^+\!-\!k_1^+}{k_1^+}\right)
        -\frac{k_2^+}{k_1^+}
\Bigg\}\;
\Bigg\{
- \log\left(\frac{k_2^+}{k_0^+}\right)
+ \log\left(\frac{k_1^+\!+\!k_2^+}{k_1^+}\right)
- \log\left(\frac{q^+}{k_1^+}\right)
\Bigg\}
\nonumber\\
&& 
-\int_{0}^{k_0^+}\!\! \frac{d{k_2^+}}{k_0^+}\;
   \left(\frac{k_0^+\!-\!k_2^+}{q^+}\right)
+ O\left(D\!-\!4\right)
\nonumber\\
&=&
- \left[\log\left(\frac{k^+_{\min}}{k_0^+}\right) +\frac{1}{2}+\frac{k_1^+}{2q^+} \right]\,
\Gamma\!\left(2\!-\! \frac{D}{2}\right)\,
\left(\frac{\overline{Q}^2}{4\pi\, \mu^2}\right)^{\frac{D}{2}-2}
+\left[2\log\left(\frac{k^+_{\min}}{k_0^+}\right)+\frac{3}{2}\right]\,
     \log\left(\frac{q^+}{k_1^+}\right)
\nonumber\\
&&
+\left[\log\left(\frac{k^+_{\min}}{k_0^+}\right)\right]^2
 +\frac{\pi^2}{6} - \textrm{Li}_2\left(-\frac{k_0^+}{k_1^+}\right)
 -2+\frac{k_0^+}{2q^+}
\nonumber\\
&&
 -\frac{k_0^+}{2q^+}
+ O\left(D\!-\!4\right)
\label{V1L_plus_V3aL_2}
\, ,
\end{eqnarray}
performing the integrations over $k_2^+$ thanks to the appendix \ref{sec:integrals}, imposing the cutoff $k_2^+>k^+_{\min}$ when needed. The last term in the expression \eqref{V1L_plus_V3aL_2} is the contribution from the $(D\!-\!4)/(D\!-\!4)$ rational term.
Similarly as in the transverse photon case, both the double logarithmic divergent term and the terms proportional to $\log(q^+/k_1^+)$ found in ${\cal V}_{1}^{L}+ {\cal V}_{3a}^{L}$ cancel against the ones present in ${\cal V}_{A}^{L}$, see eq. \eqref{VAL_eval_1}. More precisely, one obtains
\begin{eqnarray}
{\cal V}_{A}^{L}+{\cal V}_{1}^{L}+ {\cal V}_{3a}^{L}
&=&
\left[\log\left(\frac{k^+_{\min}}{k_0^+}\right) +1
    -\frac{k_1^+}{2q^+}\right]\,
  \Gamma\!\left(2\!-\! \frac{D}{2}\right)\,
  \left(\frac{\overline{Q}^2}{4\pi\, \mu^2}\right)^{\frac{D}{2}-2}
-\left[2\log\left(\frac{k^+_{\min}}{k_0^+}\right)+\frac{3}{2}\right]\,
     \log\left(\frac{\P^2+\overline{Q}^2}{\overline{Q}^2}\right)
\nonumber\\
&&
- \textrm{Li}_2\left(-\frac{k_0^+}{k_1^+}\right) -\frac{\pi^2}{6}
 +\frac{3}{2}-\frac{k_1^+}{2q^+}
 +\frac{k_1^+}{2q^+}
+ O\left(D\!-\!4\right)
\label{VAL_plus_V1L_plus_V3aL}
\, ,
\end{eqnarray}
where the last term collects all the contributions from the $(D\!-\!4)/(D\!-\!4)$ rational terms.

The other contributions to ${\cal V}^{L}$ can be obtained by interchange of the momenta $\underline{k_{0}}$ and $\underline{k_{1}}$. In eq. \eqref{VAL_plus_V1L_plus_V3aL}, all the dependence on $\k_0$ or $\k_1$ happens via $\P^2$, which is invariant under the interchange of $\underline{k_{0}}$ and $\underline{k_{1}}$. Hence, only the light-cone momenta $k_0^+$ and $k_1^+$ need to be exchanged, so that
\begin{eqnarray}
{\cal V}^{L}
&=& \bigg[ {\cal V}_{A}^{L}  +{\cal V}_{1}^{L}  +{\cal V}_{3a}^{L}\bigg]
+ \Big(k_0^+ \leftrightarrow k_1^+\Big)
\nonumber\\
&=&
\left[\log\left(\frac{k^+_{\min}}{k_0^+}\right) +\log\left(\frac{k^+_{\min}}{k_1^+}\right)
+\frac{3}{2} \right]\,
\left[
\Gamma\!\left(2\!-\! \frac{D}{2}\right)\,
\left(\frac{\overline{Q}^2}{4\pi\, \mu^2}\right)^{\frac{D}{2}-2}
-2 \log\left(\frac{\P^2+\overline{Q}^2}{\overline{Q}^2}\right)
\right]
\nonumber\\
&&
- \textrm{Li}_2\left(-\frac{k_0^+}{k_1^+}\right)
- \textrm{Li}_2\left(-\frac{k_1^+}{k_0^+}\right)
-\frac{\pi^2}{3}
 +\frac{5}{2}
 +\frac{1}{2}
 + O\left(D\!-\!4\right)
\label{VL_result_1}
\, .
\end{eqnarray}
Note that for both the longitudinal and transverse photon cases, the rational terms induced by the $D$ dependence of the coefficient in front of $\mathcal{A}_0(\Delta_2)$ give a total contribution of $1/2$ to the form factor ${\cal V}^{L}$ (respectively ${\cal V}^{T}$), see eq. \eqref{VT_result_1}. The other type of rational term, induced by the Passarino-Veltman decomposition of the tensor integral $\mathcal{B}^{ij}$, exists only in the transverse photon case. Nevertheless, the total constant term is identical in ${\cal V}^{L}$  and ${\cal V}^{T}$. Moreover, ${\cal V}^{L}$  and ${\cal V}^{T}$ differ only by the $\P^2$-dependent term.

Using the identity \eqref{prop_Li2}, the final result for ${\cal V}^{L}$ can be written as
\begin{eqnarray}
{\cal V}^{L}
&=&
\left[\log\left(\frac{k^+_{\min}}{k_0^+}\right) +\log\left(\frac{k^+_{\min}}{k_1^+}\right)
+\frac{3}{2} \right]\,
\left[
\Gamma\!\left(2\!-\! \frac{D}{2}\right)\,
\left(\frac{\overline{Q}^2}{4\pi\, \mu^2}\right)^{\frac{D}{2}-2}
-2 \log\left(\frac{\P^2+\overline{Q}^2}{\overline{Q}^2}\right)
\right]
\nonumber\\
&&
+\frac{1}{2}\; \left[\log\left(\frac{k_0^+}{k_1^+}\right)\right]^2
-\frac{\pi^2}{6}
 +3
 + O\left(D\!-\!4\right)
\label{VL_result_2}
\, .
\end{eqnarray}


\section{One-loop virtual photon wave-functions in mixed space\label{sec:FT_NLO_WF}}

For the quark-antiquark Fock state contribution, the Fourier transform into mixed space of the LFWF of transverse or longitudinal virtual photon is written (see eq. \eqref{def_3_LFWF_mix})
\begin{eqnarray}
 \widetilde{\Psi}_{\gamma_{T,L}^{*}\rightarrow q_0 \bar{q}_1}
 & \equiv & \int \frac{d^{D-2} \k_0}{(2\pi)^{D-2}}
 \int \frac{d^{D-2} \k_1}{(2\pi)^{D-2}} \;
 e^{i\k_0 \cdot \x_0 +i\k_1 \cdot \x_1}\;
 \Psi_{\gamma_{T,L}^{*}\rightarrow q_0 \bar{q}_1}
 \nonumber\\
 &=& \int \frac{d^{D-2} \k_1}{(2\pi)^{D-2}}
 \int \frac{d^{D-2} \k_2}{(2\pi)^{D-2}} \;
 e^{i\k_0 \cdot \x_0 +i\k_1 \cdot \x_1}\;
 \Psi_{\gamma_{T,L}^{*}\rightarrow q_0 \bar{q}_1}^{LO}\;
\times\;
\left[1+\left(\frac{\alpha_s\, C_F}{2\pi}\right)\; {\cal V}^{T,L}\;\right]\;
+{\cal O}(e\, \alpha_s^2)
 \, . \label{FT_WF_gamma_TL_def}
\end{eqnarray}


\subsection{Longitudinal photon case}

Inserting the expression \eqref{WF_L_LO} for the LO wave-function in eq. \eqref{FT_WF_gamma_TL_def}, one can proceed to calculate the Fourier transform in the longitudinal photon case. Using the transverse delta function present in the LO wave-function to perform the integral over $\k_1$ and changing variables from $\k_0$ to the relative momentum $\P$ defined in eq. \eqref{cv_k0_to_P}, one finds
\begin{eqnarray}
 \widetilde{\Psi}_{\gamma_{L}^{*}\rightarrow q_0 \bar{q}_1}
 &=&
 (2\pi)\delta({k_0^+}\!+\!{k_1^+}\!-\!{q^+})\;
\delta_{\alpha_{0} ,\,\alpha_{1}}\;
\overline{u_G}(0)\,\gamma^+ v_G(1)\;
  e\, e_f\,
\left(-\frac{2k_0^+ k_1^+}{(q^+)^2}\right)\; Q\;
e^{i\frac{\q}{q^+}\cdot (k_0^+\x_0+k_1^+\x_1)}\;
\nonumber\\
&& \hspace{1cm}\times\;
\mu^{2-\frac{D}{2}}\,
 \int \frac{d^{D-2} \P}{(2\pi)^{D-2}}\;
 e^{i\P \cdot \x_{01}}\;
  \frac{1}{\left[\P^2\!+\!\overline{Q}^2\right]}\;\;
\left[1+\left(\frac{\alpha_s\, C_F}{2\pi}\right)\; {\cal V}^{L}\;\right]\;
+{\cal O}(e\, \alpha_s^2)
 \, . \label{FT_WF_gamma_L_1}
\end{eqnarray}
One sees in the expression \eqref{VL_result_2} that only one term in ${\cal V}^{L}$ depends on the relative momentum $\P$ of the pair, so that the only two needed integrals to perform the Fourier transform are
\begin{eqnarray}
 \int \frac{d^{D-2} \P}{(2\pi)^{D-2}}\;
 e^{i\P \cdot \x_{01}}\;
  \frac{1}{\left[\P^2\!+\!\overline{Q}^2\right]}
 &=& \frac{1}{2\pi}\; \left(\frac{\overline{Q}}{2\pi\, |\x_{01}|}\right)^{\frac{D}{2}-2}\;\;
 \textrm{K}_{\frac{D}{2}-2}\Big(|\x_{01}|\, \overline{Q}\Big)
\label{int_FT_WF_gamma_L_LO}
\\
\int \frac{d^{D-2} \P}{(2\pi)^{D-2}}\;
 e^{i\P \cdot \x_{01}}\;
  \frac{1}{\left[\P^2\!+\!\overline{Q}^2\right]}\;
  \log\left(\frac{\P^2\!+\!\overline{Q}^2}{\overline{Q}^2}\right)
 &=& \frac{1}{2\pi}\; \left(\frac{\overline{Q}}{2\pi\, |\x_{01}|}\right)^{\frac{D}{2}-2}\;\;
 \textrm{K}_{\frac{D}{2}-2}\Big(|\x_{01}|\, \overline{Q}\Big)\;
\nonumber\\
&& \hspace{0.5cm} \times\;
 \bigg\{
 -\frac{1}{2}\; \log\left(\frac{{\x_{01}}^2\, \overline{Q}^2}{4}\right)
 +\Psi(1)
 +{\cal O}(D\!-\!4)
 \bigg\}
 \, .
\label{int_FT_WF_gamma_L_A}
\end{eqnarray}
Their calculation is outlined in the appendix \ref{sec:FT_int}. $\textrm{K}_{\alpha}(z)$ is the modified Bessel function of the second kind. Inserting the results \eqref{int_FT_WF_gamma_L_LO} and \eqref{int_FT_WF_gamma_L_A} into the expression \eqref{FT_WF_gamma_L_1}, one can rewrite the wave-function as
\begin{eqnarray}
\widetilde{\Psi}_{\gamma_{L}^{*}\rightarrow q_0 \bar{q}_1}
 &=&
 (2\pi)\delta({k_0^+}\!+\!{k_1^+}\!-\!{q^+})\;
\delta_{\alpha_{0} ,\,\alpha_{1}}\;
\overline{u_G}(0)\,\gamma^+ v_G(1)\;
  \frac{e\, e_f}{2\pi}\,
\left(-\frac{2k_0^+ k_1^+}{(q^+)^2}\right)\; Q\;
e^{i\frac{\q}{q^+}\cdot (k_0^+\x_0+k_1^+\x_1)}\;
\nonumber\\
&& \hspace{1cm}\times\;
\left(\frac{\overline{Q}}{2\pi\, |\x_{01}|\, \mu}\right)^{\frac{D}{2}-2}\;\;
 \textrm{K}_{\frac{D}{2}-2}\Big(|\x_{01}|\, \overline{Q}\Big)\;\;
\left[1+\left(\frac{\alpha_s\, C_F}{2\pi}\right)\;
\widetilde{{\cal V}}^{L}\;\right]\;
+{\cal O}(e\, \alpha_s^2)
 \, , \label{FT_WF_gamma_L_2}
\end{eqnarray}
with the NLO form factor
\begin{eqnarray}
\widetilde{{\cal V}}^{L}
&=&
\left[\log\left(\frac{k^+_{\min}}{k_0^+}\right) +\log\left(\frac{k^+_{\min}}{k_1^+}\right)
+\frac{3}{2} \right]\,
\left[
\frac{(4\pi)^{2\!-\!\frac{D}{2}}}{\left(2\!-\!\frac{D}{2}\right)}\, \Gamma\!\left(3\!-\! \frac{D}{2}\right)
+ \log\left(\frac{{\x_{01}}^2\, \mu^2}{4}\right)-2\Psi(1)
\right]
\nonumber\\
&&
+\frac{1}{2}\; \left[\log\left(\frac{k_0^+}{k_1^+}\right)\right]^2
-\frac{\pi^2}{6}
 +3
 + O\left(D\!-\!4\right)
\label{mixed_VL_result}
\, .
\end{eqnarray}
Interestingly, $\widetilde{{\cal V}}^{L}$ is independent of the photon virtuality $Q$, by contrast to its momentum-space counterpart ${{\cal V}}^{L}$. Instead, the dependence on $Q$ is now fully contained in the LO wave-function.



\subsection{Transverse photon case}

In the transverse photon case, the LO wave-function is given by eq. \eqref{WF_T_LO}. All transverse momentum dependence is extracted from the spinors and photon polarization vector thanks to the identity \eqref{quark_current_times_epsilon_mom_cons} \footnote{Remember that in light-front quantization, the good components of the spinors $u$ and $v$ are independent on the transverse momenta.}. Then, from eq. \eqref{FT_WF_gamma_TL_def}, one easily finds
\begin{eqnarray}
\widetilde{\Psi}_{\gamma_{T}^{*}\rightarrow q_0 \bar{q}_1}
 &=&
 2\pi\,\delta({k_0^+}\!+\!{k_1^+}\!-\!{q^+})\;
\delta_{\alpha_{0} ,\,\alpha_{1}}\;
  e\, e_f\,
\left(-\frac{2k_0^+ k_1^+}{q^+}\right)\;
e^{i\frac{\q}{q^+}\cdot (k_0^+\x_0+k_1^+\x_1)}\;
\nonumber\\
&& \hspace{1cm}\times\;
\varepsilon_{\lambda}^i\,
\Bigg\{
 \left(\frac{k^+_0\!-\!k^+_1}{2k_0^+ k_1^+}\right)\; \delta^{ij}\; \overline{u_G}(0)\, \gamma^+  v_G(1)
-\frac{q^+}{4k_0^+ k_1^+}\;
\overline{u_G}(0)\; \gamma^+ \left[\gamma^i, \gamma^j\right]\, v_G(1)
\Bigg\}
\nonumber\\
&& \hspace{1cm}\times\;
\mu^{2-\frac{D}{2}}\,
 \int \frac{d^{D-2} \P}{(2\pi)^{D-2}}\;
 e^{i\P \cdot \x_{01}}\;
  \frac{\P^j}{\left[\P^2\!+\!\overline{Q}^2\right]}\;\;
\left[1+\left(\frac{\alpha_s\, C_F}{2\pi}\right)\; {\cal V}^{T}\;\right]\;
+{\cal O}(e\, \alpha_s^2)
 \, . \label{FT_WF_gamma_T_1}
\end{eqnarray}
The dependence of ${\cal V}^{T}$ (see eq. \eqref{VT_result_2}) on the relative transverse momentum $\P$ is more complicated than in the longitudinal photon case. Three independent integrals are now needed to perform the Fourier transform to mixed space:
\begin{eqnarray}
 \int \frac{d^{D-2} \P}{(2\pi)^{D-2}}\;
 e^{i\P \cdot \x_{01}}\;
  \frac{\P^j}{\left[\P^2\!+\!\overline{Q}^2\right]}
 &=& i\, \x_{01}^j\; \left(\frac{\overline{Q}}{2\pi\, |\x_{01}|}\right)^{\frac{D}{2}-1}\,
 \textrm{K}_{\frac{D}{2}-1}\Big(|\x_{01}|\, \overline{Q}\Big)
 \, ,
\label{int_FT_WF_gamma_T_LO}
\end{eqnarray}
\begin{eqnarray}
&&\int \frac{d^{D-2} \P}{(2\pi)^{D-2}}\;
 e^{i\P \cdot \x_{01}}\;
  \frac{\P^j}{\left[\P^2\!+\!\overline{Q}^2\right]}\;
  \log\left(\frac{\P^2\!+\!\overline{Q}^2}{\overline{Q}^2}\right)
 = i\, \x_{01}^j\; \left(\frac{\overline{Q}}{2\pi\, |\x_{01}|}\right)^{\frac{D}{2}-1}\;\;
\nonumber\\
&& \hspace{1cm} \times\;
\Bigg\{
\bigg[-\frac{1}{2}\; \log\left(\frac{{\x_{01}}^2\, \overline{Q}^2}{4}\right)
 +\Psi(1)\bigg]\;
   \textrm{K}_{\frac{D}{2}-1}\Big(|\x_{01}|\, \overline{Q}\Big)\;
+\frac{1}{|\x_{01}|\, \overline{Q}}\;
   \textrm{K}_{0}\Big(|\x_{01}|\, \overline{Q}\Big)
 +{\cal O}(D\!-\!4)
 \Bigg\}
\label{int_FT_WF_gamma_T_A}
\end{eqnarray}
and
\begin{eqnarray}
&&\int \frac{d^{D-2} \P}{(2\pi)^{D-2}}\;
 e^{i\P \cdot \x_{01}}\;
  \frac{\P^j}{\left[\P^2\!+\!\overline{Q}^2\right]}\;\;
   \frac{\left[\P^2\!+\!\overline{Q}^2\right]}{\P^2}\;
  \log\left(\frac{\P^2\!+\!\overline{Q}^2}{\overline{Q}^2}\right)
 = 2i\, \x_{01}^j\; \left(\frac{\overline{Q}}{2\pi\, |\x_{01}|}\right)^{\frac{D}{2}-1}\;\;
\nonumber\\
&& \hspace{6cm} \times\;
\Bigg\{
\frac{1}{|\x_{01}|\, \overline{Q}}\;
   \textrm{K}_{0}\Big(|\x_{01}|\, \overline{Q}\Big)
 +{\cal O}(D\!-\!4)
 \Bigg\}
 \, .
\label{int_FT_WF_gamma_T_1}
\end{eqnarray}
The calculation of the integrals is explained in the appendix \ref{sec:FT_int}. They allow one to rewrite \eqref{FT_WF_gamma_T_1} as
\begin{eqnarray}
 \widetilde{\Psi}_{\gamma_{T}^{*}\rightarrow q_0 \bar{q}_1}
 &=&
 2\pi\,\delta({k_0^+}\!+\!{k_1^+}\!-\!{q^+})\;
\delta_{\alpha_{0} ,\,\alpha_{1}}\;
  e\, e_f\;
e^{i\frac{\q}{q^+}\cdot (k_0^+\x_0+k_1^+\x_1)}\;
\nonumber\\
&& \hspace{1cm}\times\;
(-i)\, \varepsilon_{\lambda}^i\, \x_{01}^j\;
\Bigg\{
 \left(\frac{k^+_0\!-\!k^+_1}{q^+}\right)\; \delta^{ij}\; \overline{u_G}(0)\, \gamma^+  v_G(1)
-\frac{1}{2}\;
\overline{u_G}(0)\; \gamma^+ \left[\gamma^i, \gamma^j\right]\, v_G(1)
\Bigg\}
\nonumber\\
&& \hspace{1cm}\times\;
\mu^{2-\frac{D}{2}}\,
\left(\frac{\overline{Q}}{2\pi\, |\x_{01}|}\right)^{\frac{D}{2}-1}
 \textrm{K}_{\frac{D}{2}-1}\Big(|\x_{01}|\, \overline{Q}\Big)\;\;
\left[1+\left(\frac{\alpha_s\, C_F}{2\pi}\right)\; \widetilde{{\cal V}}^{T}\;\right]\;
+{\cal O}(e\, \alpha_s^2)
 \, , \label{FT_WF_gamma_T_2}
\end{eqnarray}
with the NLO form factor
\begin{eqnarray}
\widetilde{{\cal V}}^{T}
&=&
\left[\log\left(\frac{k^+_{\min}}{k_0^+}\right) +\log\left(\frac{k^+_{\min}}{k_1^+}\right)
+\frac{3}{2} \right]\,
\left[
\frac{(4\pi)^{2\!-\!\frac{D}{2}}}{\left(2\!-\!\frac{D}{2}\right)}\,
\Gamma\!\left(3\!-\! \frac{D}{2}\right)
+ \log\left(\frac{{\x_{01}}^2\, \mu^2}{4}\right)-2\Psi(1)
\right]
\nonumber\\
&&
+\frac{1}{2}\; \left[\log\left(\frac{k_0^+}{k_1^+}\right)\right]^2
-\frac{\pi^2}{6}
 +3
 + O\left(D\!-\!4\right)
\nonumber\\
&=& \widetilde{{\cal V}}^{L}  + O\left(D\!-\!4\right)
\label{mixed_VT_result}
\, .
\end{eqnarray}
Note that the contributions proportional to
$\textrm{K}_{0}(|\x_{01}|\, \overline{Q})$ coming from the integrals
\eqref{int_FT_WF_gamma_T_A} and \eqref{int_FT_WF_gamma_T_A} cancel with each other, so that all the leftover terms come with the same Bessel function $\textrm{K}_{\frac{D}{2}-1}(|\x_{01}|\, \overline{Q})$ as in the LO contribution.

In mixed space, the NLO form factors $\widetilde{{\cal V}}^{L}$ and
$\widetilde{{\cal V}}^{T}$ for the longitudinal and transverse photon cases are equal to each other, which was not the case in momentum space.
Hence, the NLO QCD correction to the quark-antiquark Fock state wave-function
shows some universality in the mixed-space representation, since it is independent on both the photon polarization and the photon virtuality. That universality is absent from the full momentum-space representation. Hence, the dipole picture of DIS processes seems particularly convenient and physical not only at LO but also at NLO.



\begin{acknowledgments}
I thank Renaud Boussarie, Andrey Grabovsky, Lech Szymanowski and Samuel Wallon for discussions and correspondence about their results before publication.
\end{acknowledgments}


\appendix


\section{Basics of light-front quantization and light-front perturbation theory\label{sec:formalism}}

An elementary introduction to light-front quantization and light-front perturbation theory is presented in this section, both to make the paper self-contained, and to make all the conventions explicit.

\subsection{Light-front quantized free fields\label{sec:free_fields_conventions}}

Applying the procedure of light-front quantization \cite{Kogut:1969xa} (see refs. \cite{Zhang:1994ti} and \cite{Brodsky:1997de} for modern pedagogical reviews in the QCD case) to free quarks and gluon fields, one obtains the following expression for the quantized fields\footnote{The convention used in this paper to define the light-cone vectors (and coordinates) is $k^{\pm}\equiv (k^0\pm k^3)/\sqrt{2}$. The $D\!-\!2$ transverse vectors are noted in bold characters, so that for example $\k$ denotes the transverse vector made out of the transverse components of $k^{\mu}$. Individual transverse components are noted as $\k^i$, labeled by an index like $i$, $j$ and so on. Finally, since the $k^-$ are integrated over in light-front perturbation theory, it is convenient to introduce the notation $\underline{k}\equiv (k^+, \k)$ for the leftover $D\!-\!1$ components.}:
\begin{eqnarray}
\Psi_{\alpha}(x)&=&\int_{0}^{+\infty}\!\!\!\! \frac{dk^+}{(2\pi) 2k^+}
 \int \frac{d^{D-2} \k}{(2\pi)^{D-2}} \sum_{h=\pm \frac{1}{2}}
 \bigg[e^{-ik\cdot x}\, b(\underline{k},h,\alpha)\, u(\underline{k},h)
 +e^{+ik\cdot x}\, d^{\dagger}(\underline{k},h,\alpha)\, v(\underline{k},h)
 \bigg]\Bigg|_{k^-\equiv \frac{\k^2+m^2}{2k^+}}
 \label{quant_free_q}
\\
A_{a}^{\mu}(x)&=&\int_{0}^{+\infty}\!\!\!\! \frac{dk^+}{(2\pi) 2k^+}
 \int \frac{d^{D-2} \k}{(2\pi)^{D-2}} \sum_{\textrm{phys. pol. }\lambda}
 \bigg[e^{-ik\cdot x}\, a(\underline{k},\lambda,a)\, \epsilon^{\mu}_{\lambda}\!(\underline{k})
 +e^{+ik\cdot x}\, a^{\dagger}(\underline{k},\lambda,a)\, \epsilon^{\mu\, *}_{\lambda}\!(\underline{k})
 \bigg]\Bigg|_{k^-\equiv \frac{\k^2}{2k^+}}
\, .  \label{quant_free_g}
\end{eqnarray}
In these expressions, the quarks\footnote{In order to simplify notations, only one flavor of quarks is considered, so that flavor indices can be dropped. Then, the sum over quark flavors can be taken at the cross-section level, when applying this formalism, or results of the present paper, to calculate an observable.} have a mass $m$, an light-front helicity $h$, and a fundamental color index $\alpha$, whereas the gluons have a physical polarization $\lambda$ and an adjoint color index $a$.
In the case of leptons and photons, one gets expressions identical to \eqref{quant_free_q} and \eqref{quant_free_g} respectively, up to the disappearance of color indices.

The nontrivial (anti)commutation relations for the creation and annihilation operators are written
\begin{eqnarray}
\big[a(\underline{k_1},\lambda_1,a_1),a^{\dagger}(\underline{k_2},\lambda_2,a_2)\big]
&=& (2k_1^+)(2\pi)^{D\!-\!1}\delta^{(D\!-\!1)}(\underline{k_1}\!-\!\underline{k_2})\;
\delta_{\lambda_1,\lambda_2}\; \delta_{a_1,a_2}\\
\big\{b(\underline{k_1},h_1,\alpha_1),b^{\dagger}(\underline{k_2},h_2,\alpha_2)\big\}
&=& (2k_1^+)(2\pi)^{D\!-\!1}\delta^{(D\!-\!1)}(\underline{k_1}\!-\!\underline{k_2})\;
\delta_{h_1,h_2}\; \delta_{\alpha_1,\alpha_2}\\
\big\{d(\underline{k_1},h_1,\alpha_1),d^{\dagger}(\underline{k_2},h_2,\alpha_2)\big\}
&=& (2k_1^+)(2\pi)^{D\!-\!1}\delta^{(D\!-\!1)}(\underline{k_1}\!-\!\underline{k_2})\;
\delta_{h_1,h_2}\; \delta_{\alpha_1,\alpha_2}
\, .\label{com_rel_mom}
\end{eqnarray}

For simplicity, the light-cone gauge $A_{a}^{+}(x)=0$ has been used for the gluon and photon fields. In that gauge, the polarization vectors $\epsilon^{\mu}_{\lambda}\!(\underline{k})$ corresponding to physical polarizations $\lambda$ are written
\begin{eqnarray}
\epsilon^{+}_{\lambda}\!(\underline{k})&=& 0, \nonumber\\
\epsilon^{j}_{\lambda}\!(\underline{k})&=& \varepsilon^{j}_{\lambda}\ ,\nonumber\\
\epsilon^{-}_{\lambda}\!(\underline{k})&=& \frac{\k^j\, \varepsilon^{j}_{\lambda}}{k^+}\ ,
\label{4_to_2_polarization}
\end{eqnarray}
where the transverse vectors $\varepsilon_{\lambda}$ are such that
\begin{eqnarray}
\sum_{\textrm{phys. pol. }\lambda}\varepsilon^{i}_{\lambda}\, \varepsilon^{j\, *}_{\lambda}&=& -g^{ij},\nonumber\\
-g_{ij}\, \varepsilon^{i}_{\lambda_1}\, \varepsilon^{j\, *}_{\lambda_2}&=&
\varepsilon^{j}_{\lambda_1}\, \varepsilon^{j\, *}_{\lambda_2}
=\delta_{\lambda_1,\lambda_2}\, .
\label{Rel_polarizations}
\end{eqnarray}
At the level of the full polarization vectors, one has instead
\begin{eqnarray}
\sum_{\textrm{phys. pol. }\lambda} \;
 {\epsilon}_{\lambda}^{\mu}\!(\underline{k})\; {\epsilon}_{\lambda}^{\nu\, *}\!(\underline{k})\;
= -g^{\mu \nu} + \frac{k^{\mu} n^{\nu}+n^{\mu}k^{\nu} }{k^+}
\label{completeness_gluon_pol_vect}
\end{eqnarray}
where by convention, $k^-\equiv\k^2/(2k^+)$. In eq. \eqref{completeness_gluon_pol_vect}, $n^{\mu}$ is a unit vector along the light cone, chosen in such a way that for any $b^{\mu}$, one has $n^{\mu}\, b_{\mu}=b^+$.
Note that there are $D\!-\!2$ physical polarizations $\lambda$: one linear polarization for each transverse direction.

Let us introduce the usual Dirac matrices $\gamma^{\mu}$, satisfying
\begin{eqnarray}
&&\left\{ \gamma^{\mu}\, ,\, \gamma^{\nu}\right\} = 2\, g^{\mu \nu} \label{anticom_gamma_mu}
\\
&&\big(\gamma^{\mu}\big)^{\dag} = \gamma^{0}\gamma^{\mu}\gamma^{0}
\label{conjugate gamma_mu}
\, .
\end{eqnarray}
Hence, one has for example $\gamma^+ \gamma^+=0$.
In the conventional dimensional regularization \cite{'tHooft:1972fi} we are considering, there are $D$ distinct $\gamma^{\mu}$ matrices, but each of them is treated as a $4\times 4$ matrix. This implies in particular
\begin{eqnarray}
\textrm{Tr}\big[\gamma^{\mu} \gamma^{\nu}\big]&=&4\, g^{\mu \nu}
\, .
\end{eqnarray}
The $\gamma^{\mu}$ matrices also satisfy the contraction identities
\begin{eqnarray}
g_{\mu \nu} \gamma^{\mu} \gamma^{\rho} \gamma^{\nu}&=& -(D\!-\!2) \gamma^{\rho}
\\
g_{\mu \nu} \gamma^{\mu} \gamma^{\rho} \gamma^{\sigma} \gamma^{\nu}
&=& 4 g^{\rho \sigma}
+(D\!-\!4) \gamma^{\rho} \gamma^{\sigma}
\\
g_{\mu \nu} \gamma^{\mu} \gamma^{\rho} \gamma^{\sigma} \gamma^{\eta} \gamma^{\nu}
&=& -2 \gamma^{\eta}\gamma^{\sigma}\gamma^{\rho}
-(D\!-\!4) \gamma^{\rho} \gamma^{\sigma} \gamma^{\eta}
\, .
\label{5_gamma_contraction}
\end{eqnarray}
In the light-cone gauge, thanks to the expression \eqref{4_to_2_polarization} for physical polarizations vectors, one has
\begin{eqnarray}
\left\{ \gamma^+ , \slashed{\epsilon}_{\lambda}\!(\underline{k})\right\}
=\left\{ \gamma^+ , \slashed{\epsilon}_{\lambda}^*\!(\underline{k})\right\}
=0\, .
\label{anticom_gammaplus_epsilonslash}
\end{eqnarray}

The spinor fields can be decomposed into the so-called good ($G$) and bad ($B$) components. The former ones are the independent degrees of freedom which are quantized within the light-front quantization, whereas the latter are determined at each $x^+$ by constraint equations as a function of the former \cite{Kogut:1969xa}. These constraint equations are a subset of the equations of motion, which involve no derivative with respect to $x^+$, once written in light-cone coordinates. The projectors over good and bad components of a spinor field $\Psi$ are
\begin{eqnarray}
{\cal P}_{G}\equiv \frac{\gamma^-\, \gamma^+}{2} = \frac{\gamma^0\, \gamma^+}{\sqrt{2}}\ ,
\nonumber \\
{\cal P}_{B}\equiv \frac{\gamma^+\, \gamma^-}{2} = \frac{\gamma^0\, \gamma^-}{\sqrt{2}}\, ,
\end{eqnarray}
so that
\begin{eqnarray}
\textrm{Tr}\big[{\cal P}_{G}\big]&=&\textrm{Tr}\big[{\cal P}_{B}\big]=2
\, .
\end{eqnarray}
The projections of the field are written
\begin{equation}
\Psi_{G,B}   \equiv {\cal P}_{G,B}\; \Psi\ .
\end{equation}
This implies for conjugate spinor fields $\overline{\Psi}\equiv (\Psi)^\dag\, \gamma^0$ the relations
\begin{equation}
\overline{\Psi} \;  {\cal P}_{B}     = \overline{\Psi_{G}},\ \
\overline{\Psi}  \; {\cal P}_{G}     = \overline{\Psi_{B}}.
\end{equation}

The spinors $u(\underline{k},h)$ and $v(\underline{k},h)$ entering in eq. \eqref{quant_free_q} are defined as usual as the momentum-space solutions of the free Dirac equation, so that
\begin{eqnarray}
(\slashed{k}-m)\, u(\underline{k},h)&=& (\slashed{k}+m)\, v(\underline{k},h)=0\nonumber\\
\overline{u}(\underline{k},h)\,(\slashed{k}-m)
&=&\overline{v}(\underline{k},h)\,(\slashed{k}+m)=0\, ,
\label{mom_Dirac_eqs}
\end{eqnarray}
using the notation $k^-\equiv(\k^2+m^2)/(2k^+)$.
They can be decomposed into good and bad components, which are related through the constraint equations as
\begin{eqnarray}
u_B(\underline{k},h) &=& \frac{\gamma^+}{2 k^+}\, \left(\k^j \gamma^j \!+\! m \right)\, u_G(k^+,h), \label{u_bad}\nonumber\\
v_B(\underline{k},h) &=& \frac{\gamma^+}{2 k^+}\, \left(\k^j \gamma^j \!-\! m \right)\, v_G(k^+,h)\, , \label{v_bad}
\end{eqnarray}
which corresponds, for the conjugate spinors, to
\begin{eqnarray}
\overline{u_B}(\underline{k},h) &=&  \overline{u_G}(k^+,h)\, \left(\k^j \gamma^j \!+\! m \right)\,  \frac{\gamma^+}{2 k^+}\ ,  \label{ubar_bad}\nonumber\\
\overline{v_B}(\underline{k},h) &=& \overline{v_G}(k^+,h)\, \left(\k^j \gamma^j \!-\! m \right)\,\frac{\gamma^+}{2 k^+}\ . \label{vbar_bad}
\end{eqnarray}
Note that the good components depend only on $k^+$ and $h$, whereas the bad components carry as well a dependence on $\k$ and $m$. One has the completeness relations
\begin{eqnarray}
\sum_{h=\pm\frac{1}{2}} u_G(k^+,h)\;  \overline{u_G}(k^+,h)\, \gamma^+
&=& \sum_{h=\pm\frac{1}{2}} v_G(k^+,h)\;  \overline{v_G}(k^+,h)\, \gamma^+
= 2k^+\; {\cal P}_{G}\ .
\label{completeness_spinors_good}
\end{eqnarray}
at the level of the good components, and
\begin{eqnarray}
\sum_{h=\pm 1/2} u(\underline{k},h)\;\;
\overline{u}(\underline{k},h)\;  &=& \slashed{k}+m \nonumber\\
\sum_{h=\pm 1/2} v(\underline{k},h)\;\;
\overline{v}(\underline{k},h)\;  &=& \slashed{k}-m
\label{completeness_spinors_full}
\end{eqnarray}
at the level of the complete spinors (again with the notation $k^-\equiv(\k^2+m^2)/(2k^+)$).
One has also the orthogonality relations
\begin{eqnarray}
\overline{u_G}(k^{'+},h')\, \gamma^+ \,u_G(k^+,h)
&=& \overline{v_G}(k^{'+},h')\, \gamma^+ \,v_G(k^+,h)
= \sqrt{2k^+}\;  \sqrt{2k^{'+}}\;\delta_{h',h}
\nonumber\\
\overline{u_G}(k^{'+},h')\, \gamma^+ \,v_G(k^+,h)
&=& \overline{v_G}(k^{'+},h')\, \gamma^+ \,u_G(k^+,h)
= \sqrt{2k^+}\;  \sqrt{2k^{'+}}\;\delta_{h',-h}
\ .
\label{orthogonality_spinors_good}
\end{eqnarray}

In the rest of the paper, the notations
\begin{eqnarray}
u(n)\equiv u(\underline{k_n},h_n) \, ,\;  v(n)\equiv v(\underline{k_n},h_n)
\end{eqnarray}
as well as
\begin{eqnarray}
a_n\equiv a(\underline{k_n},\lambda_n,a_n) \, ,\;
b_n\equiv b(\underline{k_n},h_n,\alpha_n)  \, ,\;
d_n\equiv d(\underline{k_n},h_n,\alpha_n)
\end{eqnarray}
are often used, in order to get more compact expressions.


\subsection{Light-front wave functions in perturbation theory}

In the case of an interacting field theory in the Heisenberg picture, the physical states are the eigenstates of the light-front Hamiltonian $\hat{P}^-$, which generate the $x^+$ evolution of quantum operators. However, these eigenstates can be constructed in perturbation theory, using the interaction picture.
In practice, one first uses the classical equations of motions to write $\hat{P}^-$ in terms of the independent components of the fields, like the good components of the spinors, and eliminate the constrained components, like the bad components of spinors, which \emph{a priori} depend on interactions as well. As a result, one obtains a decomposition of the type $\hat{P}^-= \hat{T}+\hat{V}$ (see for example ref. \cite{Brodsky:1997de}), where $\hat{T}$ is the kinetic term for the independent components, and $\hat{V}$ collects the interaction terms.

Switching to the interaction picture, the $x^+$ evolution of operators is now generated by the free Hamiltonian $\hat{T}$, for example
\begin{equation}
\hat{V}_I(x^+)= e^{i \hat{T} x^+}\, \hat{V}_I(0)\, e^{-i \hat{T} x^+}
\label{evol_int_op_I_pict}\, ,
\end{equation}
for the interaction operator, and the states in the interaction picture $|i_I(x^+)\rangle$ evolve as
\begin{equation}
 \big|\, i_I(x_2^+)\big\rangle= {\cal P}\, \exp\left(  -i\int_{x_1^+}^{x_2^+} \textrm{d} x^+\, \hat{V}_I(x^+)\right)\: \big|\, i_I(x_1^+)\big\rangle ,\label{evol_state_I_pict}
\end{equation}
where ${\cal P}$ indicates the ordering along $x^+$ of the operators $\hat{V}_I(x^+)$.
By convention, states in the interaction picture are defined in such a way that they coincide at $x^+=0$ with the corresponding states in the Heisenberg picture, \emph{i.e.} $|i_I(0)\rangle \equiv |i_H\rangle$.
Hence, for $x_2^+=0$ and $x_1^+\rightarrow -\infty$, one has
\begin{equation}
 \big|\, i_H\big\rangle= {\cal P}\, \exp\left(  -i\int_{-\infty}^{0} d x^+\, \hat{V}_I(x^+)\right)\: \big|\, i_I(-\infty)\big\rangle .\label{evol_state_I_pict_2}
\end{equation}
In perturbation theory, the state $\big|\, i_I(-\infty)\big\rangle$ can be considered free, since it comes before any insertion of the interaction operator $\hat{V}_I(x^+)$. Hence, eq. \eqref{evol_state_I_pict_2} relates the interacting (a.k.a. dressed or physical) state $\big|\, i_H\big\rangle$ to its free (a.k.a. bare or asymptotic) analog $\big|\, i_I(-\infty)\big\rangle$.

One can introduce the decomposition of the identity
\begin{equation}
{ \mathbf{ 1}}=\sum_{\cal F} |{\cal F}\rangle \langle {\cal F}|,\label{identity_decomp_1}
\end{equation}
over a basis of Fock states $|{\cal F}\rangle$.
Here, $|{\cal F}\rangle$ simply represent the vacuum state $|0\rangle$ with an arbitrary number of creation operators (in momentum space) of any type acting on it, without any normalization factor. Hence, each Fock state $|{\cal F}\rangle$ is an eigenstate of the free Hamiltonian $\hat{T}$.
The sum over Fock states used in eq. \eqref{identity_decomp_1} contains summations over the number $N_t$ of partons of each type $t$ present in the Fock state, with a symmetry factor $1/N_t!$ for each parton type.
Then, the sum in eq. \eqref{identity_decomp_1} also contains for each parton present in the Fock state a summation over all of its quantum numbers and a phase-space integration
\begin{eqnarray}
\int_{0}^{+\infty}\!\!\!\! \frac{dk^+}{(2\pi) 2k^+}
 \int \frac{d^{D-2} \k}{(2\pi)^{D-2}}
 \label{phase_space_mom}
\, .
\end{eqnarray}

The LFWFs $\Phi_{i\rightarrow {\cal F}}$ are defined as the coefficients of the expansion of a physical state $\big|\, i_H\big\rangle$ over the basis of Fock states, as
\begin{equation}
 \big|\, i_H\big\rangle=  \sum_{\cal F} |{\cal F}\rangle\; \Phi_{i\rightarrow {\cal F}}
 \, ,\label{def_1_LFWF}
\end{equation}
so that
\begin{equation}
\Phi_{i\rightarrow {\cal F}} =   \langle {\cal F}  \big|\, i_H\big\rangle
= \langle {\cal F} \big|
{\cal P}\, \exp\left(  -i\int_{-\infty}^{0} d x^+\, \hat{V}_I(x^+)\right)\: \big|\, i_I(-\infty)\big\rangle
 \, ,\label{def_2_LFWF}
\end{equation}

Expanding the exponential in \eqref{def_2_LFWF}, using the relation \eqref{evol_int_op_I_pict} to extract the $x^+$ dependence from $\hat{V}_I(x^+)$, and inserting various times the relation \eqref{identity_decomp_1}, one obtains after some calculations the perturbative expansion \cite{Bjorken:1970ah}
\begin{eqnarray}
\Phi_{i\rightarrow {\cal F}}
&=& \langle {\cal F}|i_I(-\infty)\rangle
+\sum_{n=1}^{\infty}  \sum_{{\cal F}_{n\!-\!1}}\cdots \sum_{{\cal F}_0} \frac{1}{T_{{\cal F}_{0}}\!-\!T_{{\cal F}}+i\epsilon}  \, \langle {\cal F}| \hat{V}_I(0)|{\cal F}_{n-1}\rangle \,\frac{1}{T_{{\cal F}_{0}}\!-\!T_{{\cal F}_{n\!-\!1}}+i\epsilon}  \, \cdots\nonumber\\
& &  \cdots  \,\frac{1}{T_{{\cal F}_{0}}\!-\!T_{{\cal F}_{1}}+i\epsilon}  \, \langle {\cal F}_1| \hat{V}_I(0)|{\cal F}_0\rangle\;
\langle {\cal F}_0 |i_I(-\infty)\rangle
\label{LFWF_pert}
\end{eqnarray}
for the LFWFs. In eq. \eqref{LFWF_pert}, $T_{{\cal F}_{m}}$ is the eigenvalue of the free Hamiltonian $\hat{T}$ in the state $|{\cal F}_{m}\rangle$. Hence, $T_{{\cal F}_{m}}$ is the sum of $({\mathbf{k}_l}^2+m_l^2)/(2 k_l^+)$ for each parton $l$ present in the Fock state $|{\cal F}_{m}\rangle$. For simplicity, one often uses the notation $k_l^-\equiv({\mathbf{k}_l}^2+m_l^2)/(2 k_l^+)$ in this context.

In the particular case where $\big|\, i_H\big\rangle$ is a (perturbative) one-particle state, $|i_I(-\infty)\rangle$ is a one-particle Fock state.
It is then convenient to define the LFWFs in a slightly different way \cite{Bjorken:1970ah}, extracting the wave-function renormalization constant $Z_i$, as
\begin{equation}
 \big|\, i_H\big\rangle= \sqrt{Z_i}\;\bigg\{|i_I(-\infty)\rangle
  +\sum_{{\cal F} \neq i} |{\cal F}\rangle\; \Psi_{i\rightarrow {\cal F}}
 \bigg\}
 \, ,\label{def_1_LFWF_one_part}
\end{equation}
where the sum over Fock states is now excluding  $|i_I(-\infty)\rangle$. The new LFWFs (for ${\cal F}\neq i$) admit the perturbative expansion
\begin{eqnarray}
\Psi_{i\rightarrow {\cal F}}
&=&
\frac{\langle {\cal F}| \hat{V}_I(0)|i_I(-\infty)\rangle}{T_{i}\!-\!T_{{\cal F}}+i\epsilon}
+\sum_{n=2}^{\infty}  \sum_{{\cal F}_{n\!-\!1}\neq i}\cdots \sum_{{\cal F}_1\neq i} \frac{1}{T_{i}\!-\!T_{{\cal F}}+i\epsilon}  \, \langle {\cal F}| \hat{V}_I(0)|{\cal F}_{n-1}\rangle \,\frac{1}{T_{i}\!-\!T_{{\cal F}_{n\!-\!1}}+i\epsilon}  \, \cdots\nonumber\\
& &  \cdots  \,\frac{1}{T_{i}\!-\!T_{{\cal F}_{1}}+i\epsilon}  \, \langle {\cal F}_1| \hat{V}_I(0) |i_I(-\infty)\rangle
\, ,
\label{LFWF_pert_one_part}
\end{eqnarray}
where $T_{i}=k_i^-\equiv({\mathbf{k}_i}^2+m_i^2)/(2 k_i^+)$ is the eigenvalue of $\hat{T}$ corresponding to the Fock state $|i_I(-\infty)\rangle$.


\subsection{Light-front perturbation theory for QED+QCD}

In the perturbative expansions \eqref{LFWF_pert} or \eqref{LFWF_pert_one_part} of the LFWFs, it remains to specify the interaction vertices $\langle {\cal F}| \hat{V}_I(0)|{\cal F}'\rangle$, which depend of course on the considered field theory. For the calculation done in the present paper, one needs the QCD and QED sector on the Standard Model (including for simplicity only one quark flavor). The general procedure to calculate the interaction operator $\hat{V}_I(x^+)$ in the interaction picture is explained in detail in the chapter 2 of ref. \cite{Brodsky:1997de}, with the QCD case treated explicitly. It amounts to calculating the light-cone energy operator $\hat{P}^-$ and then using the constraint equations to eliminate the dependent components of the fields. Finally, the kinetic and the interaction pieces can be separated as $\hat{P}^-= \hat{T}+\hat{V}$, and the latter piece can be written in the interaction picture using the free quantized fields of section \ref{sec:free_fields_conventions}.

In QCD with one quark flavor, one gets \cite{Brodsky:1997de}
\begin{eqnarray}
&& \hat{V}^{\textrm{QCD}}_I(x^+)= \int d^{D-2} \x \int dx^- \Bigg\{
(\mu)^{2-\frac{D}{2}}\, g\, A^{a}_{\mu}(x)\, \overline{\Psi}(x)\, \gamma^{\mu}
t^{a} \Psi(x)
-(\mu)^{2-\frac{D}{2}}\, g\, f^{abc}\, A_{a}^{\mu}(x)\, \left(\partial_{\mu}A_{b}^{j}(x)\right)\, A_{c}^{j}(x)
\nonumber\\
&&
+\frac{(\mu^2)^{2-\frac{D}{2}}\, g^2}{2}\, \left[
\overline{\Psi}(x)\, \gamma^{+} t^{a} \Psi(x)
- f^{abc}\, \left(\partial_{-}A_{b}^{i}(x)\right)\, A_{c}^{i}(x)
\right]\, \frac{1}{(i\partial_{-})^2}\, \left[
\overline{\Psi}(x)\, \gamma^{+} t^{a} \Psi(x)
- f^{ade}\, \left(\partial_{-}A_{d}^{j}(x)\right)\, A_{e}^{j}(x)
\right]
\nonumber\\
&&
+\frac{(\mu^2)^{2-\frac{D}{2}}\, g^2}{2}\, \left[
\overline{\Psi_G}(x)\, \gamma^{i} t^{a} \, A_{a}^{i}(x)
\right]\,  \frac{\gamma^+}{i\partial_{-}}\, \left[
\gamma^{j} t^{b} \, A_{b}^{j}(x)
{\Psi_G}(x)
\right]
+\frac{(\mu^2)^{2-\frac{D}{2}}\, g^2}{4}\, f^{abe}f^{cde}\, A_{a}^{i}(x)\, A_{b}^{j}(x)
\, A_{c}^{i}(x)\, A_{d}^{j}(x)
\Bigg\}
\label{int_op_QCD}
\, ,
\end{eqnarray}
where fundamental color indices have been kept implicit. The first two terms correspond to the standard $q\bar{q}g$ and $ggg$ interaction vertices, the last term to the usual $gggg$ vertex, whereas the second line in eq. \eqref{int_op_QCD} collects instantaneous Coulomb interactions between quarks or gluons and the first term in the third line corresponds to an instantaneous quark exchange.

The QED case is similar, after dropping the terms containing $f^{abc}$ and replacing the fundamental color generator $t^a$ by the fractional charge $e_f$ of the considered flavor of quark (or lepton), the coupling $g$ by $e$, and obviously the gluon field by the photon field. One gets
\begin{eqnarray}
\hat{V}^{\textrm{QED}}_I(x^+)&=& \int d^{D-2} \x \int dx^- \Bigg\{
(\mu)^{2-\frac{D}{2}}\, e\, e_f\, A_{\mu}(x)\, \overline{\Psi}(x)\, \gamma^{\mu} \Psi(x)
\nonumber\\
&&
+\frac{(\mu^2)^{2-\frac{D}{2}}\, e^2\, e_f^2}{2}\, \left[
\overline{\Psi}(x)\, \gamma^{+} \Psi(x)
\right]\, \frac{1}{(i\partial_{-})^2}\, \left[
\overline{\Psi}(x)\, \gamma^{+} \Psi(x)
\right]
\nonumber\\
&&
+\frac{(\mu^2)^{2-\frac{D}{2}}\, e^2\, e_f^2}{2}\, \left[
\overline{\Psi_G}(x)\, \gamma^{i} \, A^{i}(x)
\right]\,  \frac{\gamma^+}{i\partial_{-}}\, \left[
\gamma^{j}\, A^{j}(x)
{\Psi_G}(x)
\right]
\Bigg\}
\label{int_op_QED}
\, .
\end{eqnarray}

When taking simultaneously into account QCD and QED, like in the calculation done in the present paper, one should use the interaction operator
\begin{eqnarray}
\hat{V}_I(x^+)&=& \hat{V}^{\textrm{QCD}}_I(x^+)
+\hat{V}^{\textrm{QED}}_I(x^+)
+\hat{V}^{\textrm{mixed}}_I(x^+)
\label{int_op_QCD+QED}
\, ,
\end{eqnarray}
with the mixed QCD/QED interaction terms given by
\begin{eqnarray}
\hat{V}^{\textrm{mixed}}_I(x^+)&=& \int d^{D-2} \x \int dx^- \Bigg\{
\frac{(\mu^2)^{2-\frac{D}{2}}\, g\, e\, e_f}{2}\, \left[
\overline{\Psi_G}(x)\, \gamma^{i} t^{a} \, A_{a}^{i}(x)
\right]\,  \frac{\gamma^+}{i\partial_{-}}\, \left[
\gamma^{j} \, A^{j}(x)
{\Psi_G}(x)
\right]
\nonumber\\
&&
+\frac{(\mu^2)^{2-\frac{D}{2}}\, g\, e\, e_f}{2}\, \left[
\overline{\Psi_G}(x)\, \gamma^{j}\, A^{j}(x)
\right]\,  \frac{\gamma^+}{i\partial_{-}}\, \left[
\gamma^{i} t^{a} \, A_{a}^{i}(x)
{\Psi_G}(x)
\right]
\Bigg\}
\label{int_op_mix_QCD_QED}
\, ,
\end{eqnarray}
which provides nonlocal $q\bar{q}\gamma g$ interaction vertices via an instantaneous quark exchange.

The last step is to evaluate the interaction vertices $\langle {\cal F}| \hat{V}_I(0)|{\cal F}'\rangle$ encountered in the perturbative expansion of the LFWFs of interest. For each Fock state $|{\cal F}\rangle$ and $|{\cal F}'\rangle$, this can be done by inserting in $\hat{V}_I(0)$, as written in eq. \eqref{int_op_QCD+QED}, the expressions \eqref{quant_free_q} and \eqref{quant_free_g} of the free quantized fields in terms of creation and annihilation operators, and then using the commutation relations \eqref{com_rel_mom}. In principle, one gets two types of nonzero contributions.

In the first type, each creation and annihilation operator from the free fields in $\hat{V}_I(0)$ gets contracted with one creation operator from $|{\cal F}'\rangle$ or one annihilation operator from $\langle {\cal F}|$. The contributions of this first type are the ones obtained when calculating $\langle {\cal F}| \!:\!\hat{V}_I(0)\!:\!|{\cal F}'\rangle$, with the normal ordering prescription applied to $\hat{V}_I(0)$.

The contributions of the second type are the ones providing the mismatch between
$\langle {\cal F}| \hat{V}_I(0)|{\cal F}'\rangle$ and $\langle {\cal F}| \!:\!\hat{V}_I(0)\!:\!|{\cal F}'\rangle$. They correspond to contributions to
$\langle {\cal F}| \hat{V}_I(0)|{\cal F}'\rangle$ in which at least one
creation and one annihilation operator from the free fields in $\hat{V}_I(0)$ are contracted with each other. In this way, the three-field terms in $\hat{V}_I(0)$ can generate one-point vertices for gluon and photon, and the four-field terms in $\hat{V}_I(0)$ can generate two-points vertices and zero-point vertices. In dimensional regularization, all the contributions of this second type vanish\footnote{If one is reluctant to use this argument, or if another UV regularization is used, it is useful to note the following things. First, the zero-point vertices correspond to contributions to vacuum energy, which are irrelevant as long as gravity is not included. Second, the one-gluon vertices vanish identically because they come with a zero color factor. Moreover, the one-gluon and the one-photon vertices can be shown to vanish due to rotational symmetry in the transverse plane. Hence, in the general case, only the two-points vertices coming from the internal contractions of the four-field vertices inside $\hat{V}_I(0)$ might give  nontrivial contribution, called self-inertia (see for example the discussions in refs. \cite{Mustaki:1990im} and \cite{Zhang:1993dd}).} due to the scale-less transverse integrals they contain. Hence, in dimensional regularization, the interaction operator $\hat{V}_I(0)$ can be treated as normal-ordered when calculating the vertices $\langle {\cal F}| \hat{V}_I(0)|{\cal F}'\rangle$.

Inserting the free quantized fields \eqref{quant_free_q} and \eqref{quant_free_g} into the first term (the $q\bar{q}g$ term) in $\hat{V}^{\textrm{QCD}}_I(0)$ from eq. \eqref{int_op_QCD}, one finds after straightforward calculations the following nontrivial vertices:
\begin{eqnarray}
\langle 0| a_{2} b_{1}\, V_I(0)\, b_{0}^{\dag} |0 \rangle
&=& (2\pi)^{D\!-\!1} \delta^{(D\!-\!1)}(\underline{k_{1}}\!+\!\underline{k_{2}}\!-\!\underline{k_0})\;
\mu^{2-\frac{D}{2}}\, g\, t^{a_2}_{\alpha_{1}\alpha_{0}}\;
\overline{u}(1)\; \slashed{\epsilon}_{\lambda_2}^*\!(\underline{k_2})\, {u}(0)
\nonumber\\
\langle 0| b_{1}\, V_I(0)\, b_{0}^{\dag} a_{2}^{\dag} |0 \rangle
&=& (2\pi)^{D\!-\!1} \delta^{(D\!-\!1)}(\underline{k_{1}}\!-\!\underline{k_{2}}\!-\!\underline{k_0})\;
\mu^{2-\frac{D}{2}}\, g\, t^{a_2}_{\alpha_{1}\alpha_{0}}\;
\overline{u}(1)\; \slashed{\epsilon}_{\lambda_2}\!(\underline{k_2})\, {u}(0)
\nonumber\\
\langle 0| a_{2} d_{0}\, V_I(0)\, d_{1}^{\dag} |0 \rangle
&=& (2\pi)^{D\!-\!1} \delta^{(D\!-\!1)}(\underline{k_{0}}\!+\!\underline{k_{2}}\!-\!\underline{k_1})\;
\mu^{2-\frac{D}{2}}\, g\, t^{a_2}_{\alpha_{1}\alpha_{0}}\;
(-1)\, \overline{v}(1)\; \slashed{\epsilon}_{\lambda_2}^*\!(\underline{k_2})\, {v}(0)
\nonumber\\
\langle 0|  d_{0}\, V_I(0)\, d_{1}^{\dag} a_{2}^{\dag} |0 \rangle
&=& (2\pi)^{D\!-\!1} \delta^{(D\!-\!1)}(\underline{k_{0}}\!-\!\underline{k_{2}}\!-\!\underline{k_1})\;
\mu^{2-\frac{D}{2}}\, g\, t^{a_2}_{\alpha_{1}\alpha_{0}}\;
(-1)\, \overline{v}(1)\; \slashed{\epsilon}_{\lambda_2}\!(\underline{k_2})\, {v}(0)
\nonumber\\
\langle 0| d_{0} b_{1}\, V_I(0)\, a_{2}^{\dag} |0 \rangle
&=& (2\pi)^{D\!-\!1} \delta^{(D\!-\!1)}(\underline{k_{0}}\!+\!\underline{k_{1}}\!-\!\underline{k_2})\;
\mu^{2-\frac{D}{2}}\, g\, t^{a_2}_{\alpha_{1}\alpha_{0}}\;
\overline{u}(1)\; \slashed{\epsilon}_{\lambda_2}\!(\underline{k_2})\, {v}(0)
\nonumber\\
\langle 0| a_{2}\, V_I(0)\, b_{0}^{\dag} d_{1}^{\dag} |0 \rangle
&=& (2\pi)^{D\!-\!1} \delta^{(D\!-\!1)}(\underline{k_{2}}\!-\!\underline{k_{0}}\!-\!\underline{k_1})\;
\mu^{2-\frac{D}{2}}\, g\, t^{a_2}_{\alpha_{1}\alpha_{0}}\;
\overline{v}(1)\; \slashed{\epsilon}_{\lambda_2}^*\!(\underline{k_2})\, {u}(0)
\label{q_qbar_g_vertices}
\, .
\end{eqnarray}

In the QED case, the three-point vertices are the same as in eq. \eqref{q_qbar_g_vertices}, up to the replacement of $g\, t^{a_2}_{\alpha_{1}\alpha_{0}}$ by $e\, e_f$.
Obviously, there are many more nontrivial vertices that one can obtain out of
$\hat{V}_I(0)$ in the QCD+QED case, following the same method. Instead of writing them all, let us only list the ones appearing in the calculation done in the present paper.

First, one has the $\gamma\rightarrow q\bar{q}g$ vertex
\begin{eqnarray}
\langle 0| a_{2} d_{1} b_{0'}\, V_I(0)\, a_{\gamma}^{\dag} |0 \rangle
&=& (2\pi)^{D-1}\delta^{(D-1)}(\underline{k_2}+\underline{k_1}+
\underline{k_{0'}}\!-\!\underline{q})\;
(\mu^2)^{2-\frac{D}{2}}\; \frac{e\, g}{2}\, e_f\, \left(t^{a_2}\right)_{\alpha_{0'}\,\alpha_{1}}\nonumber\\
&& \;\; \times\;\;
\overline{u}(0')\,
\left[\frac{\slashed{\epsilon}_{\lambda_2}^*\!(\underline{k_2})\; \gamma^+\,
             \slashed{\epsilon}_{\lambda}\!(\underline{q})}{(k_{0'}^++k_{2}^+)}
      -\frac{\slashed{\epsilon}_{\lambda}\!(\underline{q})\; \gamma^+\,
             \slashed{\epsilon}_{\lambda_2}^*\!(\underline{k_2})
             }{(k_{1}^++k_{2}^+)}
\right] \; v(1)
\label{gamma_2_qqbarg_vertex}
\, ,
\end{eqnarray}
obtained from $\hat{V}^{\textrm{mixed}}_I(0)$ in eq. \eqref{int_op_mix_QCD_QED}, where the incoming photon has a momentum $\underline{q}$ and a polarization $\lambda$.
The first term in the eq. \eqref{gamma_2_qqbarg_vertex} contributes to the
diagrams $A'$ and $2'$ from fig. \ref{Fig:gammaT_NLO}, whereas the second term contributes to the diagrams $1'$ and $B'$.

From the second line in eq. \eqref{int_op_QCD} for $\hat{V}^{\textrm{QCD}}_I(0)$, one obtains the $q\bar{q}\rightarrow q\bar{q}$  vertex
\begin{eqnarray}
\langle 0| d_{1} b_{0}\, V_I(0)\, b_{0'}^{\dag} d_{1'}^{\dag}  |0 \rangle
&=& (2\pi)^{D-1}\delta^{(D-1)}(\underline{k_{0'}}+\underline{k_{1'}}\!-\!
\underline{k_{0}}\!-\!\underline{k_1})\;
(\mu^2)^{2-\frac{D}{2}}\; g^2\,\nonumber\\
&& \;\; \times\;\;
\bigg\{
-\left(t^{a_2}\right)_{\alpha_{0}\,\alpha_{0'}}\, \left(t^{a_2}\right)_{\alpha_{1'}\,\alpha_{1}}\, \frac{1}{(k_{0}^+\!-\!k_{0'}^+)^2}\;
\overline{u}(0)\, \gamma^+ u(0')\;\; \overline{v}(1')\, \gamma^+ v(1)
\nonumber\\
&& \;\; \;\; \;\; \;\; \;\;\;
+\left(t^{a_2}\right)_{\alpha_{0}\,\alpha_{1}}\, \left(t^{a_2}\right)_{\alpha_{1'}\,\alpha_{0'}}\, \frac{1}{(k_{0}^+\!+\!k_{1}^+)^2}\;
\overline{u}(0)\, \gamma^+ v(1)\;\; \overline{v}(1')\, \gamma^+ u(0')
\bigg\}
\, .
\label{qqbar_to_qqbar_vertex}
\end{eqnarray}
In this expression, only the first term corresponds to an instantaneous Coulomb interaction between a quark and an antiquark in the $t$ channel. By contrast, the second term stands for a $q\bar{q}$ annihilation followed (instantaneously) by a $q\bar{q}$ pair creation annihilation, via an $s$-channel intermediate gluon.
Only the first term in eq. \eqref{qqbar_to_qqbar_vertex} contributes to the diagram $3$ of fig.  \ref{Fig:gammaT_NLO}, since the incoming $q\bar{q}$ pair is in a color singlet state.


\subsection{Fourier transform to mixed space}

In view of applications to the dipole factorization of DIS observables, or more generally to gluon saturation physics, it is convenient to switch from the full momentum-space representation to the mixed-space representation ($k^+$ and $\x$). The two are related by transverse Fourier transform as\footnote{The convention used here to define the Fourier transform is different from the one used in ref. \cite{Beuf:2011xd}, and more standard. Compare with eqs. (A1), (A2) and (A11) from ref. \cite{Beuf:2011xd}. This represents one of the few changes of conventions from ref. \cite{Beuf:2011xd} to the present paper, together with the switch to dimensional regularization and the different way of writing the interaction vertices.}
\begin{eqnarray}
a^{\dagger}(\underline{k},\lambda,a)&=&\int d^{D-2}\x\;\; e^{i\k \cdot \x}\;\;
a^{\dagger}(k^+,\x,\lambda,a)
\label{FT_a_dag}
\\
b^{\dagger}(\underline{k},h,\alpha)&=&\int d^{D-2}\x\;\; e^{i\k \cdot \x}\;\;
b^{\dagger}(k^+,\x,h,\alpha)
\label{FT_b_dag}
\\
d^{\dagger}(\underline{k},h,\alpha)&=&\int d^{D-2}\x\;\; e^{i\k \cdot \x}\;\;
d^{\dagger}(k^+,\x,h,\alpha)
\label{FT_d_dag}
\, .
\end{eqnarray}
Accordingly, the mixed-space commutation relations are
\begin{eqnarray}
\big[a(k_1^+,\x_1,\lambda_1,a_1),a^{\dagger}(k_2^+,\x_2,\lambda_2,a_2)\big]
&=& (2k_1^+)(2\pi)\delta(k_1^+\!-\!k_1^+)\;\delta^{(D\!-\!2)}(\x_1\!-\!\x_2)\;
\delta_{\lambda_1,\lambda_2}\; \delta_{a_1,a_2}\label{commute_a_adag_mix}\\
\big\{b(k_1^+,\x_1,h_1,\alpha_1),b^{\dagger}(k_2^+,\x_2,h_2,\alpha_2)\big\}
&=& (2k_1^+)(2\pi)\delta(k_1^+\!-\!k_1^+)\;\delta^{(D\!-\!2)}(\x_1\!-\!\x_2)\;
\delta_{h_1,h_2}\; \delta_{\alpha_1,\alpha_2}\label{anticommute_b_bdag_mix}\\
\big\{d(k_1^+,\x_1,h_1,\alpha_1),d^{\dagger}(k_2^+,\x_2,h_2,\alpha_2)\big\}
&=& (2k_1^+)(2\pi)\delta(k_1^+\!-\!k_1^+)\;\delta^{(D\!-\!2)}(\x_1\!-\!\x_2)\;
\delta_{h_1,h_2}\; \delta_{\alpha_1,\alpha_2}\label{anticommute_d_ddag_mix}
\, .
\end{eqnarray}
The mixed-space representation is useful in high-energy scattering on a dense target. Indeed, in that case, one can use the eikonal approximation, and each parton in the mixed-space representation scatters independently on the target simply by picking up a Wilson line in the appropriate representation or, in the QED case, an eikonal phase \cite{Bjorken:1970ah}.

Due to eqs. \eqref{FT_a_dag}, \eqref{FT_b_dag} and \eqref{FT_d_dag}, it is possible to relate a momentum-space Fock state $|{\cal F}\rangle$ to the analog mixed-space Fock state $|\widetilde{{\cal F}}\rangle$, constructed from mixed-space creation operators, by
\begin{eqnarray}
|\widetilde{{\cal F}}\rangle &=& \int \left[\prod_{l\in {\cal F}}
 \frac{d^{D-2} \k_l}{(2\pi)^{D-2}}\;  e^{-i\k_l \cdot \x_l}
\right]\; |{\cal F}\rangle \nonumber\\
|{\cal F}\rangle &=& \int \left[\prod_{l\in {\cal F}}
 d^{D-2} \x_l\;  e^{i\k_l \cdot \x_l}
\right]\; |\widetilde{{\cal F}}\rangle
\label{FT_Fock_state}
\, ,
\end{eqnarray}
with a product over each parton $l$ present in the Fock state $|{\cal F}\rangle$. Then, one has the the decomposition of the identity
\begin{equation}
{ \mathbf{ 1}}=\sum_{\widetilde{{\cal F}}} |\widetilde{{\cal F}}\rangle \langle \widetilde{{\cal F}}|,\label{identity_decomp_mix}
\end{equation}
over a basis of mixed-space Fock states. The definition of the summation in eq. \eqref{identity_decomp_mix} is the same as the one in eq. \eqref{identity_decomp_1}, up to the replacement of the phase-space integration \eqref{phase_space_mom} for each parton by
\begin{eqnarray}
\int_{0}^{+\infty}\!\!\!\! \frac{dk^+}{(2\pi) 2k^+}
 \int d^{D-2} \x
 \label{phase_space_mix}
\, .
\end{eqnarray}

Then, an arbitrary physical state $|i_H\rangle$ can be decomposed in a mixed-space Fock basis as
\begin{equation}
\big|\, i_H\big\rangle=  \sum_{\widetilde{{\cal F}}} |\widetilde{{\cal F}}\rangle\; \widetilde{\Phi}_{i\rightarrow {\cal F}}
\, ,\label{def_1_LFWF_mix}
\end{equation}
with the mixed-space LFWFs
\begin{equation}
\widetilde{\Phi}_{i\rightarrow {\cal F}} =   \langle \widetilde{{\cal F}}  \big|\, i_H\big\rangle
= \int \left[\prod_{l\in {\cal F}}
 \frac{d^{D-2} \k_l}{(2\pi)^{D-2}}\;  e^{i\k_l \cdot \x_l}
\right]\; \Phi_{i\rightarrow {\cal F}}
 \, .\label{def_2_LFWF_mix}
\end{equation}

Similarly, in the case of a one-particle state, one can define the mixed-space version of the renormalized LFWFs $\Psi_{i\rightarrow {\cal F}}$ as
\begin{equation}
\widetilde{\Psi}_{i\rightarrow {\cal F}}
= \int \left[\prod_{l\in {\cal F}}
 \frac{d^{D-2} \k_l}{(2\pi)^{D-2}}\;  e^{i\k_l \cdot \x_l}
\right]\; \Psi_{i\rightarrow {\cal F}}
 \, .\label{def_3_LFWF_mix}
\end{equation}



\section{A few relations for massless spinor bilinears\label{sec:spinor_bilin}}

A few intermediate results are derived in this section, which are useful for the calculation of the graphs encountered in this article. All the fermions are taken massless.
First, let us write component by component the current $\overline{u}(0)\; \gamma^{\mu}\;  v(1)$ as
\begin{eqnarray}
\overline{u}(0)\; \gamma^+\;  v(1)
&=&  \overline{u_G}(0)\; \gamma^+\,  v_G(1)
\label{quark_current_plus}\\
\overline{u}(0)\; \gamma^i\;  v(1)
&=& - \frac{\k_1^j}{2k^+_1}\; \overline{u_G}(0)\;
\gamma^+ \gamma^i \gamma^j\; v_G(1)
- \frac{\k_0^j}{2k^+_0}\; \overline{u_G}(0)\;
\gamma^+ \gamma^j \gamma^i\; v_G(1)\nonumber\\
&=& \bigg[\frac{\k_0^i}{2k^+_0}+ \frac{\k_1^i}{2k^+_1}\bigg]\;
\overline{u_G}(0)\; \gamma^+\; v_G(1)
+\bigg[\frac{\k_0^j}{2k^+_0}- \frac{\k_0^j}{2k^+_0}\bigg]\;
\frac{1}{2}\;\overline{u_G}(0)\; \gamma^+ \left[\gamma^i, \gamma^j\right]\, v_G(1)
\label{quark_current_perp}\\
\overline{u}(0)\; \gamma^-\;  v(1)
&=& - \frac{\k_0^i\, \k_1^j}{2k^+_0\, k^+_1}\;
\overline{u_G}(0)\;\gamma^+ \gamma^i \gamma^j\, v_G(1)
\label{quark_current_minus}
\, ,
\end{eqnarray}
using definitions and formulae from sec. \ref{sec:free_fields_conventions}.

Hence, for \emph{a priori} unrelated momenta $\underline{k_{0}}$, $\underline{k_{1}}$ and $\underline{k_{2}}$, with $k_2^- \equiv \k_2^2/(2k_2^+)$, one gets
\begin{eqnarray}
\overline{u}(0)\; \slashed{k}_{2}\;  v(1)
=  -\frac{1}{2\, k_2^+}\;
\left(\k_{2}^i \!-\!\frac{k_{2}^+}{k_{0}^+}\,\k_{0}^i\right)
\left(\k_{2}^j \!-\!\frac{k_{2}^+}{k_{1}^+}\,\k_{1}^j\right)\,
\overline{u}(0)\, \gamma^+ \gamma^{i} \gamma^{j}\,  v(1)
\label{quark_current_times_k2}
\, .
\end{eqnarray}

For \emph{a priori} unrelated momenta $\underline{k_{0}}$, $\underline{k_{1}}$ and $\underline{q}$ one also obtains from eqs. \eqref{quark_current_plus}, \eqref{quark_current_perp} and \eqref{quark_current_minus}
\begin{eqnarray}
\overline{u}(0)\; \slashed{\epsilon}_{\lambda}\!(\underline{q})\;  v(1)
=  \left[\frac{\q}{q^+}\!-\!\frac{\k_0}{2k^+_0}\!-\!\frac{\k_1}{2k^+_1} \right]
\!\cdot\! \varepsilon_{\lambda}\; \overline{u_G}(0)\, \gamma^+\,  v_G(1)
+\left[\frac{\k_1^j}{2k^+_1}\!-\!\frac{\k_0^j}{2k^+_0}\right]\;
\frac{\varepsilon_{\lambda}^i}{2}\,\overline{u_G}(0)\; \gamma^+ \left[\gamma^i, \gamma^j\right]\, v_G(1)
\label{quark_current_times_epsilon_generic}
\, .
\end{eqnarray}
On the other hand, it is straightforward to show that
\begin{eqnarray}
\overline{u}(0)\; \gamma^+ \slashed{\epsilon}_{\lambda}\!(\underline{q})\;  \gamma^j\, v(1)
&=&  \varepsilon_{\lambda}^j\;\overline{u_G}(0)\; \gamma^+\, v_G(1)
  - \frac{\varepsilon_{\lambda}^i}{2}\;\overline{u_G}(0)\; \gamma^+ \left[\gamma^i, \gamma^j\right]\, v_G(1)\label{quark_current_plus_epsilon j_com}\\
\overline{u}(0)\; \gamma^+ \gamma^j\, \slashed{\epsilon}_{\lambda}\!(\underline{q})\;  v(1)
&=&  \varepsilon_{\lambda}^j\;\overline{u_G}(0)\; \gamma^+\, v_G(1)
  + \frac{\varepsilon_{\lambda}^i}{2}\;\overline{u_G}(0)\; \gamma^+ \left[\gamma^i, \gamma^j\right]\, v_G(1)\, .
  \label{quark_current_plus j_epsilon_com}
\end{eqnarray}
Imposing now the momentum conservation relation $\underline{k_{0}}+\underline{k_{1}}=\underline{q}$, and introducing the relative momentum $\P$ of the quark as defined in eq. \eqref{cv_k0_to_P}, one can rewrite eq. \eqref{quark_current_times_epsilon_generic} as
\begin{eqnarray}
2k^+_0k^+_1\; \overline{u}(0)\; \slashed{\epsilon}_{\lambda}\!(\underline{q})\;  v(1)
=  \left(k^+_0\!-\!k^+_1\right)\; \P\!\cdot\! \varepsilon_{\lambda}\; \overline{u_G}(0)\, \gamma^+\,  v_G(1)
-\frac{q^+}{2}\; \varepsilon_{\lambda}^i\, \P^j\;
\overline{u_G}(0)\; \gamma^+ \left[\gamma^i, \gamma^j\right]\, v_G(1)
\label{quark_current_times_epsilon_mom_cons}
\, .
\end{eqnarray}
Then, combining \eqref{quark_current_times_epsilon_mom_cons} with
\eqref{quark_current_plus_epsilon j_com} and \eqref{quark_current_plus j_epsilon_com},
one obtains
\begin{eqnarray}
\P^j\;\overline{u}(0)\; \gamma^+ \slashed{\epsilon}_{\lambda}\!(\underline{q})\;  \gamma^j\, v(1)
&=&  \frac{2k^+_0k^+_1}{q^+}\;
\overline{u}(0)\; \slashed{\epsilon}_{\lambda}\!(\underline{q})\;  v(1)
+ \frac{2k^+_1}{q^+}\; \P\!\cdot\! \varepsilon_{\lambda}\;
\overline{u}(0)\, \gamma^+\,  v(1)
\label{quark_current_plus_epsilon j}\\
\P^j\;\overline{u}(0)\; \gamma^+ \gamma^j\, \slashed{\epsilon}_{\lambda}\!(\underline{q})\;  v(1)
&=&  - \frac{2k^+_0k^+_1}{q^+}\;
\overline{u}(0)\; \slashed{\epsilon}_{\lambda}\!(\underline{q})\;  v(1)
+ \frac{2k^+_0}{q^+}\; \P\!\cdot\! \varepsilon_{\lambda}\;
\overline{u}(0)\, \gamma^+\,  v(1)
\, .
  \label{quark_current_plus j_epsilon}
\end{eqnarray}



\section{Calculation of the numerators\label{sec:num_calc}}

The details of the calculation of the numerators of the various diagrams are presented in this section. More precisely, the aim is to extract and isolate the dependence on the transverse momentum running in the loop for each graph.

In order to simplify the calculations and present the results in a compact way, it is convenient to rely on Galilean invariance.
Indeed, light-front quantization does break partially the $D$-dimensional Lorentz invariance leaving intact a $D\!-\!1$-dimensional Galilean subgroup. The $D\!-\!2$-dimensional transverse space and the $x^+$ direction play respectively the role of space and time for this Galilean group, whereas $k^+$ plays the role of a Galilean mass. Hence, for each massless parton, $\k/k^+$ plays the role of a Galilean velocity.
The Galilean invariance implies that the dependence on transverse momenta of the numerator of each diagram happens only through relative Galilean velocities $\k_{i}/k_{i}^{+}-\k_{j}/k_{j}^+$. Hence, the numerators depend on one less independent transverse vector than predicted from momentum conservation only. Then, an efficient way of calculating the numerators in $D$ dimensions is to try to make such relative Galilean velocities explicit, or equivalently relative momenta $\k_{i}-(k_{i}^{+}/k_{j}^+)\k_{j}$.

\subsection{Numerator for the quark self-energy in an off-shell Fock state\label{sec:num_A}}

This section is devoted to the calculation of the numerator $\textrm{Num}_{A}$, defined in eq. \eqref{num_A_def}, for the quark self-energy loop inside an off-shell Fock state.

Using the completeness relations \eqref{completeness_spinors_full} and \eqref{completeness_gluon_pol_vect}, $\textrm{Num}_{A}$ is written
\begin{eqnarray}
\textrm{Num}_{A}&=& \left[-g_{\mu \nu} + \frac{k_{2\, \mu} n_{\nu}+n_{\mu}k_{2\, \nu} }{k_2^+}\right]
\overline{u}(0)\; \gamma^{\mu}\; \slashed{k}_{0'}\; \gamma^{\nu}\; u(0'')
\, .\label{num_A_1}
\end{eqnarray}

As obvious from eq.\eqref{WF_A_2}, momentum conservation imposes that $\underline{k_2}+ \underline{k_{0'}}=\underline{k_{0}}$. However, there is no conservation of $k^-$ at each vertex in light-front perturbation theory. This can be taken into account when eliminating $\slashed{k}_{0'}$ in favor of $\slashed{k}_{2}$ and $\slashed{k}_{0}$ by using the relation
\begin{eqnarray}
\slashed{k}_{0'} =  \slashed{k}_{0}- \slashed{k}_{2} + \left[k_2^- +k_{0'}^--k_0^-\right] \gamma^+\, .
\end{eqnarray}

On the other hand, we also have $\underline{k_{0''}}=\underline{k_{0}}$ from momentum conservation, but these two momenta are on-shell and massless, so that $k_{0''}^-=k_{0}^-$ as well. Hence, one has the identity
\begin{eqnarray}
\slashed{k}_{0}\; u(0'') =0\, .
\end{eqnarray}

Thanks to these two remarks, it is straightforward to simplify \eqref{num_A_1} into
\begin{eqnarray}
\textrm{Num}_{A}&=&
-(D\!-\!2)\; \overline{u}(0)\, \slashed{k}_{2}\, u(0'')
+(D\!-\!2)\left[k_2^+ +k_{0'}^--k_0^-\right]\; \overline{u}(0)\, \gamma^{+}\, u(0'')
\nonumber\\
&& +\frac{1}{k_2^+}\; \overline{u}(0)\, \slashed{k}_{2}\, \slashed{k}_{0}\, \gamma^{+}\, u(0'')
+\frac{1}{k_2^+}\; \overline{u}(0)\, \gamma^{+}\, \slashed{k}_{0}\,  \slashed{k}_{2}\, u(0'')
\label{num_A_2}
\end{eqnarray}

Then, anticommuting the $\slashed{k}_{0}$ to the left or to the right, one obtains
\begin{eqnarray}
\overline{u}(0)\, \gamma^{+}\, \slashed{k}_{0}\,  \slashed{k}_{2}\, u(0'') &=& 2(k_2\cdot k_0)\; \overline{u}(0)\, \gamma^{+}\, u(0'')
\end{eqnarray}
and
\begin{eqnarray}
\overline{u}(0)\, \slashed{k}_{2}\, \slashed{k}_{0}\, \gamma^{+}\, u(0'') &=& 2(k_2\cdot k_0)\; \overline{u}(0)\, \gamma^{+}\, u(0'')\, ,
\end{eqnarray}
but also
\begin{eqnarray}
\overline{u}(0)\, \slashed{k}_{2}\, \slashed{k}_{0}\, \gamma^{+}\, u(0'') &=& 2k_0^+\; \overline{u}(0)\, \slashed{k}_{2}\, u(0'')
\, ,
\end{eqnarray}
so that
\begin{eqnarray}
\overline{u}(0)\, \slashed{k}_{2}\, u(0'')
&=& \frac{(k_2\cdot k_0)}{k_0^+}\; \overline{u}(0)\, \gamma^{+}\, u(0'')
\, .
\end{eqnarray}
Hence, one gets
\begin{eqnarray}
\textrm{Num}_{A}
&=& \left\{ -(D\!-\!2)\; \frac{(k_2\cdot k_0)}{k_0^+}
+(D\!-\!2)\left[k_2^+ +k_{0'}^--k_0^-\right]
+4 \frac{(k_2\cdot k_0)}{k_2^+}
\right\}\; \overline{u}(0)\, \gamma^{+}\, u(0'')
\, .
\label{num_A_3}
\end{eqnarray}

Finally, remembering that
\begin{eqnarray}
\overline{u}(0)\, \gamma^{+}\, u(0'')
= \sqrt{2k_0^+}\; \sqrt{2k_{0''}^+}\;\; \delta_{h_{0''},\, h_0}
= 2k_0^+\; \delta_{h_{0''},\, h_0}
\, ,
\end{eqnarray}
and making the transverse momentum dependence more explicit thanks to the relations
\begin{eqnarray}
(k_2\cdot k_0)
=\frac{k_0^+}{2\, k_2^+}
\left[\k_2\!-\! \frac{k_2^+}{k_0^+}\, \k_0  \right]^2
\end{eqnarray}
and
\begin{eqnarray}
\left[k_2^+ +k_{0'}^--k_0^-\right]
= \frac{k_0^+}{2\, k_2^+(k_0^+\!-\!k_2^+)}
\left[\k_2\!-\! \frac{k_2^+}{k_0^+}\, \k_0  \right]^2
\, ,
\end{eqnarray}
one finds for the numerator $\textrm{Num}_{A}$ the expression given in eq.\eqref{num_A_result}.


\subsection{Numerator for the diagram A' for transverse photon\label{sec:num_Ap_T}}

Let us calculate the numerator $\textrm{Num}_{A'}^T$ appearing in eq.\eqref{WF_Aprime_1prime_2}, defined in eq. \eqref{num_Aprime_def}. Using the completeness relations \eqref{completeness_spinors_full} and \eqref{completeness_gluon_pol_vect}, $\textrm{Num}_{A'}^T$ is written
\begin{eqnarray}
\textrm{Num}_{A'}^T&=& \left[-g_{\mu \nu} + \frac{k_{2\, \mu} n_{\nu}+n_{\mu}k_{2\, \nu} }{k_2^+}\right]
\overline{u}(0)\; \gamma^{\mu}\, \slashed{k}_{0'}\, \gamma^+ \gamma^{\nu}
\slashed{\epsilon}_{\lambda}\!(\underline{q})\; v(1)\nonumber\\
&=&-4 (k_0^+\!-\!k_2^+)\, \overline{u}(0)\,\slashed{\epsilon}_{\lambda}\!(\underline{q})\; v(1)
-(D\!-\!4)\, \overline{u}(0)\, (\slashed{k}_{0}\!-\!\slashed{k}_{2})\, \gamma^+
\slashed{\epsilon}_{\lambda}\!(\underline{q})\; v(1)\nonumber\\
&& +\frac{1}{k_2^+}\; \overline{u}(0)\; \gamma^{+}\, (\slashed{k}_{0}\!-\!\slashed{k}_{2})\, \gamma^+ \slashed{k}_{2}
\slashed{\epsilon}_{\lambda}\!(\underline{q})\; v(1)\nonumber\\
&=&-4 (k_0^+\!-\!k_2^+)\, \overline{u}(0)\,\slashed{\epsilon}_{\lambda}\!(\underline{q})\; v(1)
+(D\!-\!4)\, \overline{u}(0)\, \slashed{k}_{2}\, \gamma^+
\slashed{\epsilon}_{\lambda}\!(\underline{q})\; v(1)\nonumber\\
&& +2\, \frac{(k_0^+\!-\!k_2^+)}{k_2^+}\; \overline{u}(0)\; \gamma^{+}\, \slashed{k}_{2} \slashed{\epsilon}_{\lambda}\!(\underline{q})\; v(1)
\, .\label{num_Aprime_1}
\end{eqnarray}
Various relations from the appendix \ref{sec:free_fields_conventions} have been used, as well as the $+$ and transverse momentum conservation relation $\underline{k_{0'}}=\underline{k_{0}}\!-\!\underline{k_{2}}$, which implies
$\slashed{k}_{0'}\, \gamma^+ = (\slashed{k}_{0}\!-\!\slashed{k}_{2})\, \gamma^+$.

Then, remarking that
\begin{eqnarray}
\overline{u}(0)\,\slashed{\epsilon}_{\lambda}\!(\underline{q})\; v(1)
=  \frac{1}{2 k_2^+}\; \overline{u}(0) \left\{\gamma^{+}\, , \slashed{k}_{2} \right\} \slashed{\epsilon}_{\lambda}\!(\underline{q})\; v(1)\, ,
\end{eqnarray}
one finds
\begin{eqnarray}
\textrm{Num}_{A'}^T&=& \left\{-2\, \frac{(k_0^+\!-\!k_2^+)}{k_2^+}+(D\!-\!4)\right\}\,
\overline{u}(0)\, \slashed{k}_{2}\, \gamma^+
\slashed{\epsilon}_{\lambda}\!(\underline{q})\; v(1)\nonumber\\
&=& \left\{-2\, \frac{k_0^+}{k_2^+}+D\!-\!2\right\}\,
\overline{u}(0)\, \left[\slashed{k}_{2}\!-\!\frac{k_2^+}{k_0^+}\, \slashed{k}_{0}\right]\, \gamma^+
\slashed{\epsilon}_{\lambda}\!(\underline{q})\; v(1)
\, .\label{num_Aprime_2}
\end{eqnarray}
Indeed, due to the adjacent $\overline{u}(0)$, $\slashed{k}_{2}$ can always be shifted by a term proportional to $\slashed{k}_{0}$. This shift is chosen in such a way that inside the square bracket, the contribution proportional to $\gamma^-$ vanishes, leaving only the terms involving $\gamma^j$. Then, one can extract dependence on $\k_2^j$, and find the result given in eq. \eqref{num_Aprime_result}.

\subsection{Numerator for the diagram 1' for transverse photon\label{sec:num_1p_T}}

For the numerator $\textrm{Num}_{1'}^T$ appearing in eq.\eqref{WF_Aprime_1prime_2}, and defined in eq. \eqref{num_1prime_def}, the completeness relations \eqref{completeness_spinors_full} and \eqref{completeness_gluon_pol_vect} give
\begin{eqnarray}
\textrm{Num}_{1'}^T&=& \left[-g_{\mu \nu} + \frac{k_{2\, \mu} n_{\nu}+n_{\mu}k_{2\, \nu} }{k_2^+}\right]
\overline{u}(0)\; \gamma^{\mu}\, \slashed{k}_{0'}\, \gamma^+
\slashed{\epsilon}_{\lambda}\!(\underline{q})  \gamma^{\nu}\; v(1)\nonumber\\
&=&2\, \overline{u}(0)\,\slashed{\epsilon}_{\lambda}\!(\underline{q})\;
 \gamma^+ (\slashed{k}_{0}\!-\!\slashed{k}_{2})   v(1)
+(D\!-\!4)\, \overline{u}(0)\, (\slashed{k}_{0}\!-\!\slashed{k}_{2})\, \gamma^+
\slashed{\epsilon}_{\lambda}\!(\underline{q})\; v(1)\nonumber\\
&& +\frac{1}{k_2^+}\; \overline{u}(0)\; \slashed{k}_{2}\, (\slashed{k}_{0}\!-\!\slashed{k}_{2})\, \gamma^+
\slashed{\epsilon}_{\lambda}\!(\underline{q})\; \gamma^+ v(1)
+\frac{1}{k_2^+}\; \overline{u}(0)\; \gamma^{+}\, (\slashed{k}_{0}\!-\!\slashed{k}_{2})\, \gamma^+
\slashed{\epsilon}_{\lambda}\!(\underline{q})\; \slashed{k}_{2}  v(1)\nonumber\\
&=&2\, \overline{u}(0)\,\slashed{\epsilon}_{\lambda}\!(\underline{q})\;
 \gamma^+ (\slashed{k}_{0}\!-\!\slashed{k}_{2})   v(1)
-(D\!-\!4)\, \overline{u}(0)\, \slashed{k}_{2}\, \gamma^+
\slashed{\epsilon}_{\lambda}\!(\underline{q})\; v(1)\nonumber\\
&& -2\, \frac{(k_0^+\!-\!k_2^+)}{k_2^+}\; \overline{u}(0)\;
\slashed{\epsilon}_{\lambda}\!(\underline{q})\;
\gamma^{+}\,  \slashed{k}_{2} v(1)\nonumber\\
&=&2\, \overline{u}(0)\,\slashed{\epsilon}_{\lambda}\!(\underline{q})\;
 \gamma^+ \left[\slashed{k}_{0}\!-\!\frac{k_0^+}{k_2^+}\, \slashed{k}_{2}\right]   v(1)
-(D\!-\!4)\, \overline{u}(0)\, \slashed{k}_{2}\, \gamma^+
\slashed{\epsilon}_{\lambda}\!(\underline{q})\; v(1)
\, .\label{num_1prime_1}
\end{eqnarray}
It is again convenient to shift $\slashed{k}_{2}$ in the second term by an amount proportional to $\slashed{k}_{0}$ in the second term, and write
\begin{eqnarray}
\textrm{Num}_{1'}^T&=&-2\, \frac{k_0^+}{k_2^+}\, \overline{u}(0)\,\slashed{\epsilon}_{\lambda}\!(\underline{q})\;
 \gamma^+ \left[\slashed{k}_{2}\!-\!\frac{k_2^+}{k_0^+}\, \slashed{k}_{0}\right]   v(1)
-(D\!-\!4)\, \overline{u}(0)\, \left[\slashed{k}_{2}\!-\!\frac{k_2^+}{k_0^+}\, \slashed{k}_{0}\right]\, \gamma^+
\slashed{\epsilon}_{\lambda}\!(\underline{q})\; v(1)
\, ,\label{num_1prime_2}
\end{eqnarray}
from which one gets the expression \eqref{num_1prime_result}.

\subsection{Numerator for the diagram 3 for transverse photon\label{sec:num_3_T}}

In the case of the numerator $\textrm{Num}_{3}^T$ appearing in eq.\eqref{WF_3_T_2}, and defined in eq. \eqref{num_3_T_def}, the completeness relations \eqref{completeness_spinors_full} give
\begin{eqnarray}
\textrm{Num}_{3}^T&=&
\overline{u}(0)\, \gamma^+ \slashed{k}_{0'} \; \slashed{\epsilon}_{\lambda}\!(\underline{q})\; \slashed{k}_{1'}\, \gamma^+ v(1)
\, .
\label{num_3_T_1}
\end{eqnarray}
Then, it is convenient to evaluate the following anticommutator:
\begin{eqnarray}
\{\slashed{k}_{0'},\slashed{\epsilon}_{\lambda}\!(\underline{q})\}
= 2\,  {k}_{0'\, \mu}\; {\epsilon}^{\mu}_{\lambda}\!(\underline{q})
=-2\, {\varepsilon}_{\lambda} \!\cdot\! \left[\k_{0'} \!-\! \frac{k_{0'}^+}{q^+}\, \q\right]
\, ,
\end{eqnarray}
so that
\begin{eqnarray}
\textrm{Num}_{3}^T&=&
-2\, {\varepsilon}_{\lambda} \!\cdot\! \left[\k_{0'} \!-\! \frac{k_{0'}^+}{q^+}\, \q\right]
\overline{u}(0)\, \gamma^+  \slashed{k}_{1'}\, \gamma^+ v(1)
-
\overline{u}(0)\, \gamma^+  \slashed{\epsilon}_{\lambda}\!(\underline{q})\; \slashed{k}_{0'} \; \slashed{k}_{1'}\, \gamma^+ v(1)
\nonumber\\
&=&
-4 k_{1'}^+\; {\varepsilon}_{\lambda} \!\cdot\! \left[\k_{0'} \!-\! \frac{k_{0'}^+}{q^+}\, \q\right]  \overline{u}(0)\, \gamma^+  v(1)
-\overline{u}(0)\, \gamma^+  \slashed{\epsilon}_{\lambda}\!(\underline{q})\; \slashed{k}_{0'}  \left(\slashed{q}\!-\!\slashed{k}_{0'}\right) \gamma^+ v(1)
\, .
\label{num_3_T_2}
\end{eqnarray}
Indeed, thanks to the $+$ and transverse momentum conservation relation $\underline{k_{1'}}=\underline{q}\!-\!\underline{k_{0'}}$, one has
$\slashed{k}_{1'}\, \gamma^+ = (\slashed{q}\!-\!\slashed{k}_{0'})\, \gamma^+$.
Since $k_{0'}$ is an on-shell momentum, the $-\slashed{k}_{0'}$ present in the parentheses in the expression \eqref{num_3_T_2} gives no contribution. Instead of dropping that $-\slashed{k}_{0'}$ term, it is convenient to change its coefficient in such a way that no term proportional to $\gamma^-$ survives inside that parenthesis.
Hence, one gets
\begin{eqnarray}
\textrm{Num}_{3}^T&=&
-4 k_{1'}^+\; {\varepsilon}_{\lambda} \!\cdot\! \left[\k_{0'} \!-\! \frac{k_{0'}^+}{q^+}\, \q\right]  \overline{u}(0)\, \gamma^+  v(1)
+\overline{u}(0)\, \slashed{\epsilon}_{\lambda}\!(\underline{q})\; \gamma^+  \slashed{k}_{0'}  \left(\slashed{q}\!-\!\frac{q^+}{{k}_{0'}^+}\, \slashed{k}_{0'}\right) \gamma^+ v(1)
\nonumber\\
&=&
-4 k_{1'}^+\; {\varepsilon}_{\lambda} \!\cdot\! \left[\k_{0'} \!-\! \frac{k_{0'}^+}{q^+}\, \q\right]  \overline{u}(0)\, \gamma^+  v(1)
+
\frac{q^+}{k_{0'}^+}
\left[\k_{0'}^j \!-\! \frac{k_{0'}^+}{q^+}\, \q^j\right]
\overline{u}(0)\, \slashed{\epsilon}_{\lambda}\!(\underline{q})\; \gamma^+  \slashed{k}_{0'}  \gamma^j \gamma^+ v(1)
\nonumber\\
&=&
-4 k_{1'}^+\; {\varepsilon}_{\lambda} \!\cdot\! \left[\k_{0'} \!-\! \frac{k_{0'}^+}{q^+}\, \q\right]  \overline{u}(0)\, \gamma^+  v(1)
- 2{q^+}
\left[\k_{0'}^j \!-\! \frac{k_{0'}^+}{q^+}\, \q^j\right]
\overline{u}(0)\, \slashed{\epsilon}_{\lambda}\!(\underline{q})\; \gamma^+  \gamma^j v(1)
\, .
\label{num_3_T_3}
\end{eqnarray}
Using the identity
\begin{eqnarray}
\overline{u}(0)\, \gamma^+  \left\{\slashed{\epsilon}_{\lambda}\!(\underline{q})\, , \gamma^j\right\} v(1)
= 2\, {\varepsilon}_{\lambda}^j\; \overline{u}(0)\, \gamma^+  v(1)
\, ,
\end{eqnarray}
it is then straightforward to obtain the expression \eqref{num_3_T_result}.


\subsection{Numerator for the vertex correction diagram 1 for transverse photon\label{sec:num_1_T}}

Let us consider now the most complicated numerator $\textrm{Num}_{1}^T$, for the diagram 1, defined by the equation \eqref{num_1_T_def}. Using the completeness relations \eqref{completeness_spinors_full} and \eqref{completeness_gluon_pol_vect}, one gets
\begin{eqnarray}
\textrm{Num}_{1}^T &=&
\bigg[-g_{\mu \nu} + \frac{{k_2}_{\mu} n_{\nu} \!+\!n_{\mu}{k_2}_{\nu} }{k_2^+}\bigg] \;\;\;
\overline{u}(0)\; \gamma^{\mu}\; \slashed{k}_{0'}\,  \slashed{\epsilon}_{\lambda}\!(\underline{q})\; \slashed{k}_{1'}\,  \gamma^{\nu}\;  v(1)
\, .
\label{num_1_T_1}
\end{eqnarray}
Then, using the fact that the $+$ and transverse components of the momentum is conserved at each vertex, one has
\begin{eqnarray}
\slashed{k}_{0'} &=&  \slashed{k}_{0}- \slashed{k}_{2} + \Big(k_{0'}^- \!+\!k_2^- \!-\!k_0^-\Big) \gamma^+
\label{mom_cons_1T_020prime}
\\
\slashed{k}_{1'} &=&  \slashed{k}_{1}+ \slashed{k}_{2} + \Big(k_{1'}^- \!-\!k_2^- \!-\!k_1^-\Big) \gamma^+
\label{mom_cons_1T_121prime}
\, ,
\end{eqnarray}
where the extra term in $\gamma^+$ accounts for the nonconservation of the $-$ component of the momentum in this formalism. Inserting these identities into the equation \eqref{num_1_T_1}, it is convenient to split the numerator $\textrm{Num}_{1}^T$ into five contributions as
\begin{eqnarray}
\textrm{Num}_{1}^T &=&
\Big(k_{0'}^- \!+\!k_2^- \!-\!k_0^-\Big)\;
\bigg[-g_{\mu \nu} + \frac{{k_2}_{\mu} n_{\nu} \!+\!n_{\mu}{k_2}_{\nu} }{k_2^+}\bigg] \;\;\;
\overline{u}(0)\; \gamma^{\mu}\; \gamma^{+}\;  \slashed{\epsilon}_{\lambda}\!(\underline{q})\; \Big(\slashed{k}_{1}\!+\! \slashed{k}_{2}\Big)\,  \gamma^{\nu}\;  v(1)
\nonumber\\
&&
+\Big(k_{1'}^- \!-\!k_2^- \!-\!k_1^-\Big)\;
\bigg[-g_{\mu \nu} + \frac{{k_2}_{\mu} n_{\nu} \!+\!n_{\mu}{k_2}_{\nu} }{k_2^+}\bigg] \;\;\;
\overline{u}(0)\; \gamma^{\mu}\;
\Big(\slashed{k}_{0}\!-\! \slashed{k}_{2}\Big)\, \slashed{\epsilon}_{\lambda}\!(\underline{q})\; \gamma^{+}\; \gamma^{\nu}\;  v(1)
\nonumber\\
&&
+\frac{1}{k_2^+}\; \overline{u}(0)\; \slashed{k}_{2}\;
\Big(\slashed{k}_{0}\!-\! \slashed{k}_{2}\Big)\, \slashed{\epsilon}_{\lambda}\!(\underline{q})\;
\Big(\slashed{k}_{1}\!+\! \slashed{k}_{2}\Big)\,
\gamma^{+}\;  v(1)
\nonumber\\
&&
+\frac{1}{k_2^+}\; \overline{u}(0)\; \gamma^{+}\;
\Big(\slashed{k}_{0}\!-\! \slashed{k}_{2}\Big)\, \slashed{\epsilon}_{\lambda}\!(\underline{q})\;
\Big(\slashed{k}_{1}\!+\! \slashed{k}_{2}\Big)\,
\slashed{k}_{2}\;  v(1)
\nonumber\\
&&
-g_{\mu \nu} \;\;
\overline{u}(0)\; \gamma^{\mu}\;
\Big(\slashed{k}_{0}\!-\! \slashed{k}_{2}\Big)\, \slashed{\epsilon}_{\lambda}\!(\underline{q})\;
\Big(\slashed{k}_{1}\!+\! \slashed{k}_{2}\Big)\,
\gamma^{\nu}\;  v(1)
\, .
\label{num_1_T_2}
\end{eqnarray}
Note that, due to the anticommutation relation \eqref{anticom_gammaplus_epsilonslash}, one gets zero if only the extra term proportional to $\gamma^+$ is picked both in \eqref{mom_cons_1T_020prime} and in \eqref{mom_cons_1T_121prime}.

In the case of the first term in \eqref{num_1_T_2}, one extracts the dependence on loop momentum out of the spinor structure following the same techniques as for the numerators calculated so far, as
\begin{eqnarray}
&&
\Big(k_{0'}^- \!+\!k_2^- \!-\!k_0^-\Big)\;
\bigg[-g_{\mu \nu} + \frac{{k_2}_{\mu} n_{\nu} \!+\!n_{\mu}{k_2}_{\nu} }{k_2^+}\bigg] \;\;\;
\overline{u}(0)\; \gamma^{\mu}\; \gamma^{+}\;  \slashed{\epsilon}_{\lambda}\!(\underline{q})\; \Big(\slashed{k}_{1}\!+\! \slashed{k}_{2}\Big)\,  \gamma^{\nu}\;  v(1)
\nonumber\\
&=&
\Big(k_{0'}^- \!+\!k_2^- \!-\!k_0^-\Big)\;
\bigg\{
2\;  \overline{u}(0)\;
\Big(\slashed{k}_{1}\!+\! \slashed{k}_{2}\Big)\,
\slashed{\epsilon}_{\lambda}\!(\underline{q})\; \gamma^{+}\;  v(1)
+(D\!-\!4)\;  \overline{u}(0)\; \gamma^{+}\;
\slashed{\epsilon}_{\lambda}\!(\underline{q})\;
\Big(\slashed{k}_{1}\!+\! \slashed{k}_{2}\Big)\,  v(1)\nonumber\\
&& \hspace{4cm}
+\frac{1}{k_2^+}\;
\overline{u}(0)\; \slashed{k}_{2}\; \gamma^{+}\;  \slashed{\epsilon}_{\lambda}\!(\underline{q})\; \Big(\slashed{k}_{1}\!+\! \slashed{k}_{2}\Big)\,
\gamma^{+}\;  v(1)
\bigg\}
\nonumber\\
&=&
\Big(k_{0'}^- \!+\!k_2^- \!-\!k_0^-\Big)\;
\bigg\{
-2\;  \overline{u}(0)
\left(\slashed{k}_{1}\!+\! \slashed{k}_{2} \!-\!\frac{(k_{1}^+ \!+\!k_{2}^+)}{k_{0}^+}\,\slashed{k}_{0} \right)\,
\gamma^{+}\; \slashed{\epsilon}_{\lambda}\!(\underline{q})\;  v(1)
+(D\!-\!4)\;  \overline{u}(0)\; \gamma^{+}\;
\slashed{\epsilon}_{\lambda}\!(\underline{q})\;
\left(\slashed{k}_{2} \!-\!\frac{k_{2}^+}{k_{1}^+}\,\slashed{k}_{1}\right)\,  v(1)
\nonumber\\
&& \hspace{4cm}
+2\, \frac{(k_{1}^+ \!+\!k_{2}^+)}{k_2^+}\;
\overline{u}(0) \left(\slashed{k}_{2} \!-\!\frac{k_{2}^+}{k_{0}^+}\,\slashed{k}_{0}\right)\,
\gamma^{+}\;  \slashed{\epsilon}_{\lambda}\!(\underline{q})\; v(1)
\bigg\}
\nonumber\\
&=&
\Big(k_{0'}^- \!+\!k_2^- \!-\!k_0^-\Big)\;
\bigg\{
-2\;
\left(\k_{1}^j\!+\! \k_{2}^j \!-\!\frac{(k_{1}^+ \!+\!k_{2}^+)}{k_{0}^+}\,\k_{0}^j \right)
\overline{u}(0) \gamma^{+} \gamma^{j}\; \slashed{\epsilon}_{\lambda}\!(\underline{q})\;  v(1)
-(D\!-\!4)\;
\left(\k_{2}^j \!-\!\frac{k_{2}^+}{k_{1}^+}\,\k_{1}^j\right)\,
\overline{u}(0)\; \gamma^{+}\;
\slashed{\epsilon}_{\lambda}\!(\underline{q})\;
\gamma^{j}\;  v(1)
\nonumber\\
&& \hspace{4cm}
+2\, \frac{(k_{1}^+ \!+\!k_{2}^+)}{k_2^+}\;
\left(\k_{2}^j \!-\!\frac{k_{2}^+}{k_{0}^+}\,\k_{0}^j\right)\,
\overline{u}(0) \gamma^{+} \gamma^{j}\;  \slashed{\epsilon}_{\lambda}\!(\underline{q})\; v(1)
\bigg\}
\nonumber\\
&=&
\frac{k_0^+}{k_2^+(k_0^+\!-\! k_2^+)}\,
\left(\k_{2} \!-\!\frac{k_{2}^+}{k_{0}^+}\,\k_{0}\right)^2
\left(\k_{2}^j \!-\!\frac{k_{2}^+}{k_{1}^+}\,\k_{1}^j\right)
\bigg\{
\frac{k_1^+}{k_2^+}\;
\overline{u}(0) \gamma^{+} \gamma^{j}\; \slashed{\epsilon}_{\lambda}\!(\underline{q})\;  v(1)
-\frac{(D\!-\!4)}{2}\;
\overline{u}(0)\; \gamma^{+}\;
\slashed{\epsilon}_{\lambda}\!(\underline{q})\;
\gamma^{j}\;  v(1)
\bigg\}
\, .
\label{num_1_T_term1}
\end{eqnarray}
The calculation of the second term in the equation \eqref{num_1_T_2} is analog, and gives
\begin{eqnarray}
&&
\Big(k_{1'}^- \!-\!k_2^- \!-\!k_1^-\Big)\;
\bigg[-g_{\mu \nu} + \frac{{k_2}_{\mu} n_{\nu} \!+\!n_{\mu}{k_2}_{\nu} }{k_2^+}\bigg] \;\;\;
\overline{u}(0)\; \gamma^{\mu}\;
\Big(\slashed{k}_{0}\!-\! \slashed{k}_{2}\Big)\, \slashed{\epsilon}_{\lambda}\!(\underline{q})\; \gamma^{+}\; \gamma^{\nu}\;  v(1)
\nonumber\\
&=&
-\frac{k_1^+}{k_2^+(k_1^+\!+\! k_2^+)}\,
\left(\k_{2} \!-\!\frac{k_{2}^+}{k_{1}^+}\,\k_{1}\right)^2
\left(\k_{2}^j \!-\!\frac{k_{2}^+}{k_{0}^+}\,\k_{0}^j\right)
\bigg\{
\frac{k_0^+}{k_2^+}\;
\overline{u}(0) \gamma^{+}\; \slashed{\epsilon}_{\lambda}\!(\underline{q})\;  \gamma^{j} v(1)
+\frac{(D\!-\!4)}{2}\;
\overline{u}(0)\; \gamma^{+}\gamma^{j}\;
\slashed{\epsilon}_{\lambda}\!(\underline{q})\; v(1)
\bigg\}
\, .
\label{num_1_T_term2}
\end{eqnarray}

The third term in the equation \eqref{num_1_T_2} can be calculated as
\begin{eqnarray}
&&
\frac{1}{k_2^+}\; \overline{u}(0)\; \slashed{k}_{2}\;
\Big(\slashed{k}_{0}\!-\! \slashed{k}_{2}\Big)\, \slashed{\epsilon}_{\lambda}\!(\underline{q})\;
\Big(\slashed{k}_{1}\!+\! \slashed{k}_{2}\Big)\,
\gamma^{+}\;  v(1)
\nonumber\\
&=&
\frac{1}{k_2^+}\; \overline{u}(0)\; \big\{\slashed{k}_{2},\slashed{k}_{0}\big\}\, \slashed{\epsilon}_{\lambda}\!(\underline{q})\;
\bigg(\big\{\slashed{k}_{1}\!+\! \slashed{k}_{2},\gamma^{+} \big\}
-\gamma^{+} \Big(\slashed{k}_{1}\!+\! \slashed{k}_{2}\Big)
\bigg)
v(1)
\nonumber\\
&=&
2\, \frac{(k_0 \!\cdot\! k_2)}{k_2^+}\; \bigg\{
2 (k_1^+\!+\! k_2^+)\; \overline{u}(0)\, \slashed{\epsilon}_{\lambda}\!(\underline{q})\, v(1)
- \overline{u}(0)\, \slashed{\epsilon}_{\lambda}\!(\underline{q})\,
 \gamma^{+}  \left(\slashed{k}_{2} \!-\!\frac{k_{2}^+}{k_{1}^+}\,\slashed{k}_{1}\right)  v(1)
\bigg\}
\nonumber\\
&=&
\frac{k_0^+}{(k_2^+)^2}\,
\left(\k_{2} \!-\!\frac{k_{2}^+}{k_{0}^+}\,\k_{0}\right)^2
\bigg\{
2 (k_1^+\!+\! k_2^+)\; \overline{u}(0)\, \slashed{\epsilon}_{\lambda}\!(\underline{q})\, v(1)
-\left(\k_{2}^j \!-\!\frac{k_{2}^+}{k_{1}^+}\,\k_{1}^j\right)
\overline{u}(0)\, \gamma^{+}\,
\slashed{\epsilon}_{\lambda}\!(\underline{q})\,
\gamma^{j}\,  v(1)
\bigg\}
\, .
\label{num_1_T_term3}
\end{eqnarray}
Similarly, the fourth term in equation \eqref{num_1_T_2} gives
\begin{eqnarray}
&&
\frac{1}{k_2^+}\; \overline{u}(0)\; \gamma^{+}\;
\Big(\slashed{k}_{0}\!-\! \slashed{k}_{2}\Big)\, \slashed{\epsilon}_{\lambda}\!(\underline{q})\;
\Big(\slashed{k}_{1}\!+\! \slashed{k}_{2}\Big)\,
\slashed{k}_{2}\;  v(1)
\nonumber\\
&=&
\frac{k_1^+}{(k_2^+)^2}\,
\left(\k_{2} \!-\!\frac{k_{2}^+}{k_{1}^+}\,\k_{1}\right)^2
\bigg\{
2 (k_0^+\!-\! k_2^+)\; \overline{u}(0)\, \slashed{\epsilon}_{\lambda}\!(\underline{q})\, v(1)
+\left(\k_{2}^j \!-\!\frac{k_{2}^+}{k_{0}^+}\,\k_{0}^j\right)
\overline{u}(0)\, \gamma^{+}\gamma^{j}\,
\slashed{\epsilon}_{\lambda}\!(\underline{q})\, v(1)
\bigg\}
\, .
\label{num_1_T_term4}
\end{eqnarray}

Finally, the fifth term in equation \eqref{num_1_T_2} requires one to proceed in a slightly different way, due to the absence of an explicit $\gamma^+$ factor in the spinor structure. Nevertheless, its calculation can be performed as
\begin{eqnarray}
&&
-g_{\mu \nu} \;\;
\overline{u}(0)\; \gamma^{\mu}\;
\Big(\slashed{k}_{0}\!-\! \slashed{k}_{2}\Big)\, \slashed{\epsilon}_{\lambda}\!(\underline{q})\;
\Big(\slashed{k}_{1}\!+\! \slashed{k}_{2}\Big)\,
\gamma^{\nu}\;  v(1)
\nonumber\\
&=&
2\; \overline{u}(0)\,
\Big(\slashed{k}_{1}\!+\! \slashed{k}_{2}\Big)\, \slashed{\epsilon}_{\lambda}\!(\underline{q})\,
\Big(\slashed{k}_{0}\!-\! \slashed{k}_{2}\Big)\,
 v(1)
+ (D\!-\!4)\,
\overline{u}(0)\,
\Big(\slashed{k}_{0}\!-\! \slashed{k}_{2}\Big)\, \slashed{\epsilon}_{\lambda}\!(\underline{q})\,
\Big(\slashed{k}_{1}\!+\! \slashed{k}_{2}\Big)\,
 v(1)
\nonumber\\
&=&
2\; \overline{u}(0)\; \slashed{k}_{1}\, \slashed{\epsilon}_{\lambda}\!(\underline{q})\;
\slashed{k}_{0}\, v(1)
- 2\; \overline{u}(0)\; \slashed{k}_{1}\, \slashed{\epsilon}_{\lambda}\!(\underline{q})\;
\slashed{k}_{2}\, v(1)
+ 2\; \overline{u}(0)\; \slashed{k}_{2}\, \slashed{\epsilon}_{\lambda}\!(\underline{q})\;
\slashed{k}_{0}\, v(1)
- (D\!-\!2)\,  \overline{u}(0)\; \slashed{k}_{2}\, \slashed{\epsilon}_{\lambda}\!(\underline{q})\;
\slashed{k}_{2}\, v(1)
\nonumber\\
&=&
2\; \overline{u}(0)\, \bigg(
\left\{\slashed{k}_{1}, \slashed{\epsilon}_{\lambda}\!(\underline{q})\right\}\, \slashed{k}_{0}
- \slashed{\epsilon}_{\lambda}\!(\underline{q})\,
  \left\{\slashed{k}_{1},\slashed{k}_{0}\right\}
\bigg)\, v(1)
- 2\; \overline{u}(0)\, \bigg(
\left\{\slashed{k}_{1}, \slashed{\epsilon}_{\lambda}\!(\underline{q})\right\}\, \slashed{k}_{2}
- \slashed{\epsilon}_{\lambda}\!(\underline{q})\,
  \left\{\slashed{k}_{1},\slashed{k}_{2}\right\}
\bigg)\, v(1)
\nonumber\\
&& \hspace{1cm}
+ 2\; \overline{u}(0)\, \bigg(
\slashed{k}_{2}\, \left\{\slashed{\epsilon}_{\lambda}\!(\underline{q}), \slashed{k}_{0}\right\}\,
-  \left\{\slashed{k}_{2},\slashed{k}_{0}\right\}\,
   \slashed{\epsilon}_{\lambda}\!(\underline{q})
\bigg)\, v(1)
- (D\!-\!2)\,  \overline{u}(0)\, \left\{\slashed{k}_{2}\, \slashed{\epsilon}_{\lambda}\!(\underline{q})\right\}\,
\slashed{k}_{2}\, v(1)
\nonumber\\
&=&
4 \bigg[-(k_0 \!\cdot\! k_1) + (k_1 \!\cdot\! k_2)
      -(k_0 \!\cdot\! k_2)
\bigg]\; \overline{u}(0)\; \slashed{\epsilon}_{\lambda}\!(\underline{q})\;  v(1)
+ 4 \bigg[-{k_1}_{\mu}+{k_0}_{\mu} -\frac{(D\!-\!2)}{2}\, {k_2}_{\mu}
\bigg]\; \epsilon_{\lambda}^{\mu}\!(\underline{q})\;
\overline{u}(0)\; \slashed{k}_{2}\;  v(1)
\nonumber\\
&=&
2 \left[-\frac{k_0^+}{k_2^+}\,
\left(\k_{2} \!-\!\frac{k_{2}^+}{k_{0}^+}\,\k_{0}\right)^2
 + \frac{k_1^+}{k_2^+}\,
\left(\k_{2} \!-\!\frac{k_{2}^+}{k_{1}^+}\,\k_{1}\right)^2
      -\frac{(q^+)^2}{k_0^+ k_1^+}\,
\left(\k_{0} \!-\!\frac{k_{0}^+}{q^+}\,\q\right)^2
\right]\; \overline{u}(0)\; \slashed{\epsilon}_{\lambda}\!(\underline{q})\;  v(1)
\nonumber\\
&& \hspace{1cm}
+ 4 \left[ \left(\k_1^i\!-\!\frac{k_1^+}{q^+}\, \q^i\right)
   -\left(\k_0^i\!-\!\frac{k_0^+}{q^+}\, \q^i\right)
   +\frac{(D\!-\!2)}{2}\, \left(\k_2^i -\frac{k_2^+}{q^+}\, \q^i\right)
\right]\;
\varepsilon_{\lambda}^{i}\;
\overline{u}(0)\; \slashed{k}_{2}\;  v(1)
\nonumber\\
&=&
2 \left[-\frac{k_0^+}{k_2^+}\,
\left(\k_{2} \!-\!\frac{k_{2}^+}{k_{0}^+}\,\k_{0}\right)^2
 + \frac{k_1^+}{k_2^+}\,
\left(\k_{2} \!-\!\frac{k_{2}^+}{k_{1}^+}\,\k_{1}\right)^2
      -\frac{(q^+)^2}{k_0^+ k_1^+}\,
\left(\k_{0} \!-\!\frac{k_{0}^+}{q^+}\,\q\right)^2
\right]\; \overline{u}(0)\; \slashed{\epsilon}_{\lambda}\!(\underline{q})\;  v(1)
\nonumber\\
&& \hspace{0.5cm}
+\, \frac{\varepsilon_{\lambda}^{i}}{k_2^+}\;
 \left[4 \left(\k_0^i \!-\!\frac{k_0^+}{q^+}\, \q^i\right)
   -(D\!-\!2)\, \left(\k_2^i \!-\!\frac{k_2^+}{q^+}\, \q^i\right)
\right]\;
\left(\k_{2}^j \!-\!\frac{k_{2}^+}{k_{0}^+}\,\k_{0}^j\right)
\left(\k_{2}^l \!-\!\frac{k_{2}^+}{k_{1}^+}\,\k_{1}^l\right)\,
\overline{u}(0)\, \gamma^+ \gamma^{j} \gamma^{l}\,  v(1)
\, ,
\label{num_1_T_term5}
\end{eqnarray}
where, in the last step, the relation \eqref{quark_current_times_k2} has been used.

All in all, collecting the results \eqref{num_1_T_term1}, \eqref{num_1_T_term2}, \eqref{num_1_T_term3}, \eqref{num_1_T_term4} and \eqref{num_1_T_term5} for the five terms in the expression \eqref{num_1_T_2} of the numerator $\textrm{Num}_{1}^T$, one obtains the expression \eqref{num_1_T_result}.



\subsection{Numerator for the vertex correction diagram 1 for longitudinal photon\label{sec:num_1_L}}

Finally, there remains to calculate the numerator $\textrm{Num}_{1}^L$, for the diagram 1 in the longitudinal photon case, defined by the equation \eqref{num_1_L_def}. Using the completeness relations \eqref{completeness_spinors_full} and \eqref{completeness_gluon_pol_vect}, one gets
\begin{eqnarray}
\textrm{Num}_{1}^L &=&
\bigg[-g_{\mu \nu} + \frac{{k_2}_{\mu} n_{\nu} \!+\!n_{\mu}{k_2}_{\nu} }{k_2^+}\bigg] \;\;\;
\overline{u}(0)\; \gamma^{\mu}\; \slashed{k}_{0'}\,  \gamma^{+}\; \slashed{k}_{1'}\,  \gamma^{\nu}\;  v(1)
\, .
\label{num_1_L_1}
\end{eqnarray}
The momentum flow is obviously the same as in the transverse photon case. Then, using the momentum conservation relations \eqref{mom_cons_1T_020prime} and \eqref{mom_cons_1T_121prime}, one writes
\begin{eqnarray}
\textrm{Num}_{1}^L &=&
\frac{1}{k_2^+}\; \overline{u}(0)\; \slashed{k}_{2}\;
\Big(\slashed{k}_{0}\!-\! \slashed{k}_{2}\Big)\, \gamma^{+}\;
\Big(\slashed{k}_{1}\!+\! \slashed{k}_{2}\Big)\,
\gamma^{+}\;  v(1)
\nonumber\\
&&
+\frac{1}{k_2^+}\; \overline{u}(0)\; \gamma^{+}\;
\Big(\slashed{k}_{0}\!-\! \slashed{k}_{2}\Big)\, \gamma^{+}\;
\Big(\slashed{k}_{1}\!+\! \slashed{k}_{2}\Big)\,
\slashed{k}_{2}\;  v(1)
\nonumber\\
&&
-g_{\mu \nu} \;\;
\overline{u}(0)\; \gamma^{\mu}\;
\Big(\slashed{k}_{0}\!-\! \slashed{k}_{2}\Big)\, \gamma^{+}\;
\Big(\slashed{k}_{1}\!+\! \slashed{k}_{2}\Big)\,
\gamma^{\nu}\;  v(1)
\, .
\label{num_1_L_2}
\end{eqnarray}
The term in the first line of eq. \eqref{num_1_L_2} is easily calculated as
\begin{eqnarray}
\frac{1}{k_2^+}\; \overline{u}(0)\; \slashed{k}_{2}\;
\Big(\slashed{k}_{0}\!-\! \slashed{k}_{2}\Big)\, \gamma^{+}\;
\Big(\slashed{k}_{1}\!+\! \slashed{k}_{2}\Big)\,
\gamma^{+}\;  v(1)
&=&
\frac{1}{k_2^+}\; \overline{u}(0)\; \Big\{\slashed{k}_{2} ,
\slashed{k}_{0}\Big\}\, \gamma^{+}\;
\Big\{\Big(\slashed{k}_{1}\!+\! \slashed{k}_{2}\Big),
\gamma^{+}\Big\}\;  v(1)
\nonumber\\
&=&
4\, \frac{(k_1^+\!+\!k_2^+)}{k_2^+}\; (k_2\cdot k_0)\;
\overline{u}(0)\, \gamma^{+}\, v(1)
\nonumber\\
&=&
2\, \frac{k_0^+(k_1^+\!+\!k_2^+)}{(k_2^+)^2}\;
\left[\k_2\!-\! \frac{k_2^+}{k_0^+}\, \k_0\right]^2\;
\overline{u}(0)\, \gamma^{+}\, v(1)
\label{num_1_L_1st_line}
\end{eqnarray}
In the same way, one gets for the term in the second line of eq. \eqref{num_1_L_2}
\begin{eqnarray}
\frac{1}{k_2^+}\; \overline{u}(0)\; \gamma^{+}\;
\Big(\slashed{k}_{0}\!-\! \slashed{k}_{2}\Big)\, \gamma^{+}\;
\Big(\slashed{k}_{1}\!+\! \slashed{k}_{2}\Big)\,
\slashed{k}_{2}\;  v(1)
&=&
2\, \frac{k_1^+(k_0^+\!-\!k_2^+)}{(k_2^+)^2}\;
\left[\k_2\!-\! \frac{k_2^+}{k_1^+}\, \k_1\right]^2\;
\overline{u}(0)\, \gamma^{+}\, v(1)
\label{num_1_L_2nd_line}
\end{eqnarray}
Finally, for the term in the third line, one uses the identity \eqref{5_gamma_contraction} and then extracts the transverse momentum dependence by appropriate shifts as done with the previously considered numerators, as
\begin{eqnarray}
&&
-g_{\mu \nu} \;\;
\overline{u}(0)\; \gamma^{\mu}\;
\Big(\slashed{k}_{0}\!-\! \slashed{k}_{2}\Big)\, \gamma^{+}\;
\Big(\slashed{k}_{1}\!+\! \slashed{k}_{2}\Big)\,
\gamma^{\nu}\;  v(1)
\nonumber\\
&=& 2\, \overline{u}(0)\; \Big(\slashed{k}_{1}\!+\! \slashed{k}_{2}\Big)\,
    \gamma^{+}\; \Big(\slashed{k}_{0}\!-\! \slashed{k}_{2}\Big)\, v(1)
    +(D\!-\!4)\, \overline{u}(0)\;
\Big(\slashed{k}_{0}\!-\! \slashed{k}_{2}\Big)\, \gamma^{+}\;
\Big(\slashed{k}_{1}\!+\! \slashed{k}_{2}\Big)\, v(1)
\nonumber\\
&=&
2\, \overline{u}(0)\; \left[\slashed{k}_{1}\!+\! \slashed{k}_{2}
    \!-\!\frac{(k_1^+\!+\!k_2^+)}{k_0^+}\, \slashed{k}_{0}\right]\,
    \gamma^{+}\; \left[\slashed{k}_{0}\!-\! \slashed{k}_{2}
    \!-\!\frac{(k_0^+\!-\!k_2^+)}{k_1^+}\, \slashed{k}_{1}\right]\, v(1)
    -(D\!-\!4)\, \overline{u}(0)\;
\left[\slashed{k}_{2}\!-\!\frac{k_2^+}{k_0^+}\, \slashed{k}_{0}\right]\, \gamma^{+}\;
\left[\slashed{k}_{2}\!-\!\frac{k_2^+}{k_1^+}\, \slashed{k}_{1}\right]\, v(1)
\nonumber\\
&=&
\Bigg\{2\, \left[\k_{1}^i\!+\! \k_{2}^i
                  \!-\!\frac{(k_1^+\!+\!k_2^+)}{k_0^+}\, \k_{0}^i\right]\,
    \left[\k_{0}^j\!-\! \k_{2}^j
          \!-\!\frac{(k_0^+\!-\!k_2^+)}{k_1^+}\, \k_{1}^j\right]
\nonumber\\
&& \hspace{3cm}
-(D\!-\!4)\, \left[\k_{2}^i\!-\!\frac{k_2^+}{k_0^+}\, \k_{0}^i\right]\,
   \left[\k_{2}^j\!-\!\frac{k_2^+}{k_1^+}\, \k_{1}^j\right]
\Bigg\}\;
 \overline{u}(0)\, \gamma^{i}\gamma^{+}\gamma^{j}\, v(1)
\label{num_1_L_3rd_line}
\end{eqnarray}
Collecting the three contributions \eqref{num_1_L_1st_line}, \eqref{num_1_L_2nd_line} and \eqref{num_1_L_3rd_line}, one finds for the numerator $\textrm{Num}_{1}^L $ the expression given in eq. \eqref{num_1_L_result}.


\section{Passarino-Veltman tensor reduction on the Light-Front \label{sec:PassVelt}}

Passarino and Veltman showed in Ref. \cite{Passarino:1978jh} that Feynman graphs including loops in gauge theories are easier to calculate by first dealing with the tensor structure of the integrands, in order to rewrite each graph as a linear combination of a small number of scalar integrals.
The method they explicitly constructed to perform the tensor reduction crucially relies on Lorentz invariance. Hence, it is valid only when a Lorentz-invariant UV regulator is used, like in dimensional regularization, but not in other cases, like with a naive UV cutoff.

In this appendix, we recall the Passarino-Veltman method, following closely the original the derivation, but adapting it to the case of interest. Indeed, since Light-Front perturbation theory is used in the present study, one wants to simplify transverse momentum integrals in $D-2$ dimensions using rotational invariance instead of full momentum integrals in $D$ dimensions using Lorentz invariance.

The transverse integrals encountered in the present study are of the form
\begin{eqnarray}
\mathcal{A}_0(\Delta)&\equiv & 4\pi\, (\mu^2)^{2-\frac{D}{2}}
\int \frac{d^{D-2} \K}{(2\pi)^{D-2}}\;
\frac{1}{\left[\K^2 +\Delta\right]} \label{def_A0}
\\
\mathcal{B}_0(\Delta_1,\Delta_2,\L)&\equiv & 4\pi\, (\mu^2)^{2-\frac{D}{2}}
\int \frac{d^{D-2} \K}{(2\pi)^{D-2}}\;
\frac{1}{\left[\K^2 +\Delta_1\right]\left[(\K+\L)^2 +\Delta_2\right]}
\label{def_B0}
\\
\mathcal{B}^i(\Delta_1,\Delta_2,\L)&\equiv & 4\pi\, (\mu^2)^{2-\frac{D}{2}}
\int \frac{d^{D-2} \K}{(2\pi)^{D-2}}\;
\frac{\K^i}{\left[\K^2 +\Delta_1\right]\left[(\K+\L)^2 +\Delta_2\right]}
\label{def_Bi}
\\
\mathcal{B}^{ij}(\Delta_1,\Delta_2,\L)&\equiv & 4\pi\, (\mu^2)^{2-\frac{D}{2}}
\int \frac{d^{D-2} \K}{(2\pi)^{D-2}}\;
\frac{\K^i\, \K^j}{\left[\K^2 +\Delta_1\right]\left[(\K+\L)^2 +\Delta_2\right]}
\label{def_Bij}\, .
\end{eqnarray}
The goal is to express the integrals \eqref{def_Bi} and \eqref{def_Bij} in terms of the scalars ones, \eqref{def_A0} and \eqref{def_B0}. First, due to rotational symmetry in $D-2$ dimensions, it is clear that \eqref{def_Bi} and \eqref{def_Bij} are of the form
\begin{eqnarray}
\mathcal{B}^i(\Delta_1,\Delta_2,\L)&= & \L^i\; \mathcal{B}_1(\Delta_1,\Delta_2,\L^2)
\label{Bi_PV_app}
\\
\mathcal{B}^{ij}(\Delta_1,\Delta_2,\L)&= & \L^i \L^j\; \mathcal{B}_{21}(\Delta_1,\Delta_2,\L^2)
+ \delta^{ij}\; \mathcal{B}_{22}(\Delta_1,\Delta_2,\L^2)
\label{Bij_PV_app}
\, ,
\end{eqnarray}
where $\mathcal{B}_1$, $\mathcal{B}_{21}$ and $\mathcal{B}_{22}$ are scalar integrals which remain to be determined. Contracting the relation \eqref{Bi_PV_app} with $\L^i$, one gets
\begin{eqnarray}
\L^2\; \mathcal{B}_1(\Delta_1,\Delta_2,\L^2)&=&
4\pi\, (\mu^2)^{2-\frac{D}{2}}
\int \frac{d^{D-2} \K}{(2\pi)^{D-2}}\;
\frac{\L\!\cdot\!\K}{\left[\K^2 +\Delta_1\right]\left[(\K+\L)^2 +\Delta_2\right]}
\label{B1 1}\, .
\end{eqnarray}
Then, making the replacement
\begin{eqnarray}
\L\!\cdot\!\K &=& \frac{1}{2}\,
\bigg\{ \Big[(\K+\L)^2+\Delta_2\Big] - \Big[\K^2+\Delta_1\Big]
    +\Big[\Delta_1 \!-\!\Delta_2\!-\!\L^2\Big]
\bigg\}
\label{scalar_prod_identity}
\end{eqnarray}
in the integrand, one easily finds that
\begin{eqnarray}
 \mathcal{B}_1(\Delta_1,\Delta_2,\L^2)&=&
\frac{1}{2\, \L^2}\,
\bigg\{ \mathcal{A}_0(\Delta_1) - \mathcal{A}_0(\Delta_2)
    +\Big[\Delta_1 \!-\!\Delta_2\!-\!\L^2\Big]\, \mathcal{B}_0(\Delta_1,\Delta_2,\L)
\bigg\}
\label{B1 result}\, .
\end{eqnarray}

In the case of the tensor integral $\mathcal{B}^{ij}$, one needs two relations, in order to determine both $\mathcal{B}_{21}(\Delta_1,\Delta_2,\L^2)$ and $\mathcal{B}_{22}(\Delta_1,\Delta_2,\L^2)$.
First, contracting the relation \eqref{Bij_PV_app} with $\L^j$ and using the identity \eqref{scalar_prod_identity}, one finds
\begin{eqnarray}
\Big[ \L^2\, \mathcal{B}_{21}(\Delta_1,\Delta_2,\L^2)
 + \mathcal{B}_{22}(\Delta_1,\Delta_2,\L^2)\Big]\, \L^i
 &=&
 4\pi\, (\mu^2)^{2-\frac{D}{2}}
\int \frac{d^{D-2} \K}{(2\pi)^{D-2}}\;
\frac{1}{2}\,
\bigg\{
\frac{\K^i}{\left[\K^2 +\Delta_1\right]}
-
\frac{(\K^i+\L^i)-\L^i}{\left[(\K+\L)^2 +\Delta_2\right]}\nonumber\\
 & &
+\Big[\Delta_1 \!-\!\Delta_2\!-\!\L^2\Big]\;
\frac{\K^i}{\left[\K^2 +\Delta_1\right]\left[(\K+\L)^2 +\Delta_2\right]}
\bigg\}\nonumber\\
&=& \frac{1}{2}\,
\bigg\{
\mathcal{A}_0(\Delta_2)\, \L^i
+\Big[\Delta_1 \!-\!\Delta_2\!-\!\L^2\Big]\;
\mathcal{B}^i(\Delta_1,\Delta_2,\L)
\bigg\}
\, ,
\end{eqnarray}
dropping the terms which vanish identically due to rotational symmetry. Thus, one has
\begin{eqnarray}
\L^2\, \mathcal{B}_{21}(\Delta_1,\Delta_2,\L^2)
 + \mathcal{B}_{22}(\Delta_1,\Delta_2,\L^2)
&=& \frac{1}{2}\,
\bigg\{
\mathcal{A}_0(\Delta_2)
+\Big[\Delta_1 \!-\!\Delta_2\!-\!\L^2\Big]\;
\mathcal{B}_1(\Delta_1,\Delta_2,\L^2)
\bigg\}
\label{Bij rel_1}\, .
\end{eqnarray}
On the other hand, contracting the relation \eqref{Bij_PV_app} with $\delta^{ij}$, it is straightforward to find
\begin{eqnarray}
\L^2\, \mathcal{B}_{21}(\Delta_1,\Delta_2,\L^2)
 + (D\!-\!2)\, \mathcal{B}_{22}(\Delta_1,\Delta_2,\L^2)
&=&
\mathcal{A}_0(\Delta_2)
-\Delta_1 \;
\mathcal{B}_0(\Delta_1,\Delta_2,\L)
\label{Bij rel_2}\, .
\end{eqnarray}
For $D\neq 3$, the solution of the system of equations \eqref{Bij rel_1} and \eqref{Bij rel_2} is then
\begin{eqnarray}
(D\!-\!3)\, \L^2\, \mathcal{B}_{21}(\Delta_1,\Delta_2,\L^2)
&=&
\frac{(D\!-\!4)}{2}\,\mathcal{A}_0(\Delta_2)
+\Delta_1 \;
\mathcal{B}_0(\Delta_1,\Delta_2,\L)
+\frac{(D\!-\!2)}{2}\, \Big[\Delta_1 \!-\!\Delta_2\!-\!\L^2\Big]\;
\mathcal{B}_1(\Delta_1,\Delta_2,\L^2)
\label{B21 result}\\
 (D\!-\!3)\, \mathcal{B}_{22}(\Delta_1,\Delta_2,\L^2)
&=&
\frac{1}{2}\,\mathcal{A}_0(\Delta_2)
-\Delta_1 \;
\mathcal{B}_0(\Delta_1,\Delta_2,\L)
-\frac{1}{2}\,\Big[\Delta_1 \!-\!\Delta_2\!-\!\L^2\Big]\;
\mathcal{B}_1(\Delta_1,\Delta_2,\L^2)
\label{B22 result}\, .
\end{eqnarray}
Hence, both  the vector integral $\mathcal{B}^i(\Delta_1,\Delta_2,\L)$ \eqref{def_Bi} and the tensor integral $\mathcal{B}^{ij}(\Delta_1,\Delta_2,\L)$ \eqref{def_Bij} can indeed be written as a linear combination of the scalar integrals $\mathcal{A}_0(\Delta_1)$, $\mathcal{A}_0(\Delta_2)$ and $\mathcal{B}_{0}(\Delta_1,\Delta_2,\L)$.

In principle, there is no problem to generalize this method to more complicated integrals \cite{Passarino:1978jh}, with more than two denominators, more complicated numerators, and even for multiple integrals for calculations beyond NLO, even though it becomes rapidly quite cumbersome to use explicitly by hand.

\section{Calculation of the scalar master integrals \label{sec:scalInt}}

The explicit calculation of the integral $\mathcal{A}_0(\Delta)$ defined in eq. \eqref{def_A0} is a straightforward exercise and gives
\begin{eqnarray}
\mathcal{A}_0(\Delta)
&=& \Gamma\left(2\!-\! \frac{D}{2}\right)\;
    \left[\frac{\Delta}{4\pi\,\mu^2}\right]^{\frac{D}{2}-2}\label{A0_exact_result}\\
    & = & \frac{(4\pi)^{2-\frac{D}{2}}}{\left(2\!-\! \frac{D}{2}\right)}\; \Gamma\left(3\!-\! \frac{D}{2}\right)
- \log\left(\frac{\Delta}{\mu^2}\right)
+ O\left(D\!-\!4\right)\, ,
\label{A0_expand}
\end{eqnarray}
keeping the universal constants together with the $D\rightarrow 4$ pole, by analogy with the $\overline{MS}$ scheme for UV renormalization.

The integral $\mathcal{B}_0(\Delta_1,\Delta_2,\L)$ defined in eq. \eqref{def_B0} can be simplified thanks to the introduction of a Feynman parameter, as
\begin{eqnarray}
\mathcal{B}_0(\Delta_1,\Delta_2,\L)
&=& \frac{1}{\left[4\pi\,\mu^2\right]^{\frac{D}{2}-2}}\;
\Gamma\left(3\!-\! \frac{D}{2}\right)\;
\int_{0}^{1} dx
\Big[x(1\!-\!x)\L^2 +(1\!-\!x)\Delta_1 +x\Delta_2\Big]^{\frac{D}{2}-3}\\
& = &
\int_{0}^{1} dx\;
\frac{1}{\Big[x(1\!-\!x)\L^2 +(1\!-\!x)\Delta_1 +x\Delta_2\Big]}
+ O\left(D\!-\!4\right)\, .
\label{B0_expand}
\end{eqnarray}
Note that the integral $\mathcal{B}_0(\Delta_1,\Delta_2,\L)$ is UV finite for $D=4$, whereas the integral $\mathcal{A}_0(\Delta)$ has a logarithmic UV divergence.

In the case of the diagram 1 for transverse or longitudinal photon (calculated in sections \ref{sec:diag_1_T} and \ref{sec:diag_1_L} respectively), the value of the parameters $\Delta_1$, $\Delta_2$ and $\L$ are given in eqs. \eqref{def_Delta1} \eqref{def_Delta2} and \eqref{def_L} respectively, so that
\begin{eqnarray}
\mathcal{A}_0(\Delta_1)
&=& \Gamma\left(2\!-\! \frac{D}{2}\right)\;
\left(\frac{\P^2+\overline{Q}^2}{4\pi\, \mu^2}\right)^{\frac{D}{2}-2}
- \log\left(\frac{k_2^+}{k_0^+}\right)
- \log\left(\frac{k_0^+\!-\!k_2^+}{k_0^+}\right)
- \log\left(\frac{q^+}{k_1^+}\right)
+ O\left(D\!-\!4\right)\label{A0D1_result}\\
\mathcal{A}_0(\Delta_2)
&=& \Gamma\left(2\!-\! \frac{D}{2}\right)\;
\left(\frac{\overline{Q}^2}{4\pi\, \mu^2}\right)^{\frac{D}{2}-2}
- \log\left(\frac{k_1^+\!+\!k_2^+}{k_1^+}\right)
- \log\left(\frac{k_0^+\!-\!k_2^+}{k_0^+}\right)
+ O\left(D\!-\!4\right)\label{A0D2_result}
\, .
\end{eqnarray}
Moreover,
\begin{eqnarray}
\Big[x(1\!-\!x)\L^2 +(1\!-\!x)\Delta_1 +x\Delta_2\Big]
&=& \left(\frac{k_0^+\!-\!k_2^+}{k_0^+}\right)\;
\left[ x \left(\frac{k_0^+\!-\!k_2^+}{k_0^+}\right) + \frac{q^+ k_2^+}{k_0^+ k_1^+} \right]\; \Big[(1\!-\!x)\P^2 + \overline{Q}^2\Big]\, .
\end{eqnarray}
This factorization simplifies the evaluation of the integral over the Feynman parameter in \eqref{B0_expand}, and one finally obtains
\begin{eqnarray}
\mathcal{B}_0(\Delta_1,\Delta_2,\L)
&=& \frac{k_0^+}{(k_0^+\!-\!k_2^+)}\;
\frac{1}{\left[ \left(\frac{k_1^+\!+\!k_2^+}{k_1^+}\right)\P^2 + \left(\frac{k_0^+\!-\!k_2^+}{k_0^+}\right)\overline{Q}^2 \right]}\;
\Bigg\{
\log\left(\frac{\P^2+\overline{Q}^2}{\overline{Q}^2}\right)
\nonumber\\
&& \hspace{2cm}
- \log\left(\frac{k_2^+}{k_0^+}\right)
+ \log\left(\frac{k_1^+\!+\!k_2^+}{k_1^+}\right)
- \log\left(\frac{q^+}{k_1^+}\right)
\Bigg\}
+ O\left(D\!-\!4\right)\, .
\label{B0_result}
\end{eqnarray}


\section{Some useful integrals \label{sec:integrals}}

In this appendix are collected for reference the integrals which appear repeatedly when performing integrations over the gluon light-cone momentum $k_2^+$. For $0<\xi_{\min}<1$ and $R>-1$, one has
\begin{eqnarray}
\int_{0}^{1} \frac{d\xi}{\xi}\; \log(1\!+\!R\, \xi) &=& -\textrm{Li}_2\left(-R\right)
\label{Li2_int}
\\
\int_{0}^{1} d\xi\; \log(1\!+\!R\, \xi) &=& \frac{(R+1)}{R}\, \log(R+1)  -1
\\
\int_{0}^{1} d\xi\; \xi\; \log(1\!+\!R\, \xi) &=& \frac{(R+1)(R-1)}{2R^2}\, \log(R+1) +\frac{1}{2R} -\frac{1}{4}
\\
\int_{0}^{1} \frac{d\xi}{\xi}\; \log(1\!-\!\xi) &=& -\frac{\pi^2}{6}
\label{zeta2_int}
\\
\int_{0}^{1} d\xi\; \log(1\!-\!\xi) &=& -1
\\
\int_{0}^{1} d\xi\; \xi\; \log(1\!-\!\xi) &=& -\frac{3}{4}
\\
\int_{\xi_{\min}}^{1} \frac{d\xi}{\xi}\; \log(\xi) &=& -\frac{1}{2}\, \Big[\log(\xi_{\min})\Big]^2
\\
\int_{0}^{1} d\xi\; \log(\xi) &=& -1
\\
\int_{0}^{1} d\xi\; \xi\; \log(\xi) &=& -\frac{1}{4}
\, .
\end{eqnarray}
All these integrals are straightforward to calculate by standard techniques, apart from \eqref{Li2_int} which is equivalent to the definition \eqref{def_Li2} of the dilogarithm function $\textrm{Li}_2$, and
\eqref{zeta2_int}, which can be calculated by first expanding $\log(1\!-\!\xi)$ as a power series for small $\xi$, then integrating term by term, and finally recognizing $-\zeta(2)=-\pi^2/6$ in the resulting series.


\section{Fourier transforms to mixed space \label{sec:FT_int}}

Using the Schwinger trick
\begin{eqnarray}
\frac{1}{A^{\beta}}=\frac{1}{\Gamma(\beta)}\; \int_{0}^{+\infty}d\tau\; \tau^{\beta-1}\; e^{-\tau\, A}
\, ,
\label{Schwinger_trick}
\end{eqnarray}
valid for $A>0$ and $\beta>0$, one can transform the integral over $\P$ in eqs. \eqref{int_FT_WF_gamma_L_LO} and \eqref{int_FT_WF_gamma_T_LO} into a Gaussian integral. After performing it, one gets
\begin{eqnarray}
 \int \frac{d^{D-2} \P}{(2\pi)^{D-2}}\;
 e^{i\P \cdot \x_{01}}\;
  \frac{1}{\left[\P^2\!+\!\overline{Q}^2\right]}
 &=&  \left(4\pi\right)^{1-\frac{D}{2}}\;
 \int_{0}^{+\infty}d\tau\; \tau^{1-\frac{D}{2}}\; e^{-\tau\, \overline{Q}^2}\;
    e^{-\frac{{\x_{01}}^2}{4\tau}}
\label{int_FT_WF_gamma_L_LO_1}
\nonumber\\
 \int \frac{d^{D-2} \P}{(2\pi)^{D-2}}\;
 e^{i\P \cdot \x_{01}}\;
  \frac{\P^j}{\left[\P^2\!+\!\overline{Q}^2\right]}
 &=& \frac{i}{2}\, \x_{01}^j\;
  \left(4\pi\right)^{1-\frac{D}{2}}\;
 \int_{0}^{+\infty}d\tau\; \tau^{-\frac{D}{2}}\; e^{-\tau\, \overline{Q}^2}\;
    e^{-\frac{{\x_{01}}^2}{4\tau}}
 \, .
\label{int_FT_WF_gamma_T_LO_1}
\end{eqnarray}
Then, the results \eqref{int_FT_WF_gamma_L_LO} and \eqref{int_FT_WF_gamma_T_LO} are obtained thanks to the formula
\begin{eqnarray}
\int_{0}^{+\infty}d\tau\; \tau^{\nu-1}\;
    e^{-\frac{B}{\tau}}\; e^{-\tau\, C}
&=& 2 \left(\frac{B}{C}\right)^{\frac{\nu}{2}}\;
      \textrm{K}_{-\nu}\Big(2\sqrt{BC}\Big)
\label{BesselK_int_rep}
\end{eqnarray}
valid for $B>0$ and $C>0$, found in ref. \cite{gradshteyn2000table}.

Using the standard trick
\begin{eqnarray}
\log(A)=\lim_{\alpha\rightarrow 0} \partial_{\alpha}\; A^{\alpha}
\label{log_trick}
\end{eqnarray}
as well, one can rewrite the integral in eq. \eqref{int_FT_WF_gamma_L_A} as
\begin{eqnarray}
&& \int \frac{d^{D-2} \P}{(2\pi)^{D-2}}\;
 e^{i\P \cdot \x_{01}}\;
  \frac{1}{\left[\P^2\!+\!\overline{Q}^2\right]}\;
  \log\left(\frac{\P^2\!+\!\overline{Q}^2}{\overline{Q}^2}\right)
 = \left(4\pi\right)^{1-\frac{D}{2}}\;
     \lim_{\alpha\rightarrow 0} \partial_{\alpha}
     \left\{
     \frac{\left[\overline{Q}^2\right]^{-\alpha}}{\Gamma(1-\alpha)}\;
     \int_{0}^{+\infty}d\tau\; \tau^{1-\frac{D}{2}-\alpha}\; e^{-\tau\, \overline{Q}^2}\;
    e^{-\frac{{\x_{01}}^2}{4\tau}}
     \right\}
\nonumber\\
&& \hspace{1cm}= \left(4\pi\right)^{1-\frac{D}{2}}\;
     \lim_{\alpha\rightarrow 0} \partial_{\alpha}
     \left\{
     \frac{\left[\overline{Q}^2\right]^{-\alpha}}{\Gamma(1-\alpha)}\;
     2
     \left(\frac{2\overline{Q}}{|\x_{01}|}\right)^{\frac{D}{2}-2+\alpha}
 \textrm{K}_{\frac{D}{2}-2+\alpha}\Big(|\x_{01}|\, \overline{Q}\Big)
     \right\}
\nonumber\\
&& \hspace{1cm}= \frac{1}{2\pi}\; \left(\frac{\overline{Q}}{2\pi|\x_{01}|}\right)^{\frac{D}{2}-2}\;
\left\{
    \left[-\frac{1}{2}\; \log\left(\frac{{\x_{01}}^2\,
            \overline{Q}^2}{4}\right)+\Psi(1)
    \right]\;
    \textrm{K}_{\frac{D}{2}-2}\Big(|\x_{01}|\, \overline{Q}\Big)
  +\lim_{\alpha\rightarrow 0} \partial_{\alpha}
     \textrm{K}_{\frac{D}{2}-2+\alpha}\Big(|\x_{01}|\, \overline{Q}\Big)
     \right\}
\, .
\label{int_FT_WF_gamma_L_A_bis}
\end{eqnarray}
However, for $z>0$, $\textrm{K}_{\nu}(z)$ is an analytic function of $\nu$. Moreover, it is even in $\nu$: $\textrm{K}_{\nu}(z)=\textrm{K}_{-\nu}(z)$. Hence, for $D\rightarrow 4$, one has
\begin{eqnarray}
\lim_{\alpha\rightarrow 0} \partial_{\alpha}
     \textrm{K}_{\frac{D}{2}-2+\alpha}\Big(|\x_{01}|\, \overline{Q}\Big)
= \lim_{\alpha\rightarrow 0} \partial_{\alpha}
     \textrm{K}_{\alpha}\Big(|\x_{01}|\, \overline{Q}\Big)
     + O\left(D\!-\!4\right)
=   O\left(D\!-\!4\right)
\, ,
\end{eqnarray}
which gives the result \eqref{int_FT_WF_gamma_L_A}.

Following the same steps, the integral in eq. \eqref{int_FT_WF_gamma_T_A} can be written as
\begin{eqnarray}
&& \int \frac{d^{D-2} \P}{(2\pi)^{D-2}}\;
 e^{i\P \cdot \x_{01}}\;
  \frac{\P^j}{\left[\P^2\!+\!\overline{Q}^2\right]}\;
  \log\left(\frac{\P^2\!+\!\overline{Q}^2}{\overline{Q}^2}\right)
  \nonumber\\
&=& \frac{i}{2}\, \x_{01}^j\;\left(4\pi\right)^{1-\frac{D}{2}}\;
     \lim_{\alpha\rightarrow 0} \partial_{\alpha}
     \left\{
     \frac{\left[\overline{Q}^2\right]^{-\alpha}}{\Gamma(1-\alpha)}\;
     2
     \left(\frac{2\overline{Q}}{|\x_{01}|}\right)^{\frac{D}{2}-1+\alpha}
 \textrm{K}_{\frac{D}{2}-1+\alpha}\Big(|\x_{01}|\, \overline{Q}\Big)
     \right\}
\nonumber\\
&=& i\, \x_{01}^j\; \left(\frac{\overline{Q}}{2\pi|\x_{01}|}\right)^{\frac{D}{2}-1}\;
\left\{
    \left[-\frac{1}{2}\; \log\left(\frac{{\x_{01}}^2\,
            \overline{Q}^2}{4}\right)+\Psi(1)
    \right]\;
    \textrm{K}_{\frac{D}{2}-1}\Big(|\x_{01}|\, \overline{Q}\Big)
  +\lim_{\alpha\rightarrow 0} \partial_{\alpha}
     \textrm{K}_{\frac{D}{2}-1+\alpha}\Big(|\x_{01}|\, \overline{Q}\Big)
     \right\}
\, .
\label{int_FT_WF_gamma_T_A_bis}
\end{eqnarray}
The last term can be evaluated thanks to the formula (see ref. \cite{gradshteyn2000table})
\begin{eqnarray}
z\, \textrm{K}_{\nu+1}(z)=\nu\, \textrm{K}_{\nu}(z)
-z\partial_z \textrm{K}_{\nu}(z)
\, .
\label{recurrence_BesselK}
\end{eqnarray}
Indeed, it gives
\begin{eqnarray}
\lim_{\alpha\rightarrow 0} \partial_{\alpha}
     \textrm{K}_{\frac{D}{2}-1+\alpha}(z)
&=& \frac{1}{z}\, \textrm{K}_{\frac{D}{2}-2}(z)
    +\left[\frac{\frac{D}{2}\!-\!2}{z}-\partial_z\right]\,
     \lim_{\alpha\rightarrow 0} \partial_{\alpha}
     \textrm{K}_{\frac{D}{2}-2+\alpha}(z)\nonumber\\
&=& \frac{1}{z}\, \textrm{K}_{0}(z)
    +    O\left(D\!-\!4\right)\, .
  \label{rec_BesselK_T_A}
\end{eqnarray}
Inserting \eqref{rec_BesselK_T_A} into \eqref{int_FT_WF_gamma_T_A_bis}, one finds the result announced in eq. \eqref{int_FT_WF_gamma_T_A}.

The last Fourier transform needed is the integral in eq. \eqref{int_FT_WF_gamma_T_1}. Using the tricks \eqref{log_trick} and then \eqref{Schwinger_trick} for each denominator, and performing the Gaussian integral, one finds
\begin{eqnarray}
&& \int \frac{d^{D-2} \P}{(2\pi)^{D-2}}\;
 e^{i\P \cdot \x_{01}}\;
  \frac{\P^j}{\left[\P^2\!+\!\overline{Q}^2\right]}\;
  \frac{\left[\P^2\!+\!\overline{Q}^2\right]}{\P^2}\;
  \log\left(\frac{\P^2\!+\!\overline{Q}^2}{\overline{Q}^2}\right)
  \nonumber\\
&=& \frac{i}{2}\, \x_{01}^j\;\left(4\pi\right)^{1-\frac{D}{2}}\;
     \lim_{\alpha\rightarrow 0} \partial_{\alpha}
     \left\{
     \frac{\left[\overline{Q}^2\right]^{-\alpha}}{\Gamma(-\alpha)}\;
     \int_{0}^{+\infty}d\tau\; \tau^{-\alpha-1}\; e^{-\tau\, \overline{Q}^2}\;
     \int_{0}^{+\infty}d\sigma\; (\tau+\sigma)^{-\frac{D}{2}}
    e^{-\frac{{\x_{01}}^2}{4(\tau+\sigma)}}
    \right\}
  \nonumber\\
&=&    \frac{i}{2}\, \x_{01}^j\;\left(4\pi\right)^{1-\frac{D}{2}}\;
     \lim_{\alpha\rightarrow 0} \partial_{\alpha}
     \left\{
     \frac{\left[\overline{Q}^2\right]^{-\alpha}}{\Gamma(-\alpha)}\;
     \int_{0}^{+\infty}d\tau\; \tau^{-\alpha-1}\; e^{-\tau\, \overline{Q}^2}\;
     \int_{\tau}^{+\infty}d\eta\; \eta^{-\frac{D}{2}}
    e^{-\frac{{\x_{01}}^2}{4\eta}}
    \right\}
  \nonumber\\
&=&    \frac{i}{2}\, \x_{01}^j\;\left(4\pi\right)^{1-\frac{D}{2}}\;
     \lim_{\alpha\rightarrow 0} \partial_{\alpha}
     \Bigg\{
     \frac{\left[\overline{Q}^2\right]^{-\alpha}}{\Gamma(-\alpha)}\;
     \int_{0}^{+\infty}d\tau\; \tau^{-\alpha-1}\; e^{-\tau\, \overline{Q}^2}\;
     \Bigg[
     \int_{0}^{+\infty}d\eta\; \eta^{-\frac{D}{2}}
    e^{-\frac{{\x_{01}}^2}{4\eta}}
 \nonumber\\
&&  \hspace{9cm}
    -\int_{0}^{\tau}d\eta\; \eta^{-\frac{D}{2}}
    e^{-\frac{{\x_{01}}^2}{4\eta}}
    \Bigg]
    \Bigg\}
  \nonumber\\
&=&    \frac{i}{2}\, \x_{01}^j\;\left(4\pi\right)^{1-\frac{D}{2}}\;
     \lim_{\alpha\rightarrow 0} \partial_{\alpha}
     \Bigg\{\frac{\left[\overline{Q}^2\right]^{-\alpha}}{\Gamma(-\alpha)}\;
      \frac{\Gamma(-\alpha)}{\left[\overline{Q}^2\right]^{-\alpha}}\;
     \Gamma\left(\frac{D}{2}\!-\!1\right)\,
     \left(\frac{{\x_{01}}^2}{4}\right)^{1-\frac{D}{2}}\;
 \nonumber\\
&&  \hspace{4cm}
-\frac{\left[\overline{Q}^2\right]^{-\alpha}}{\Gamma(-\alpha)}\;
     \int_{0}^{+\infty}d\tau\; \tau^{-\alpha-1}\; e^{-\tau\, \overline{Q}^2}\;
      \int_{0}^{\tau}d\eta\; \eta^{-\frac{D}{2}}
    e^{-\frac{{\x_{01}}^2}{4\eta}}
    \Bigg\}
  \nonumber\\
&=&    \frac{i}{2}\, \x_{01}^j\;\left(4\pi\right)^{1-\frac{D}{2}}\;
     \lim_{\alpha\rightarrow 0} \partial_{\alpha}
     \Bigg\{\frac{\alpha}{\Gamma(1-\alpha)}\; \left[\overline{Q}^2\right]^{-\alpha}\;
     \int_{0}^{+\infty}d\tau\; \tau^{-\alpha-1}\; e^{-\tau\, \overline{Q}^2}\;
      \int_{0}^{\tau}d\eta\; \eta^{-\frac{D}{2}}
    e^{-\frac{{\x_{01}}^2}{4\eta}}
    \Bigg\}
\, .
\label{int_FT_WF_gamma_T_1_bis}
\end{eqnarray}
For $\overline{Q}>0$ and $|\x_{01}|>0$, the exponential factors make the double integral over $\tau$ and $\eta$ convergent for any value of $\alpha$ and $D$. Hence, that double integral is an analytic function in both $\alpha$ and $D$, so that
\begin{eqnarray}
&& \int \frac{d^{D-2} \P}{(2\pi)^{D-2}}\;
 e^{i\P \cdot \x_{01}}\;
  \frac{\P^j}{\left[\P^2\!+\!\overline{Q}^2\right]}\;
  \frac{\left[\P^2\!+\!\overline{Q}^2\right]}{\P^2}\;
  \log\left(\frac{\P^2\!+\!\overline{Q}^2}{\overline{Q}^2}\right)
  \nonumber\\
&=&
   \frac{i}{2}\, \x_{01}^j\;\left(4\pi\right)^{1-\frac{D}{2}}\;
     \Bigg\{
     \int_{0}^{+\infty}\frac{d\tau}{\tau}\;  e^{-\tau\, \overline{Q}^2}\;
      \int_{0}^{\tau}\frac{d\eta}{\eta^2}\;
    e^{-\frac{{\x_{01}}^2}{4\eta}}
    +  O\left(D\!-\!4\right)
    \Bigg\}
  \nonumber\\
&=&
   \frac{i}{2}\, \x_{01}^j\;\left(4\pi\right)^{1-\frac{D}{2}}\;
     \Bigg\{\frac{8}{{\x_{01}}^2}\;
     \textrm{K}_{0}\Big(|\x_{01}|\, \overline{Q}\Big)
    +  O\left(D\!-\!4\right)
    \Bigg\}
\, ,
\label{int_FT_WF_gamma_T_1_ter}
\end{eqnarray}
which is equivalent with the result given in eq. \eqref{int_FT_WF_gamma_T_1}.


\bibliography{MaBiblioHEQCD}


\end{document}